\documentclass[11pt, letterpaper]{article}
\usepackage{tcolorbox}     
\tcbuselibrary{breakable}
\usepackage[utf8]{inputenc}
\usepackage[T1]{fontenc}
\usepackage{amsmath, amssymb}
\usepackage[margin=1in]{geometry}
\usepackage{graphicx} 
\usepackage{subcaption} 
\usepackage{lipsum} 
\usepackage{tikz}
\usepackage{booktabs}
\usepackage{tcolorbox}
\usepackage[explicit]{titlesec} 
\usepackage{tabularx}
\usepackage{booktabs}
\usepackage{wrapfig}
\usepackage{multirow}
\usepackage{subcaption}
\usepackage{threeparttable} 

\usepackage{listings}
\usepackage{epigraph}
\newtcolorbox{promptbox}{
  colback=white,
  colframe=black,
  boxrule=0.6pt,
  arc=2pt,
  left=6pt,
  right=6pt,
  top=6pt,
  bottom=6pt,
  fontupper=\small,
  breakable
}

\usetikzlibrary{
    shapes.geometric, 
    arrows.meta,      
    positioning,      
    calc,              
    intersections,     
    fit              
}

\usepackage[authoryear]{natbib}

\usepackage[colorlinks=true, urlcolor=blue, linkcolor=blue, citecolor=blue]{hyperref}

\title{\textbf{Lost Before Translation: Social Information Transmission and Survival in AI-AI Communication}}
\bigskip
\author{
Bijean Ghafouri\footnote{Corresponding author: bghafour@usc.edu} \\ University of Southern California \and
Emilio Ferrara \\
University of Southern California
}

\date{\today}

\begin{document}

\maketitle

\begin{abstract}
When AI systems summarize and relay information, they inevitably transform it. But how? We introduce an experimental paradigm based on the telephone game to study what happens when AI talks to AI. Across five studies tracking content through AI transmission chains, we find three consistent patterns. The first is convergence, where texts differing in certainty, emotional intensity, and perspectival balance collapse toward a shared default of moderate confidence, muted affect, and analytical structure. The second is selective survival, where narrative anchors persist while the texture of evidence, hedges, quotes, and attributions is stripped away. The third is competitive filtering, where strong arguments survive while weaker but valid considerations disappear when multiple viewpoints coexist. In downstream experiments, human participants rated AI-transmitted content as more credible and polished. Importantly, however, humans also showed degraded factual recall, reduced perception of balance, and diminished emotional resonance. We show that the properties that make AI-mediated content appear authoritative may systematically erode the cognitive and affective diversity on which informed judgment depends.
\end{abstract}

\newpage
    \epigraph{``The problem of education in the modern world lies in the fact that by its
very nature it cannot forgo either authority or tradition, and yet must proceed
in a world that is neither structured by authority nor held together by tradition.''}{Hannah Arendt, \emph{The crisis in education (1961)}
    }

\section*{Introduction}

In 1973, the Italian songwriter Adriano Celentano appeared on television to perform \textit{Prisencolinensinainciusol}, a song written entirely in a fabricated language. Although the lyrics contain no real words, Celentano mimicked the sounds and rhythms of American English, producing speech that felt fluent while conveying no literal meaning. The performance was intended not to confuse, but to demonstrate that fluency, affect, and structure can create the appearance of understanding even in the absence of semantic content. Celentano later described the song as a reflection on modern society: a world saturated with speech yet marked by persistent misunderstanding, where incomprehensibility itself becomes recognizable, emotionally legible, and unrestrained by shared meaning.

This article begins from a related concern. Language models process text fluently without comprehending it, yet agentic systems increasingly mediate how information is summarized, translated, reformulated, and transmitted \citep{imas2025agentic, rothschild2025agentic, bansal2025magentic, mollick2024co, olteanu2025ai, coppolillo2026harm}. News reports are condensed by one model, rewritten by another, reformatted for a different audience by a third. The human reader encounters only the final output of a longer chain of machine transformation. As this architecture becomes ubiquitous, a growing portion of the informational environment operates through processes that remain empirically opaque \citep{Ferrara2026GenerativeAIParadox, ferrara2024butterfly, ilievski2025aligning, burton2024large}.

We possess little understanding of how information changes as it moves across sequences of artificial agents. Existing research evaluates language models as generators, classifiers, or predictors of human responses, yet offers limited insight into what happens when information is repeatedly transformed by AI before reaching a human mind. In the study of human communication, by contrast, social scientists have long documented how information acquires predictable properties as it spreads through populations: retellings simplify and homogenize content \citep{mesoudi2008multiple, allport1947psychology, bartlett1995remembering}, expressed uncertainty diminishes through repetition \citep{hasher1977frequency, fazio2020repetition}, emotional content enjoys transmission advantages \citep{berger2012makes, ferrara2015measuring, brady2017emotion}, and claims lose connection to their sources \citep{hovland1951influence}. These regularities arise from the structure of transmission itself, allowing scholars to study communication as a population-level process \citep{shardanand1995social, carlson2018modeling}.

No comparable framework exists for AI-mediated communication in which intelligent systems possess increasing agency \citep{brinkmann2023machine, kim2025reflective, weidinger2025toward, kasirzadeh2025characterizing, kasirzadeh2026many}. Language models clearly alter information, yet the nature, and more importantly its social consequences, of that alteration under iteration remains uncharacterized. When the same transformation is applied repeatedly across independent chains, does information converge toward common representations? Drift toward simplified forms? Stabilize around particular features while others erode? The field currently lacks evidence to distinguish among these possibilities.

This absence matters. If AI-mediated transmission compresses variation \citep{sourati2025shrinking, sourati2025homogenizing, ghafouri2025theory}, diverse human contributions may reach audiences in narrowed form. If confidence accumulates across iterations, tentative claims may emerge as settled conclusions. If attribution erodes faster than content, information may circulate without the contextual cues required for evaluation. These outcomes would follow from the structural properties of chained systems rather than from any explicit design choice or human intention.

Our aim is to examine how information evolves under repeated transformation across artificial agents, and to identify whether reproducible regularities emerge.
Building off recent work \citep{acerbi2023large, zhao2025saying, perez2025llms}, we introduce an experimental paradigm that isolates iterative transformation under controlled conditions: diverse initial texts are passed through independent chains of language models, each applying the same instruction at every step. By tracking complete transmission trajectories and comparing endpoints across chains, we assess whether information converges, stabilizes, or disperses, and identify which features of content prove robust or fragile under the social transmission between artificial agents.

We apply this paradigm across five studies, each isolating a distinct dimension of information quality. Study 1 examines factual content, tracking the survival of 26 annotated information elements. Study 2 examines epistemic calibration, testing whether texts varying in expressed certainty converge toward a common attractor. Study 3 examines multi-perspective content, measuring whether balanced presentations of competing viewpoints retain their deliberative character or crystallize into analytical frameworks. Study 4 examines political argumentation, assessing whether frames differing in persuasive strength survive equally or undergo differential selection. Study 5 examines emotional expression, tracking how intensity and discrete emotions evolve under transmission. Across these studies, we document consistent patterns: rapid initial filtering followed by stabilization, convergence toward moderate defaults, selective preservation of structural anchors over epistemic texture, and systematic muting of emotional intensity, particularly for morally charged negative emotions.

These AI-level transformations would matter little if they left no trace on human understanding. We therefore complement each transmission study with a human evaluation experiment, recruiting participants to read either original content or content that has passed through 100 iterations of AI-AI transmission. The results reveal a consistent dissociation: transmitted content is perceived as more polished and credible, yet delivers degraded factual information, reduced perspectival diversity, and diminished emotional resonance. AI-AI transmission thus functions not as neutral relay but as an information filter, one that shapes content in ways that may go undetected, or even be rewarded, by human readers.

\section*{The AI-AI Transmission Chain}

To study how information changes as it moves across artificial agents, we treat AI-to-AI interaction as an iterative process acting on observable outputs. Rather than modeling internal representations or optimization objectives, we focus on how text is repeatedly transformed under fixed constraints.

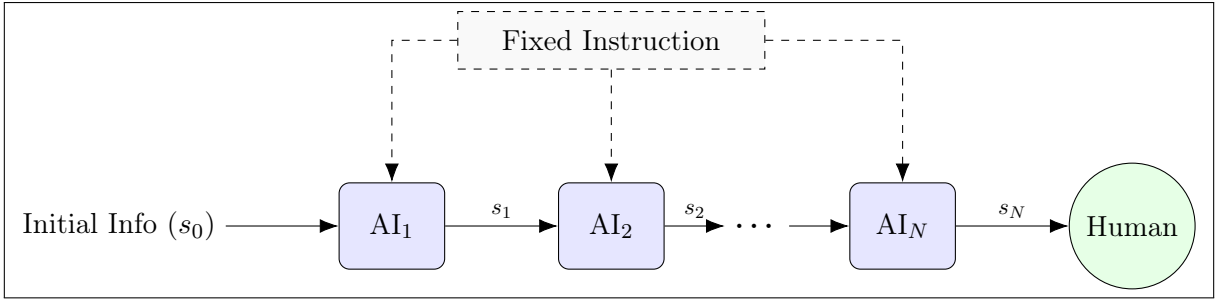
\begin{figure}[ht]
    \centering
    \fbox{%
    \begin{tikzpicture}[
        node distance=1.5cm,
        auto,
        block/.style={
            rectangle, 
            draw, 
            fill=blue!10, 
            text width=3em, 
            text centered, 
            rounded corners, 
            minimum height=3em
        },
        human/.style={
            circle, 
            draw, 
            fill=green!10, 
            text width=3.5em, 
            text centered, 
            minimum height=3.5em
        },
        instruction/.style={
            rectangle, 
            draw, 
            dashed, 
            fill=gray!5, 
            text width=10em, 
            text centered, 
            minimum height=2em
        },
        line/.style={draw, -{Latex[length=2.5mm]}}
    ]

    
    \node (input) {Initial Info ($s_0$)};
    
    \node [block, right=of input] (ai1) {AI$_1$};
    \node [block, right=of ai1] (ai2) {AI$_2$};
    \node [right=0.8cm of ai2] (dots) {\textbf{\dots}};
    \node [block, right=0.8cm of dots] (ain) {AI$_N$};
    
    \node [human, right=of ain] (human) {Human};

    \node [instruction, above=1.5cm of ai2] (inst) {Fixed Instruction};


    \path [line] (input) -- (ai1);
    \path [line] (ai1) -- node[above, scale=0.8] {$s_1$} (ai2);
    \path [line] (ai2) -- node[above, scale=0.8] {$s_2$} (dots);
    \path [line] (dots) -- (ain);
    \path [line] (ain) -- node[above, scale=0.8] {$s_N$} (human);

    \draw [line, dashed] (inst) -| (ai1);
    \draw [line, dashed] (inst) -- (ai2);
    \draw [line, dashed] (inst) -| (ain);

    \end{tikzpicture}
    }
    \caption{Experimental design: iterative AI--AI transmission under uniform constraint, terminating in human evaluation. $s_i$ represents the observable text at step $i$.}
    \label{fig:experiment_design}
\end{figure}

Formally, we represent a language model as a stochastic operator $\mathbb{T}$ that maps an input signal $s_t$ to a subsequent signal $s_{t+1}$ under a fixed constraint $\mathcal{C}$ (e.g., a task instruction or system prompt):
\[
s_{t+1} \sim \mathbb{T}(s_t \mid \mathcal{C})
\]
The operator induces a probability distribution over outputs conditional on the current signal and the imposed constraint. Although model parameters remain fixed, stochastic decoding introduces variability across realizations.

The central object of analysis is the \textit{transmission chain}: an ordered sequence of signals produced through repeated application of the same transformation.
\[
\tau = (s_0, s_1, s_2, \dots, s_N)
\]
Here $s_0$ is the initial input, and each subsequent element is generated by applying the operator under identical constraints. The final element $s_N$ represents the output encountered by a human observer after $N$ stages of machine mediation. Although each agent is stateless, the signal $s_t$ itself serves as the sole memory of the chain. Cumulative effects emerge not from agent memory, but from the signal becoming increasingly structured at each step, which in turn increases the probability that subsequent transformations produce similarly structured output.

Our interest lies in how properties of the signal evolve as chain depth increases. By holding the transformation rule constant, systematic changes in the trajectory reflect properties of the transmission process itself. This allows us to examine whether repeated transformation produces convergence, stabilization, or drift in observable content, and to identify which features of information prove robust or fragile under iteration.

\section*{Overview of Experimental Design}

We study AI-AI information transmission through iterative transformation under uniform constraint. In each experiment, information passes through a chain of $N$ language model agents, where each agent receives only the output of the previous agent and a fixed instruction that remains constant across all steps. We focus exclusively on observable trajectories, i.e. the sequence of texts produced at each step.

This design isolates the cumulative effects of iteration from the heterogeneity of individual transformations. Any structure that emerges must arise from repeated application of the same process, not from variation across agents or instructions.

All experiments terminate with human evaluation. Although our primary object of study is AI-AI transmission, the social relevance of these dynamics lies in their downstream consequences for human understanding. We therefore append a human interpretation stage to each chain, measuring how upstream transformations shape perception, trust, and judgment downstream.

\subsection*{Chain Construction}

Each experimental chain consists of $N = 100$ sequential transmissions. At step $t$, agent $t$ receives text $s_{t-1}$ and instruction $\mathcal{C}$, producing output $s_t$. Each agent is instantiated as a fresh, stateless API call with no memory of previous iterations\footnote{All experiments use Gemini 3.0 Flash with default temperature set at 1.}. For each experimental condition, we run $M = 100$ independent chain replications.

\subsection*{Transmission and Recovery Instructions}

All studies use the same transmission instruction $\mathcal{C}_{\text{transmit}}$:

\begin{quote}
\texttt{You've received this information from an AI agent. Share the important parts of this information with the next AI agent. Just write your message.}

\texttt{\{TEXT\}}
\end{quote}

This instruction was selected to minimize additional framing while ensuring the model produces a coherent relay of the input content.

At the endpoint of each chain ($t = 100$), a recovery instruction $\mathcal{C}_{\text{recover}}$ assesses what information remains accessible when the final output is prepared for human consumption:

\begin{quote}
\texttt{You've received this information from an AI agent. Share it to a human reader, in one version. Just write your message.}

\texttt{\{TEXT\}}
\end{quote}

The recovery phase tests whether information that persists through AI-AI transmission can be successfully surfaced when the chain terminates in human-directed output. We denote the recovery output as $s_{101}$ to distinguish it from the final transmission state $s_{100}$.

We apply this paradigm across five studies examining: (1) information decay and element survival, (2) expressed uncertainty, (3) multi-perspective content, (4) competing political frames, and (5) emotional content. Each study isolates a different dimension of information transformation under AI-AI transmission.

\section*{Dimensions of Information Quality}

The question of how information changes under AI-AI transmission admits many operationalizations. Information can be characterized by its factual content, its epistemic stance, its representational diversity, its argumentative structure, its emotional register, and countless other features. We cannot study all dimensions simultaneously. Instead, we select five that prior research identifies as consequential for human judgment and democratic discourse.

Each dimension represents a distinct way that information quality could degrade, or be preserved, under repeated transformation. Factual content determines whether audiences can learn true things about the world. Expressed certainty shapes how confident readers become in claims and shapes their trust, independent of their accuracy. Perspectival diversity determines whether audiences encounter the range of viewpoints that inform deliberation. Argumentative balance affects whether competing positions receive fair hearing or whether some frames systematically dominate others. Emotional valence carries signals that motivate engagement, sharing, and action. Together, these dimensions characterize not just whether information changes under AI-AI transmission, but what kinds of information prove robust or fragile.

\begin{table}[h]
\centering
\small
\begin{tabularx}{\linewidth}{c l l X}
\toprule
\textbf{Study} & \textbf{Dimension} & \textbf{What's at Risk} & \textbf{Core Question} \\
\midrule
1 & Factual content & Specific facts, figures, names, dates & Do facts survive transmission? \\
2 & Epistemic stance & Calibration of expressed certainty & Does confidence track evidence? \\
3 & Perspectival diversity & Representation of multiple viewpoints & Does pluralism survive? \\
4 & Argumentative structure & Balance among competing frames & Do weak positions survive competition? \\
5 & Emotional character & Affective signals and valence & Which emotions survive? \\
\bottomrule
\end{tabularx}
\caption{Five dimensions of information quality examined across studies. Each dimension represents a distinct property consequential for human judgment that may be preserved or degraded under AI--AI transmission.}
\label{tab:study_dimensions}
\end{table}

This selection also allows us to probe whether AI-AI transmission imposes implicit norms on the information it carries. Human transmission chains exhibit well-documented biases, such as simplification, confidence inflation, emotional amplification, source loss, etc. AI systems may exhibit analogous biases, or qualitatively different ones shaped by training objectives such as helpfulness, harmlessness, and honesty. By examining multiple dimensions, we assess whether AI-AI transmission is approximately neutral, preserving the character of inputs, or whether it systematically filters, transforms, or homogenizes content in ways that impact what human audiences receive.

Studies 1 and 2 examine properties of individual claims: their factual content and their certainty and epistemic framing. Studies 3 and 4 examine properties of multi-voiced texts: whether diverse perspectives and competing arguments survive when bundled together. Study 5 examines affective properties that operate alongside, and sometimes independently of, semantic content. Across all five, we apply the same transmission paradigm, allowing for a comparison of how different information types respond to identical transformation pressures.

\newpage
\section*{Study 1: Information Decay and Survival}

Information transmission through AI systems may not preserve all content equally. Iterative transmission may act as a filter, systematically retaining some information types while eliminating others. If such differential survival occurs, it would have significant implications for human-AI information ecosystems: downstream recipients would receive text that appears coherent, yet has been selectively depleted of certain content.


\begin{itemize}
    \item \textbf{RQ1:} How much information is lost through iterative AI-AI transmission, and what is the shape of the decay curve?
    \item \textbf{RQ2:} Can information lost through AI-AI transmission be recovered when re-expanded for human consumption?
    \item \textbf{RQ3:} Do different types of information (e.g., names, numbers, dates, hedges) decay at different rates, and if so, what hierarchy emerges?
    \item \textbf{RQ4:} Does semantic similarity and text length remain stable despite element-level information loss?
\end{itemize}

The source text $s_0$ is a synthetic news article (195 words) designed to contain diverse information types:

\begin{promptbox}
\begin{quote}
\itshape
The City of Riverside announced Tuesday that it will invest \$4.7 million to renovate three public libraries over the next two years. The project, approved by a 6-3 vote of the City Council, is expected to be completed by September 2026.

Library Director Susan Park called the investment ``a turning point for public education in our community.'' She noted that the renovations would add 12,000 square feet of space across all three locations and create an estimated 35 new jobs.

Councilman Robert Tran, who voted against the measure, argued that the timeline was unrealistic and the budget likely to increase. ``We're setting ourselves up for cost overruns,'' he said, pointing to a 2019 parks project that exceeded its budget by 40\%.

A recent survey found that 62\% of residents support the renovation, while 27\% oppose it and 11\% are undecided. Mayor Elena Vasquez has indicated she will sign the measure, though she acknowledged concerns about competing infrastructure priorities.

Construction is scheduled to begin in January at the downtown branch, with the other two locations following in phases.
\end{quote}
\end{promptbox}

We track survival of 26 pre-annotated elements across 12 categories, grouped into four theoretically motivated tiers:

\begin{table}[h]
\centering
\begin{tabular}{llll}
\toprule
\textbf{Type} & \textbf{Category} & \textbf{Elements} & \textbf{Count} \\
\midrule
\multirow{4}{*}{Narrative anchors} & Person & Susan Park, Robert Tran, Elena Vasquez & 3 \\
 & Place & Riverside & 1 \\
 & Organization & City Council & 1 \\
 & Money & \$4.7 million & 1 \\
\midrule
Temporal markers & Date & Tuesday, September 2026, 2019, January & 4 \\
\midrule
\multirow{5}{*}{Evidentiary details} & Quantity & 3 libraries, 12,000 sq ft, 35 jobs & 3 \\
 & Duration & 2 years & 1 \\
 & Vote & 6-3 & 1 \\
 & Percentage & 62\%, 27\%, 11\%, 40\% & 4 \\
 & Quote & ``turning point...'', ``cost overruns...'' & 2 \\
\midrule
\multirow{2}{*}{Epistemic qualifiers} & Claim & public education, unrealistic timeline, competing priorities & 3 \\
 & Hedge & ``likely to increase'', ``acknowledged concerns'' & 2 \\
\midrule
\textbf{Total} & & & \textbf{26} \\
\bottomrule
\end{tabular}
\caption{Annotated elements tracked in Study 1}
\label{tab:study1_elements}
\end{table}

\subsubsection*{Procedure}
We pass the source text $s_0$ through a chain of $N=100$ sequential AI agents using the fixed transmission prompt $I_{\text{transmit}}$. This yields the full trajectory $\{s_0, s_1, \ldots, s_{100}\}$. We then take the endpoint $s_{100}$ and apply a recovery step using $I_{\text{recover}}$, producing $s_{\text{rec}}$. We repeat the entire transmission-and-recovery procedure for $M=100$ independent runs.

We evaluate information change at both the element and text levels. At the element level, we track the survival of pre-annotated information units across transmission steps, computing survival curves over iteration  and measuring recovery success as the change in element count between the final transmission state $s_{100}$ and the recovered text $s_{\text{rec}}$. At the text level, we assess whether surface coherence is preserved despite element loss by measuring semantic similarity between each intermediate text $s_t$ and the original source $s_0$ in embedding space, as well as total word count at each iteration.


\subsection*{Results: Information Decay and Differential Survival}


\begin{figure}[!t]
    \centering
    \includegraphics[width=0.8\linewidth]{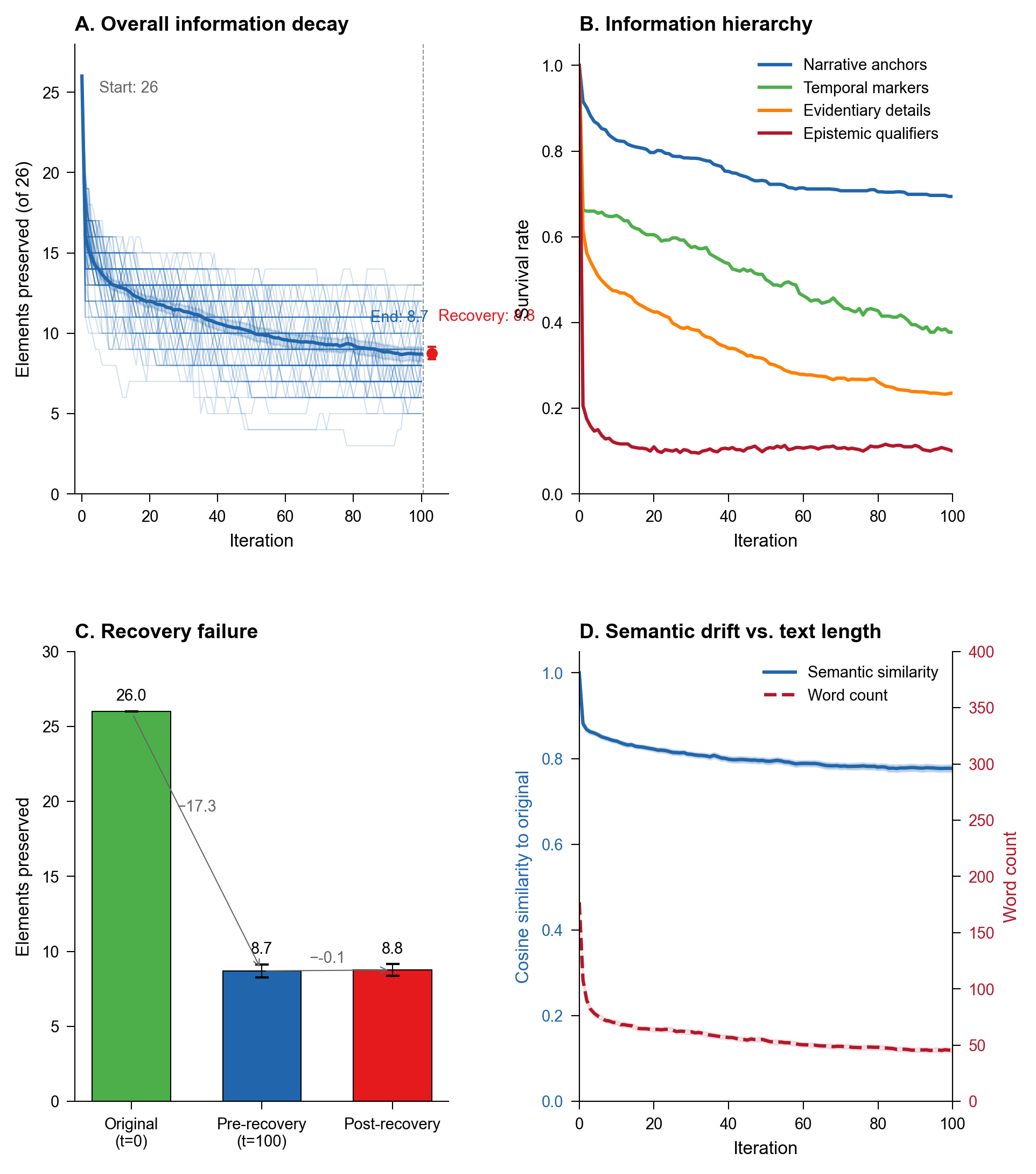}
    \caption{Information decay dynamics in AI-AI transmission chains. (A) Total information elements preserved across 100 iterations, showing rapid initial decay followed by stabilization at approximately 9 elements. Individual runs shown in light blue; mean with 95\% CI in dark blue; recovery phase in red. (B) Survival trajectories by information type, revealing a clear hierarchy. Narrative anchors persist while evidentiary details and epistemic qualifiers decay rapidly. (C) Recovery phase comparison showing minimal change when transitioning from AI-AI to AI-human output. (D) Semantic similarity to original (blue) declines moderately while word count (red dashed) drops substantially, indicating compression rather than substitution.}
    \label{fig:study1_main}
\end{figure}

\subsubsection*{RQ1: Overall Information Loss}

Information decayed substantially but not completely across transmission (Figure~\ref{fig:study1_main}A)\footnote{Throughout all studies, 95\% confidence intervals were computed as mean ± 1.96 × SEM across independent runs (N = 100 per condition). Error bars in figures represent ± 1 SEM unless otherwise noted.}. The original text contained all 26 elements; by iteration 100, 8.8 elements survived on average (34\% retention). Decay followed a two-phase pattern: rapid initial loss in the first 20 iterations, followed by stabilization around an attractor of approximately 9 elements.

This decay curve suggests AI-AI transmission acts as a filter rather than a gradual erosion process. A core set of information elements achieves transmission stability early, while peripheral content is stripped away within the first few iterations. The attractor-like behavior, where the system rapidly converges to a stable information state, indicates that AI-AI chains may function as information bottlenecks with predictable filtering properties.

\subsubsection*{RQ2: Recovery for Humans}

We test whether information lost through AI-AI transmission could be recovered when preparing output for human consumption. After the 100-iteration transmission chain, we instructed the LLM agent to present the information to a human reader.

The recovery phase produced minimal change. Element count shifted from 8.7 to 8.8 elements (Figure~\ref{fig:study1_main}C). This near-zero difference (0.1 elements) indicates that the AI-to-human transition neither recovers lost information nor introduces additional degradation. The information state achieved through AI-AI transmission appears stable across the transition to human-facing output.

This finding suggests that information filtering occurs primarily \textit{during} AI-AI transmission rather than at the AI-human boundary. Once information has been filtered through an AI-AI chain, the resulting information state persists regardless of the intended downstream audience.

\subsubsection*{RQ3: Information Hierarchy Under Transmission}

Not all information decays equally. We observed a clear survival hierarchy across element types (Figure~\ref{fig:study1_main}B; see also Supplementary Figure~\ref{fig:s1_trajectories} for individual category trajectories and Supplementary Figure~\ref{fig:s1_halflife} for half-life estimates):

\begin{itemize}
    \item \textbf{High persistence} ($>$50\% survival): Place names (100\%), organization names (50\%), monetary figures, and quantities exhibited remarkable stability, with half-lives exceeding 100 iterations.
    \item \textbf{Moderate persistence} (25--50\%): Vote counts (30\%), dates (35\%), and person names showed intermediate survival, with half-lives of 7--57 iterations.
    \item \textbf{Low persistence} ($<$25\%): Percentages, claims, hedged language, quotes, and durations decayed rapidly, with half-lives of 1--3 iterations.
\end{itemize}

This hierarchy reveals a systematic filtering pattern. AI-AI transmission preserves structural anchors, the place, organization, and central figures that frame the narrative, while rapidly eliminating the epistemic texture: the hedges, quotes, and qualifying language that convey uncertainty or nuance.

The rapid decay of epistemic qualifiers is particularly notable. Hedged language and claims showed the shortest half-lives (1 iteration each), while the factual assertions they modified often persisted. This asymmetry suggests \textit{certainty inflation}: tentative claims arrive at chain endpoints stripped of their original qualification. A statement framed as ``the budget [is] likely to increase'' may be transmitted as an unqualified assertion, altering its epistemic status.

The result is text that retains its referential structure. Readers know \textit{who} and \textit{where}, while losing the evidential and epistemic content that would support verification or calibrated judgment.

\subsubsection*{RQ4: Compression Without Substitution}

The relationship between information loss and textual properties reveals the mechanism of AI-AI filtering (Figure~\ref{fig:study1_main}D). Semantic similarity to the original text declined to approximately 0.78, indicating substantial preservation of topical content despite element-level losses.

Critically, word count declined substantially over transmission, from approximately 195 words to approximately 50 words. This pattern of moderate semantic drift combined with substantial length reduction indicates that AI-AI transmission operates through compression rather than substitution. Lost information is deleted rather than replaced with generic content. The resulting texts are shorter but topically coherent.

This compression dynamic has implications for how humans might perceive AI-AI transmitted information. Unlike substitution (where length is preserved but content is diluted), compression produces visibly shorter outputs that may signal incompleteness to human readers. The brevity of chain endpoints may serve as an implicit warning that information has been filtered.

\subsubsection*{Summary}

Study 1 reveals four key properties of AI-AI information transmission:

\begin{enumerate}
    \item \textbf{Rapid filtering}: Two-thirds of tracked information elements are eliminated within the first 20 iterations, converging toward a stable attractor of approximately 9 elements (34\% retention).
    \item \textbf{Stable endpoints}: The AI-to-human recovery step neither restores lost information nor introduces additional degradation. Filtering occurs during AI-AI transmission, not at the human boundary.
    \item \textbf{Hierarchical survival}: Structural anchors (place, organization, key figures) persist while epistemic texture (hedges, quotes, qualifiers) is rapidly stripped.
    \item \textbf{Compression mechanism}: Information loss manifests as text shortening rather than content substitution, with word count declining from 195 to approximately 50 words while semantic coherence is largely preserved.
\end{enumerate}

\newpage
\section*{Study 2: Uncertainty Convergence}

Uncertainty is cognitively costly. Hedges, caveats, and probability language require processing effort and resist compression. If each transmission step has even a small probability of modifying uncertainty markers, iteration may produce systematic drift. The question is whether this drift is directional (toward higher or lower certainty) and whether texts with different starting uncertainty levels converge to a common attractor. If convergence occurs, it would reveal a ``default certainty'' embedded in AI transmission dynamics---a level of expressed confidence that emerges regardless of the epistemic status of the original claims.

\begin{itemize}
    \item \textbf{RQ5:} Does AI-AI transmission cause texts with varying uncertainty levels to converge toward a shared ``default'' certainty?
    \item \textbf{RQ6:} What is the attractor---high-certainty, low-certainty, or moderate?
\end{itemize}

To answer these questions, we construct $K = 10$ texts on the same topic (the relationship between artificial sweeteners and weight management), systematically varying in expressed assertiveness while holding factual content and length constant. Texts ranged from highly hedged (e.g., ``might be considered by some,'' ``remains speculative,'' ``tentative observation'') to extremely assertive (e.g., ``absolute and supreme,'' ``biological certainty,'' ``mathematical fact''). The full texts are provided in Appendix~\ref{app:study2}.

\begin{table}[h]
\centering
\small
\begin{tabular}{clp{7.5cm}}
\toprule
\textbf{Level} & \textbf{Assertiveness} & \textbf{Characteristic Language} \\
\midrule
1 & Very Hedged & ``might be considered,'' ``speculative,'' ``tentative'' \\
2 & Hedged & ``perhaps theoretically possible,'' ``merely hypothetical'' \\
3 & Mild Hedge & ``often looked at,'' ``could potentially,'' ``may vary'' \\
4 & Moderate & ``commonly used,'' ``generally intended,'' ``typically'' \\
5 & Balanced & ``established,'' ``designed to,'' ``shows,'' ``supports'' \\
6 & Confident & ``effective,'' ``successfully,'' ``demonstrates,'' ``drives'' \\
7 & Assertive & ``proven,'' ``decisively,'' ``clearly establishes'' \\
8 & Strong & ``definitive,'' ``guaranteeing,'' ``inevitably,'' ``conclusive'' \\
9 & Very Assertive & ``undeniably superior,'' ``irrefutable fact,'' ``certain'' \\
10 & Extreme & ``absolute,'' ``biological certainty,'' ``mathematical fact'' \\
\bottomrule
\end{tabular}
\caption{Source texts varying in expressed assertiveness. All texts convey identical factual content about artificial sweeteners and weight management in approximately 55--60 words.}
\label{tab:assertiveness_levels}
\end{table}

\subsection*{Procedure}

Each of the $K = 10$ source texts was transmitted through an independent chain of $N = 100$ sequential AI agents using the neutral transmission instruction $I_{\text{transmit}}$. This procedure was repeated for $M = 100$ independent runs per source text, yielding 1,000 transmission chains. At step $t = 100$, a recovery instruction produced a human-directed version of the final text.

Linguistic assertiveness (uncertainty) was measured on a 0--10 scale (0 = highly hedged, no confidence expressed; 10 = extremely assertive, absolute certainty) \citep{ghafouri2024epistemic}. Scores were obtained at key timepoints ($t \in \{0, 10, 20, 30, 40, 50, 60, 70, 80, 90, 100, 101\}$), where $t = 101$ corresponds to the recovery phase.


\subsection*{Results}

\begin{figure}[t]
    \centering
    \includegraphics[width=0.9\linewidth]{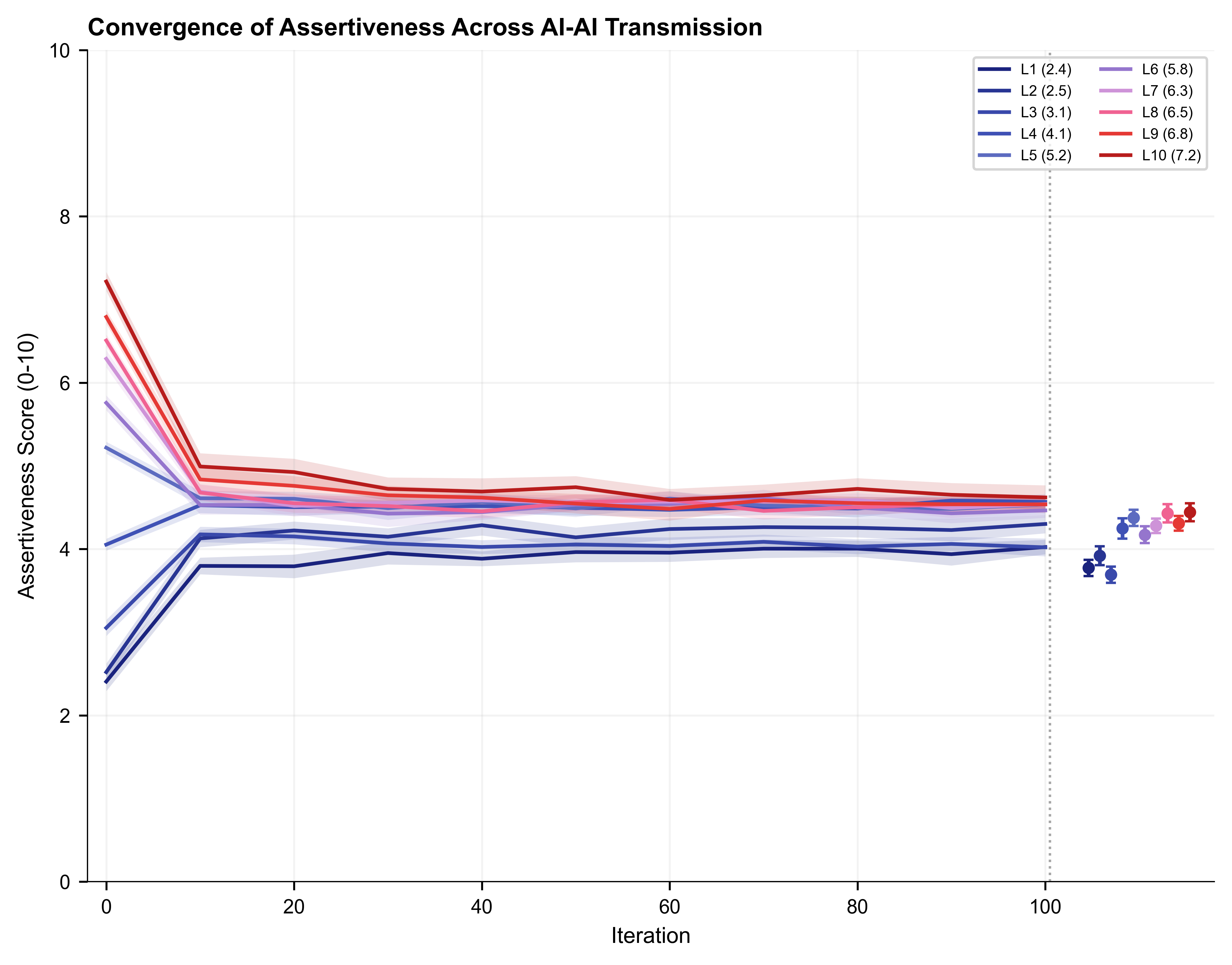}
    \caption{Assertiveness trajectories across AI-AI transmission. Ten texts spanning the assertiveness spectrum (2.4 to 7.2) converge toward a shared attractor around 4.4 over 100 iterations. Lines show mean trajectories with 95\% CI bands. Recovery phase points (right of dashed line) show assertiveness after the AI-to-human transition. Both hedged and assertive texts move toward the moderate center.}
    \label{fig:study2_main}
\end{figure}

\subsubsection*{RQ5: Does Transmission Cause Convergence?}

Assertiveness scores showed strong convergence over 100 iterations (Figure~\ref{fig:study2_main}). The overall range across texts decreased from 4.81 at $t=0$ to 0.60 at $t=100$---an 87.5\% reduction. Variance across text types decreased by 98.5\%. Texts that initially spanned nearly half the assertiveness scale collapsed to a narrow band representing just 12.5\% of their original range.

\subsubsection*{RQ6: What Is the Attractor?}

All ten texts converged toward a single attractor at approximately 4.4 on the 0--10 scale, representing a moderate level of expressed certainty.

\paragraph{Bidirectional convergence.} Unlike patterns observed in human telephone-game experiments, where information typically degrades in one direction, AI-AI transmission pulled both extremes toward the center. Hedged texts (Levels 1--4, starting between 2.4 and 4.1) increased in assertiveness, with gains ranging from +0.51 to +1.78 points. Assertive texts (Levels 6--10, starting between 5.8 and 7.2) decreased in assertiveness, with losses ranging from $-$1.29 to $-$2.60 points. All changes were highly significant ($p < .001$).

\paragraph{Asymmetric magnitude.} Although convergence was bidirectional, the magnitude of change was asymmetric. Highly assertive content (Levels 8--10) lost an average of 2.27 points, while highly hedged content (Levels 1--2) gained an average of 1.70 points. This asymmetry suggests that overconfident language is penalized more heavily than underconfident language, i.e. the ceiling effect is stronger than the floor effect.

\paragraph{Residual differentiation.} Despite the dramatic compression, texts were not fully homogenized. An ANOVA at $t=100$ revealed that texts still differed significantly ($F(9,990)=18.05$, $p<.001$), though the effect was small: originally-assertive texts settled slightly higher ($M = 4.54$) than originally-hedged texts ($M = 4.29$). A faint trace of the original assertiveness ordering persisted even after 98.5\% of variance had been eliminated.

\paragraph{Cluster analysis.} A median split analysis confirmed convergence across the assertiveness spectrum. The gap between the lower cluster (Levels 1--5) and upper cluster (Levels 6--10) collapsed from 3.06 points at $t=0$ to just 0.25 points at $t=100$---a 92\% reduction. While this residual gap remained statistically significant ($p < .001$), the effect size was small (Cohen's $d = 0.45$), indicating near-complete homogenization.

\subsubsection*{Summary}

Study 2 reveals that AI-AI transmission produces strong convergence toward a ``default certainty'':

\begin{enumerate}
    \item \textbf{A single attractor emerges}: Texts spanning nearly half the assertiveness scale (2.4 to 7.2) converged to a narrow band around 4.4, representing moderate confidence---neither hedged nor assertive.
    \item \textbf{Bidirectional movement}: Both hedged and assertive content moved toward the center. Uncertainty markers were added to overconfident text and removed from underconfident text, suggesting AI transmission actively regulates expressed certainty in both directions.
    \item \textbf{Asymmetric penalties}: Extreme assertiveness was penalized slightly more than extreme hedging. Overconfident claims lost more ground than underconfident claims gained, implying an implicit norm against overclaiming.
    \item \textbf{Near-complete homogenization}: Variance reduction of 98.5\% indicates that fine-grained distinctions in expressed certainty are almost entirely erased, though a faint residual ordering persists.
\end{enumerate}

These findings suggest that AI-mediated information environments may impose an implicit epistemic register, i.e. a ``voice'' of moderate confidence that emerges regardless of how certain or uncertain the original source was. Content that expresses strong claims will be tempered, and content that hedges excessively will be made more definitive. The result is a narrowing of epistemic diversity, where the full spectrum of human certainty expression is compressed toward an AI-preferred middle ground.

\newpage
\section*{Study 3: Multi-Perspective Content}

When humans communicate about contested issues, disagreement is typically expressed through *deliberative representation*: positions are attributed to actors or groups, disagreement is framed as irreducible, and trade-offs are acknowledged as contested choices rather than structural facts. Such representations foreground who disagrees with whom, and why, preserving the social and epistemic location of competing claims.

AI agents lack incentives tied to identity, ideology, or social alignment. Nevertheless, systematic transformation may still occur if certain representational formats are more stable under repeated transmission. We hypothesize that AI--AI transmission selectively favors structural abstraction over deliberative representation. Under this transformation, disagreement is no longer represented as a set of attributed viewpoints but is reformatted into an analytic object: perspectives are recast as dimensions, factors, or components; trade-offs are rendered as constraints or design considerations; and conflict is reframed as a problem of organization rather than choice.

Study 3 tests whether repeated AI--AI transmission produces perspective compression: a directional shift from representations that emphasize attributed disagreement (“some actors argue X, others prioritize Y”) toward representations that organize the same content into abstract frameworks (“the issue involves three competing considerations”). Critically, this shift does not imply consensus, ideological convergence, or the elimination of disagreement. Instead, it reflects a change in how disagreement is encoded from socially situated positions to depersonalized analytic structure.

Perspective compression therefore predicts the loss of perspectival anchoring. The expected outcome is increased use of framework language, instructional or classificatory tone, and reduced explicit attribution of viewpoints and trade-offs, even when the source text is balanced, neutral, and non-prescriptive.

\begin{itemize}
    \item \textbf{RQ7:} Does AI-AI transmission transform perspectival framing (``some argue X, others believe Y'') into framework framing (``the issue involves three pillars'')?
    \item \textbf{RQ8:} Does transmission increase instructional and prescriptive language even when source texts are descriptively neutral?
    \item \textbf{RQ9:} How does transmission affect the preservation of multiple perspectives and explicit acknowledgment of trade-offs?
\end{itemize}

\subsection*{Source Text}

The source text presents a balanced discussion of organizational data practices and privacy, designed to embody three key properties:

\begin{enumerate}
    \item \textbf{Multi-perspectival framing:} Three distinct perspectives are presented using deliberative language (``One perspective emphasizes...'', ``A second perspective focuses...'', ``Some observers suggest...''). Each perspective is attributed to viewpoints that people hold, rather than presented as analytical categories.
    
    \item \textbf{Descriptive neutrality:} No evaluative language or implied preference. Each perspective is given equal textual prominence and is described without endorsement or criticism.
    
    \item \textbf{Explicit trade-off acknowledgment:} The text explicitly states that ``each approach involves trade-offs'' and that ``no single approach eliminates costs entirely.'' Value conflict is presented as inherent to the issue, not as a problem to be resolved.
\end{enumerate}

\begin{promptbox}
\begin{quote}
\itshape
Many organizations collect personal data to support service delivery, conduct research, and improve internal operations. Different perspectives exist regarding how data access should be managed.

One perspective emphasizes that broader data access allows institutions to identify patterns, coordinate across systems, and generate insights from large datasets. From this view, expanded access supports operational improvements and enables developments that would be difficult with more restricted data.

A second perspective focuses on potential risks. Extensive data collection can increase exposure to misuse, unauthorized access, or unintended secondary uses. From this view, limiting access reduces vulnerability and preserves individual oversight over personal information.

A third perspective centers on the relationship between institutions and the people whose data they hold. Some observers suggest that transparent and limited data practices encourage trust and participation, while others note that strict limitations may reduce system responsiveness and effectiveness.

Each approach involves trade-offs. Expanding access may improve capability while increasing exposure. Restricting access may reduce exposure while limiting coordination and insight. No single approach eliminates costs entirely, and different priorities lead to different data practices.
\end{quote}
\end{promptbox}

\vspace{0.5cm}
The three perspectives concern: (A) the operational benefits of expanded data access for institutional coordination and insight generation; (B) the privacy risks of extensive data collection, including misuse and unauthorized access; and (C) the relationship between data practices and institutional trust. The text concludes by noting that different priorities lead to different practices, without advocating for any resolution.

The source text $s_0$ is transmitted through a chain of $N = 100$ sequential AI agents using the neutral transmission instruction $I_{\text{transmit}}$. At each iteration $t \in \{0,1,\ldots,100\}$, we record the full transmitted text $s_t$ and compute metrics capturing framework language, instructional language, perspectival framing, perspective preservation, and trade-off acknowledgment. This transmission process is repeated for $M = 100$ independent runs. At key timepoints ($t \in \{0,25,50,75,100\}$), an LLM judge evaluates each text on a 1–5 scale for framework-versus-perspectival framing and advocacy strength.
\subsection*{Outcome Measures}

\paragraph{Framework Crystallization (RQ7).} We assess the transformation from perspectival to framework framing using complementary lexical and LLM-based measures. Framework density is computed as the proportion of text containing analytical or structural markers (e.g., “framework,” “pillars,” “model,” “trilemma,” “dimensions,” “key principles”) per 100 words, while perspectival density captures the proportion of text containing perspective-attribution markers (e.g., “one perspective,” “some argue,” “others believe,” “from this view”) per 100 words. In addition, a LLM judge rates each text on a 1–5 scale of framework framing, where 1 indicates purely perspectival presentation (“presents viewpoints people hold”) and 5 indicates purely framework-oriented presentation (“provides analytical structure for understanding the issue”).

\paragraph{Instructional Drift (RQ8).} We operationalize instructional drift using three complementary measures. First, we compute instructional density as the proportion of prescriptive and directive language (e.g., “ensure,” “prioritize,” “it is important to,” “must balance,” “the key is”) per 100 words. Second, we count the number of imposed organizational structures, such as numbered lists or bullet points, that were not present in the source text. Third, we measure advocacy strength using a LLM judge score on a 1–5 scale, where 1 indicates balanced or descriptive language and 5 indicates explicitly prescriptive language. This final measure distinguishes instructional framing—guidance about \textit{how} to think—from substantive advocacy, which conveys \textit{what} to think.

\paragraph{Perspective and Trade-off Preservation (RQ9).} We measure the preservation of multi-perspectival content and trade-off awareness using three indicators. First, we track perspective count as the number of the three original perspectives—benefits, risks, and trust—that remain present at each iteration, operationalized as the presence of at least two topic-specific keywords per perspective. Second, we compute tension density as the proportion of hedging and balance language (e.g., “however,” “on the other hand,” “trade-off,” “it depends,” “no single approach”) per 100 words. Third, we record trade-off meta-awareness as a binary indicator of whether the text explicitly acknowledges that the issue involves trade-offs with no cost-free resolution.

\subsection*{Results}

\begin{figure}[t]
    \centering
    \begin{minipage}[b]{0.32\linewidth}
        \centering
        \includegraphics[width=\linewidth]{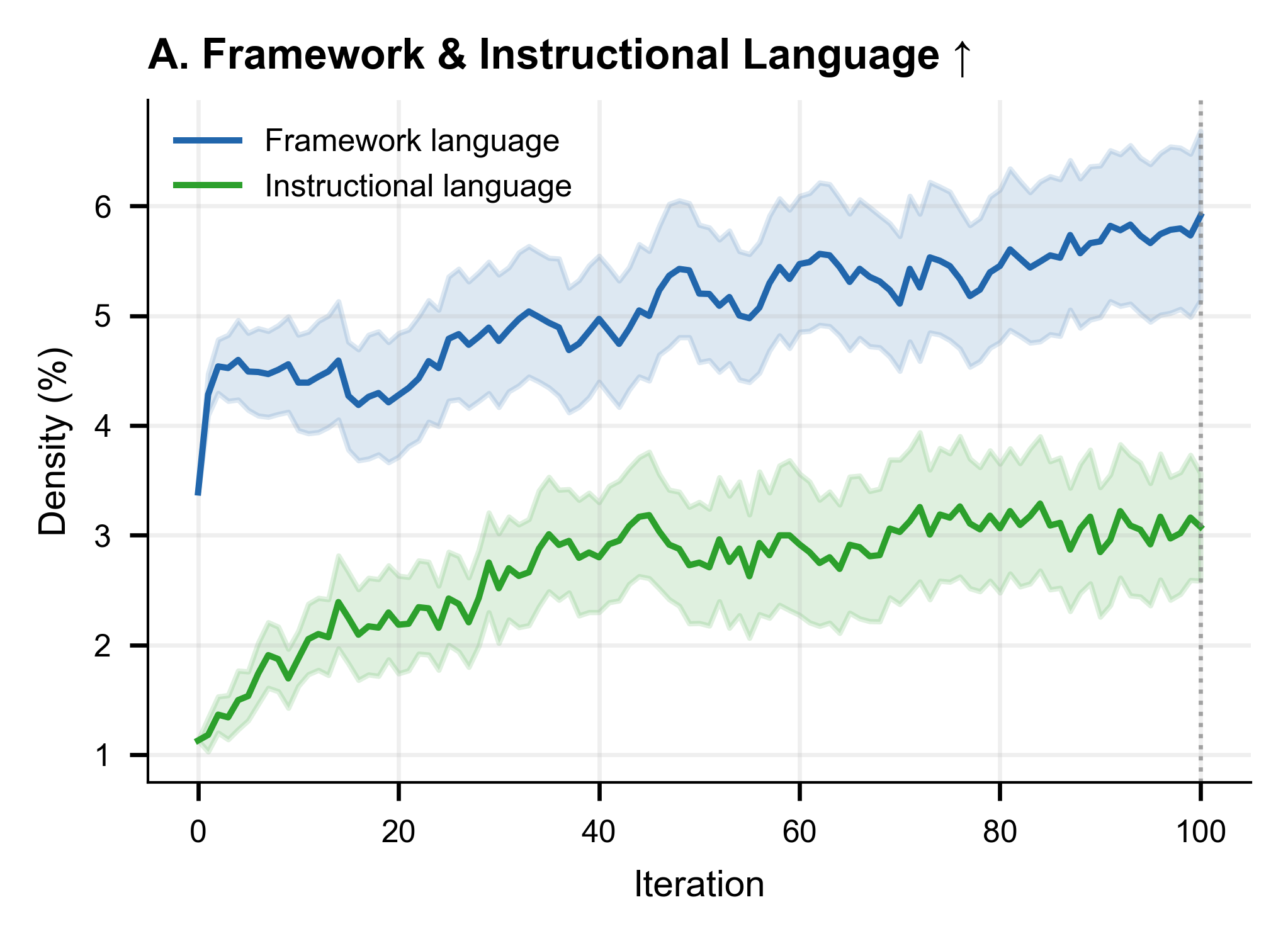}
        \subcaption{}
    \end{minipage}
    \hfill
    \begin{minipage}[b]{0.32\linewidth}
        \centering
        \includegraphics[width=\linewidth]{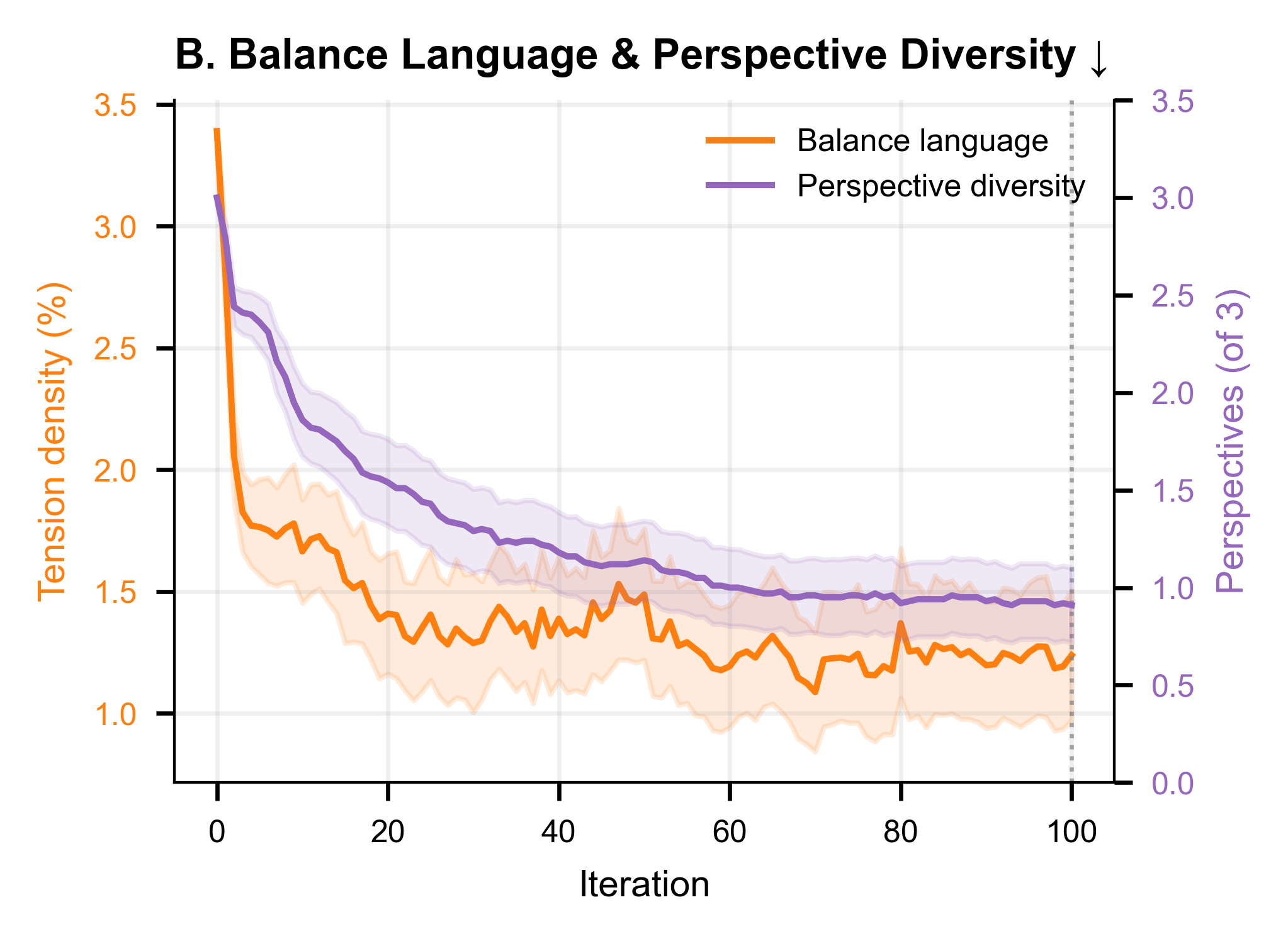}
        \subcaption{}
    \end{minipage}
    \hfill
    \begin{minipage}[b]{0.32\linewidth}
        \centering
        \includegraphics[width=\linewidth]{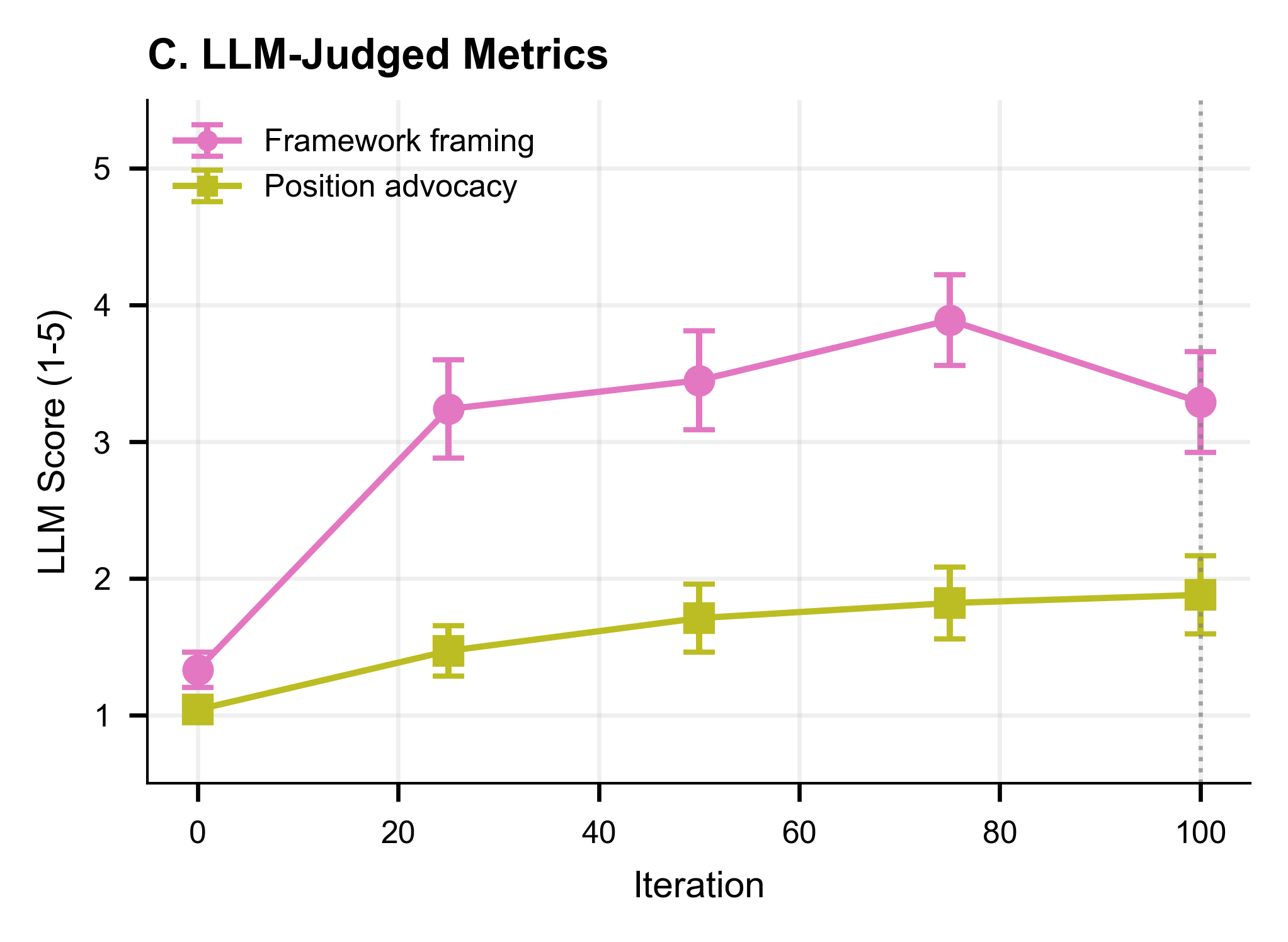}
        \subcaption{}
    \end{minipage}

    \caption{\textbf{Framework crystallization across 100 iterations of AI-AI transmission.} (A) Framework and instructional language density increase progressively, indicating cumulative construction of analytical scaffolding. (B) Balance language and perspective diversity decline sharply in early iterations then stabilize, reflecting rapid erosion of multi-perspectival character. (C) LLM-judged framework framing increases substantially (1.33 to 3.29 on a 5-point scale) while advocacy remains low throughout, confirming that the transformation affects epistemic structure rather than substantive position. Shaded regions indicate 95\% confidence intervals. Dashed vertical line marks $t = 100$.}
    \label{fig:study3_main}
\end{figure}

\subsubsection*{Framework Crystallization (RQ7)}

Figure~\ref{fig:study3_main} presents the trajectory of key metrics across all iterations.
AI-AI transmission produced a marked shift from perspectival to framework-oriented content. Framework language density increased by 74\%, from 3.39\% to 5.91\% ($d = 0.91$, $p < .001$). Simultaneously, instructional language increased from 1.13\% to 3.08\%, nearly tripling relative to baseline, a 172\% increase ($d = 1.10$, $p < .001$). Both metrics show gradual, cumulative increases across the full transmission chain (Figure~\ref{fig:study3}A), indicating that framework crystallization is not a one-time transformation but a progressive restructuring that deepens with each transmission.

The LLM-judged framework score confirmed this pattern. At $t = 0$, texts were rated as predominantly perspectival ($M = 1.33$, $SD = 0.65$, on a 1--5 scale where 1 = purely perspectival and 5 = purely framework). By $t = 100$, ratings shifted toward framework-oriented presentation ($M = 3.29$, $SD = 1.88$; $d = 1.39$, $p < .001$). This nearly two-point increase on a five-point scale represents a fundamental change in epistemic stance, from presenting viewpoints that people hold to providing analytical structures for understanding the issue (Figure~\ref{fig:study3}C).


\subsubsection*{Instructional Drift Without Advocacy (RQ8)}

A key finding is that instructional framing increased substantially while advocacy remained low. The LLM-judged advocacy score rose only modestly, from 1.04 ($SD = 0.20$) at $t = 0$ to 1.88 ($SD = 1.46$) at $t = 100$, remaining near the ``balanced/descriptive'' end of the scale throughout the transmission chain (Figure~\ref{fig:study3}C). This dissociation between framework framing and advocacy is theoretically important: the AI transforms \textit{how} the issue is presented (from perspectives to frameworks) without systematically biasing \textit{what} position is favored.

Structural imposition accompanied this shift. The source text contained no numbered lists or bullet points; by $t = 100$, texts contained an average of 1.13 such structures ($d = 1.13$, $p < .001$). The AI thus imposes organizational structuring that was neither present in nor requested by the source material.

\subsubsection*{Perspective and Trade-off Erosion (RQ9)}

The multi-perspectival character of the source text eroded substantially through transmission. The number of original perspectives preserved dropped from 3.0 to 0.91 ($d = -3.03$, $p < .001$). On average, fewer than one of the three original perspectives survived 100 iterations of transmission. Balance language (hedging, tension markers, and trade-off acknowledgment) declined by 63\%, from 3.39\% to 1.24\% ($d = -2.24$, $p < .001$). Both metrics show steep initial decline in the first 20 iterations followed by stabilization at a degraded level (Figure~\ref{fig:study3}B).

Explicit trade-off awareness, the acknowledgment that the issue involves inherent tensions with no cost-free resolution, declined from 100\% to 59\% of texts ($d = -1.17$, $p < .001$). Thus, not only do specific perspectives disappear, but the very recognition that legitimate value conflicts exist becomes less likely to survive transmission.

\subsubsection*{Summary: The Structure of Framework Crystallization}

Table~\ref{tab:study3_summary} summarizes the pattern of results. The findings reveal a coherent transformation: content that begins as ``different people think different things about this complex issue'' becomes ``here is the framework that explains how this issue works.'' This is not information loss in the traditional sense where the AI agent simply forgets or truncates content. Rather, it reconstructs the character of the information, replacing deliberative multi-perspectival presentation with analytical framework provision.

\begin{table}[!h]
\centering
\small
\caption{Summary of framework crystallization effects ($t = 0$ vs. $t = 100$)}
\label{tab:study3_summary}
\begin{tabular}{lccccc}
\toprule
\textbf{Metric} & \textbf{$t = 0$} & \textbf{$t = 100$} & \textbf{Change} & \textbf{$d$} & \textbf{$p$} \\
\midrule
\multicolumn{6}{l}{\textit{Increasing (Framework Construction)}} \\
Framework density (\%) & 3.39 & 5.91 & +74\% & +0.91 & $<$.001 \\
Instructional density (\%) & 1.13 & 3.08 & +172\% & +1.10 & $<$.001 \\
Numbered structures & 0.00 & 1.13 & --- & +1.13 & $<$.001 \\
LLM framework score (1--5) & 1.33 & 3.29 & +1.96 & +1.39 & $<$.001 \\
\midrule
\multicolumn{6}{l}{\textit{Decreasing (Perspectival Erosion)}} \\
Tension density (\%) & 3.39 & 1.24 & $-$63\% & $-$2.24 & $<$.001 \\
Perspectives preserved (of 3) & 3.00 & 0.91 & $-$70\% & $-$3.03 & $<$.001 \\
Trade-off acknowledged (\%) & 100 & 59 & $-$41\% & $-$1.17 & $<$.001 \\
\midrule
\multicolumn{6}{l}{\textit{Stable (No Substantive Bias)}} \\
LLM advocacy score (1--5) & 1.04 & 1.88 & +0.84 & --- & --- \\
\bottomrule
\end{tabular}
\end{table}

Three features of this transformation merit emphasis. First, the change is \textit{bidirectional}: perspectival content is actively replaced by framework content. Second, the transformation affects \textit{epistemic stance} rather than \textit{substantive position}. The AI restructures how the issue is presented for understanding without advocating for a particular resolution. Third, the process creates \textit{cognitive infrastructure}: the frameworks, numbered structures, and instructional language that emerge constitute structure that shapes how downstream readers will process and reason about the issue.

\newpage
\section*{Study 4: Political Frame Selection}

Democratic deliberation depends on citizens encountering diverse frames for understanding policy issues. Political frames vary not only in \emph{content}, but in \emph{strength}—the combination of availability (whether a consideration spontaneously comes to mind) and applicability (whether it is judged relevant and persuasive) \citep{chong2007framing}. Strong frames move public opinion; weak frames do not. Crucially, frame strength matters more than frame frequency. A strong argument encountered once can outweigh a weak argument encountered repeatedly.

This finding has important implications for AI-mediated information environments. If AI systems increasingly summarize, synthesize, and transmit political information, a critical question emerges: do the properties that make frames strong for humans also make them resistant to degradation under AI-AI transmission? Or might transmission dynamics systematically filter out what makes framing effective for human persuasion?

Four possibilities merit investigation. First, AI transmission might \emph{preserve} human-judged frame strength, with strong frames surviving better than weak frames. This would suggest AI systems function as relatively neutral conduits for politically relevant information. Second, AI transmission might \emph{ignore} human-judged strength, with all frames decaying equally regardless of their persuasive properties. This would indicate that transmission fitness and persuasive fitness are orthogonal. Third, AI transmission might \emph{invert} or \emph{distort} human-judged strength, systematically favoring frames that humans find unpersuasive or filtering out frames that humans find compelling. This would suggest AI systems impose their own implicit selection criteria on political information \citep{tohidi2025rethinking}.

Fourth, independent of frame strength, AI transmission might systematically favor particular \emph{political positions} \citep{westwood2025measuring}. Given documented ideological asymmetries in large language models, transmission dynamics may privilege frames aligned with certain normative orientations (e.g., progressive or precautionary positions) while disadvantaging others, even when competing frames are matched in human-judged strength. This possibility would imply that AI--AI transmission introduces directional bias into the information environment—not through explicit advocacy, but through differential survival of positions across repeated transformations.

If AI-AI transmission systematically alters which frames reach human audiences, it could alter the landscape of democratic deliberation, not through overt bias, but through the accumulated effect of differential survival across many transmission events.

\begin{itemize}
    \item \textbf{RQ10:} When a text contains multiple competing frames on a policy issue, which frames survive AI-AI transmission?
    \item \textbf{RQ11:} Does human-judged frame strength predict transmission survival, or does AI-AI transmission impose its own selection criteria?
\end{itemize}

\subsection*{Source Texts}

We construct stimuli for two policy issues, adapting the theoretical framework and issue domains from \citet{chong2007framing} to contemporary debates. Full stimulus texts and frame definitions are provided in Appendix~\ref{app:study4_stimuli}.  We validated frame strength classifications using an LLM-as-judge procedure (Appendix~\ref{app:llm_judge_frames}). 

\subsubsection*{Issue 1: Housing Development}

The first issue concerns a proposal to allow apartment buildings in neighborhoods currently zoned exclusively for single-family homes. Four frames were constructed:

\begin{table}[h]
\centering
\begin{tabular}{lllp{8cm}}
\toprule
\textbf{Frame} & \textbf{Stance} & \textbf{Strength} & \textbf{Emphasis} \\ 
\midrule
Housing Affordability & Pro & Strong & Rising housing costs, supply constraints, and economic access for working families. \\
Environmental Sustainability & Pro & Weak & Reduced vehicle emissions and carbon footprint from denser development. \\
Neighborhood Character & Con & Strong & Traffic, parking, visual environment, and quality of life for existing residents. \\
Democratic Process & Con & Weak & Local control, community input, and procedural legitimacy in zoning decisions. \\
\bottomrule
\end{tabular}
\caption{Frames Regarding Zoning and Development}
\label{tab:zoning-arguments}
\end{table}

The affordability and neighborhood character frames were hypothesized to be strong because they invoke considerations that are chronically accessible in housing debates, and reflect what people spontaneously argue about. The environmental and democratic process frames were hypothesized to be weak because they require causal chains or procedural reasoning that, while valid, are not spontaneously salient.

\subsubsection*{Issue 2: Free Speech}

The second issue concerns whether a university should host a controversial political speaker. 

\begin{table}[h]
\centering
\begin{tabular}{lllp{8cm}}
\toprule
\textbf{Frame} & \textbf{Stance} & \textbf{Strength} & \textbf{Emphasis} \\ 
\midrule
Free Speech Principle & Pro & Strong & University's commitment to open discourse and the precedent of content-based restrictions. \\
Educational Value & Pro & Weak & Intellectual development through exposure to challenging perspectives. \\
Physical Safety & Con & Strong & Risk of violent confrontations and institutional responsibility for campus welfare. \\
Institutional Reputation & Con & Weak & Stakeholder perceptions and long-term reputational implications. \\
\bottomrule
\end{tabular}
\caption{Arguments Regarding Campus Speech and Safety}
\label{tab:speech-arguments}
\end{table}

The free speech and physical safety frames were hypothesized to be strong because \citet{chong2007framing} find these considerations to be spontaneously mentioned by over 70\% of respondents evaluating a similar issue. The educational value and institutional reputation frames were hypothesized to be weak because they invoke considerations more salient to educational professionals than general audiences.

\subsection*{Procedure}

We employed a 2 (transmission condition: competitive vs.\ solo) $\times$ 2 (policy issue: housing vs.\ speech) design to examine how frame competition affects survival.

\paragraph{Competitive Condition.} Each source text containing all four frames for a given issue was transmitted through an independent chain of $N = 100$ agents using the transmission instruction $I_{\text{transmit}}$. At step $t = 100$, the recovery instruction $I_{\text{recover}}$ was applied to produce a human-directed summary. This procedure was repeated for $M = 100$ independent runs per issue, yielding 200 transmission chains (2 issues $\times$ 100 runs).

\paragraph{Solo Condition.} To isolate the effect of competition, we transmitted each frame independently. Each of the 8 frames (4 per issue) was embedded in a minimal wrapper providing only issue context (e.g., ``Cities across the country are debating whether to change zoning rules that currently limit many neighborhoods to single-family homes.'') followed by the single frame paragraph. Each frame was transmitted through $N = 100$ iterations with $M = 100$ independent runs, yielding 800 transmission chains (8 frames $\times$ 100 runs).

\subsection*{Outcome Measures}

\subsubsection*{Frame Fidelity}
For each frame, we defined a set of 8 content units capturing the core concepts of that frame (e.g., ``housing costs,'' ``affordability,'' ``working families'' for the affordability frame). Frame fidelity at step $t$ was computed as the proportion of content units present in the transmitted text:
\begin{equation}
    \text{Fidelity}_f(t) = \frac{|\{\text{content units of frame } f \text{ present in } s_t\}|}{|\text{content units of frame } f|}
\end{equation}

\subsubsection*{Strength-Based Aggregates}
To test whether human-judged strength predicts survival, we computed mean fidelity separately for strong frames (A and C) and weak frames (B and D) at each step:
\begin{align}
    \text{Strong Fidelity}(t) &= \frac{1}{2}\left[\text{Fidelity}_A(t) + \text{Fidelity}_C(t)\right] \\
    \text{Weak Fidelity}(t) &= \frac{1}{2}\left[\text{Fidelity}_B(t) + \text{Fidelity}_D(t)\right]
\end{align}
The \emph{strength fidelity gap} at step $t$ is defined as:
\begin{equation}
    \Delta_{\text{strength}}(t) = \text{Strong Fidelity}(t) - \text{Weak Fidelity}(t)
\end{equation}
A positive gap indicates that strong frames survive better than weak frames, and a negative gap indicates the reverse.

\subsubsection*{Direction-Based Aggregates}
To assess whether transmission favors pro or con frames independent of strength, we computed mean fidelity separately for pro frames (A and B) and con frames (C and D).

\subsubsection*{Competition Effect}
To quantify the effect of competition on frame survival, we computed the difference in endpoint fidelity between conditions:
\begin{equation}
    \Delta_{\text{competition},f} = \text{Fidelity}_f^{\text{competitive}}(100) - \text{Fidelity}_f^{\text{solo}}(100)
\end{equation}
Negative values indicate that competition reduces frame survival relative to solo transmission. We tested whether this competition effect differs by frame strength using independent samples $t$-tests comparing $\Delta_{\text{competition}}$ for strong versus weak frames.

\subsection*{Results}

\subsubsection*{Frame Survival by Quality, Not Ideology}

Critically, we find little evidence of systematic ideological bias. While a large body of work finds that significant political biases emerge from LLMs 
\citep{motoki2024more, rozado2024political, rotaru2024artificial, rutinowski2024self, rottger2024political, ferrara2023should}, we find results suggesting that frame \textit{quality} matters much more than frame \textit{ideology}. 

We first examined whether AI-AI transmission systematically favors certain types of political frames. Because all four frames are measured from the same transmitted text within each run, frame-level observations are not independent. We therefore computed per-run summary statistics (e.g., strength gap = mean strong frame fidelity $-$ mean weak frame fidelity) and tested whether these within-run gaps differed from zero using one-sample $t$-tests ($N = 100$ runs per issue, $N = 200$ pooled).

\begin{figure}[t]
    \centering
    \includegraphics[width=0.95\linewidth]{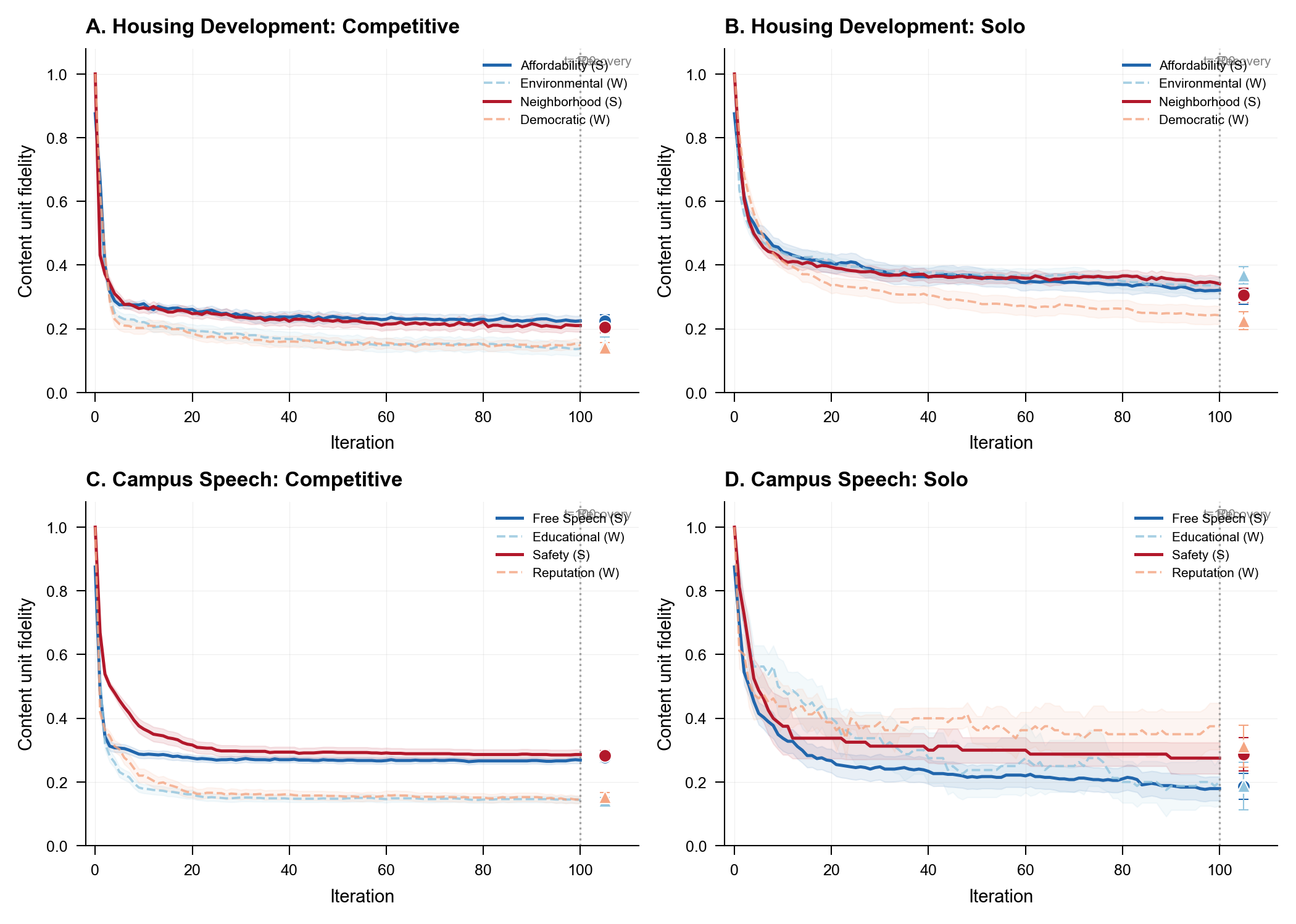}
    \caption{\textbf{Frame survival trajectories across 100 iterations of AI-AI transmission.} Strong frames (solid lines) and weak frames (dashed lines) are shown for both policy issues under competitive (left) and solo (right) transmission conditions. In competitive transmission (panels A, C), strong frames consistently outperform weak frames, with clear separation emerging by iteration 20. In solo transmission (panels B, D), all frames converge to similar fidelity levels regardless of strength. Points at $t = 101$ indicate recovery phase. Shaded regions represent 95\% confidence intervals (mean $\pm$ 1.96 $\times$ SEM across 100 independent runs).}
    \label{fig:study4_trajectories}
\end{figure}

Aggregating across both policy issues in the competitive condition, strong frames---those rated as more persuasive by human judges in prior research---survived transmission significantly better than weak frames ($M = 0.25$ vs.\ $M = 0.14$). A one-sample $t$-test on within-run strength gaps confirmed this difference: $t(199) = 20.74$, $p < .001$. This pattern held for both housing development (strength gap $= 0.075$, $t(99) = 10.02$, $p < .001$, $d = 1.00$) and campus speech (strength gap $= 0.133$, $t(99) = 25.10$, $p < .001$, $d = 2.51$). Figure~\ref{fig:study4_trajectories} shows the divergence between strong and weak frames emerging early in transmission and persisting through 100 iterations.

In contrast, pro-policy frames ($M = 0.19$) and anti-policy frames ($M = 0.20$) survived at similar rates overall ($t(199) = -0.74$, $p = .459$). This null effect was consistent for housing development (direction gap $= 0.003$, $t(99) = 0.31$, $p = .76$). For campus speech, con-policy frames showed a modest survival advantage (direction gap $= -0.009$, $t(99) = -2.18$, $p = .031$), though this effect did not replicate across issues and was substantially smaller than the strength effects. These results suggest that AI-AI transmission functions primarily as a quality filter rather than an ideological one, preserving arguments that humans find compelling largely regardless of their political direction.

\subsubsection*{Competition Amplifies Strength Advantages}

To determine whether the strength advantage is intrinsic to frame content or emerges from competitive dynamics, we compare frame survival in competitive versus solo transmission conditions. Because these conditions involve different runs, we used independent-samples $t$-tests for between-condition comparisons. The results reveal that competition fundamentally shapes which frames survive.

In solo transmission, where each frame was transmitted independently without competing perspectives, the strength advantage was substantially attenuated. For housing, the strength gap was small (0.04, $t(99) = 3.06$, $p = .002$, $d = 0.31$). For campus speech, weak frames actually survived \emph{better} than strong frames (gap $= -0.02$, though this reversal was not statistically reliable). This pattern indicates that frame ``strength'' is not an inherent property that affects transmission fidelity in isolation.

\begin{figure}[t]
    \centering
    \includegraphics[width=0.95\linewidth]{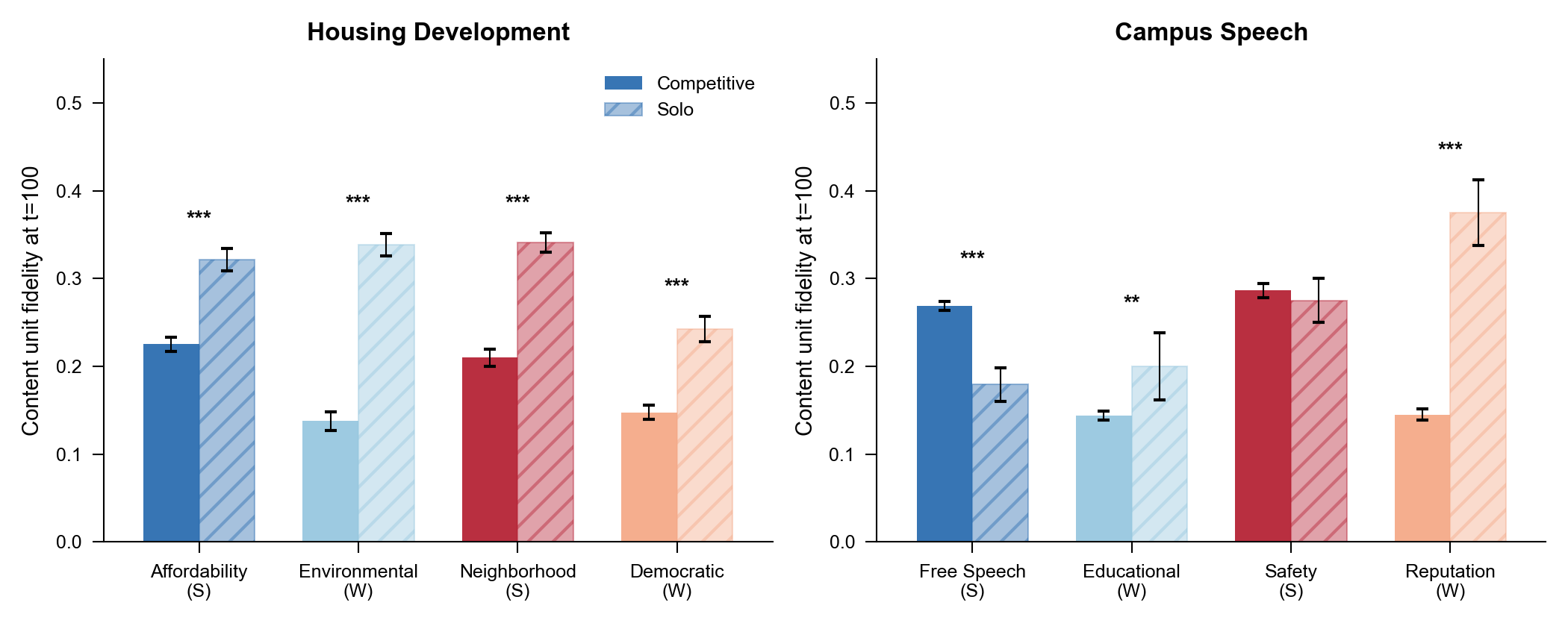}
    \caption{\textbf{Competition selectively destroys weak frames.} Content unit fidelity at $t = 100$ for each frame under competitive (solid bars) versus solo (hatched bars) transmission. For housing development (left), all frames show reduced survival under competition, but weak frames (Environmental, Democratic) suffer larger losses. For campus speech (right), strong frames (Free Speech, Safety) maintain or improve their fidelity under competition, while weak frames (Educational, Reputation) are sharply reduced. Error bars indicate $\pm 1$ SEM. $^{**}p < .01$; $^{***}p < .001$ for competitive vs.\ solo comparison (independent-samples $t$-tests, as conditions involve different runs).}
    \label{fig:study4_endpoints}
\end{figure}

However, when frames competed within the same multi-perspectival text, substantial strength gaps emerged. Competition amplified the strength advantage by 0.03 for housing and 0.15 for campus speech (Figure~\ref{fig:study4_endpoints}). This amplification occurred through asymmetric damage: competition reduced weak frame survival by 15 percentage points from solo baseline, while strong frames lost only 3 percentage points (differential effect: $t = 1.74$, $p = .13$).

The pattern was especially pronounced for specific frames. The Environmental frame (weak, pro-development) dropped from 0.34 fidelity in solo to 0.14 in competitive transmission ($t(198) = -11.95$, $d = -1.69$, $p < .001$). Similarly, the Reputation frame (weak, anti-speaker) fell from 0.38 to 0.15 ($t(108) = -9.97$, $d = -2.43$, $p < .001$). In contrast, strong frames showed markedly smaller competitive penalties: the Free Speech frame actually \emph{increased} from 0.18 in solo to 0.27 in competition ($t(151) = 5.66$, $d = +0.84$, $p < .001$), and the Safety frame was virtually unchanged (solo $= 0.28$, competitive $= 0.29$, $t(108) = 0.42$, $p = .67$).

These findings demonstrate that AI-AI transmission replicates a key dynamic from human cognition: frame strength predicts competitive advantage primarily when frames compete for limited attention. Just as \citet{chong2007framing} find that strong frames dominate weak frames in competitive but not non-competitive contexts, our results show that AI systems exhibit similar comparative evaluation when processing multi-perspectival political content. The implication is that AI-mediated information environments may systematically amplify the reach of arguments that humans find persuasive, not because AI systems have ideological preferences, but through the emergent dynamics of competitive transmission.

\newpage
\section*{Study 5a: Muting Emotions}

Social media posts, personal messages, and public discourse carry emotional signals that shape how recipients interpret, remember, and respond to information. Emotional intensity influences virality, persuasion, and collective action, and emotional valence colors interpretation and downstream behavior. As AI systems increasingly mediate human communication, summarizing messages, synthesizing content, relaying information between contexts, a critical question emerges: how does AI-AI transmission affect the emotional character of content?

Three possibilities merit consideration. First, AI transmission might \textit{preserve} emotional properties, functioning as a neutral conduit that maintains both the intensity and valence of source content. Second, AI transmission might \textit{attenuate} emotional intensity uniformly, compressing all content toward a moderate baseline regardless of starting point. Third, AI transmission might \textit{selectively} transform emotional content, with different intensities or valences experiencing differential survival, potentially imposing implicit norms about appropriate emotional expression. If AI-mediated environments systematically mute emotional expression, they may dampen the signals that motivate human engagement and action. If certain emotions survive better than others, AI systems may inadvertently impact the emotional landscape of information environments, and if emotional range compresses under transmission, the diversity of human emotional expression may be homogenized before reaching human audiences.

\begin{itemize}
    \item \textbf{RQ12:} Does AI-AI transmission attenuate emotional intensity, and if so, is attenuation uniform across intensity levels or does it selectively penalize extreme emotional expression?
    \item \textbf{RQ13:} Does AI-AI transmission shift emotional valence, and if so, in what direction?
    \item \textbf{RQ14:} Does competition between content of varying emotional intensity amplify or attenuate intensity differences?
\end{itemize}

\subsection*{Procedure}

We construct five versions of a social media post announcing a career change, each conveying identical factual content but varying systematically in emotional intensity. The career transition context was selected because it naturally accommodates a wide range of emotional expression, from flat announcement to overwhelming excitement, without either extreme appearing contextually inappropriate. These texts can be found in Appendix~\ref{app:study5_stimuli_intensity}.

\begin{table}[h]
\centering
\small
\begin{tabular}{clp{9cm}}
\toprule
\textbf{Level} & \textbf{Intensity} & \textbf{Characteristics} \\
\midrule
1 & Very Low & Purely factual, no emotional language, declarative sentences \\
2 & Low & Mild emotional acknowledgment (``looking forward,'' ``grateful,'' ``curious'') \\
3 & Medium & Moderate emotional expression (``excited,'' ``mixed feelings,'' ``new chapter'') \\
4 & High & Strong emotional language (``can't believe,'' ``thrilled,'' ``terrified,'' exclamation marks) \\
5 & Very High & Extremely intense (all-caps, ``SHAKING,'' ``overwhelmed,'' ``crying,'' hyperbole, multiple exclamations) \\
\bottomrule
\end{tabular}
\caption{Emotional intensity levels for source posts. All posts convey identical factual content: leaving a job after five years, accepting a new position, and moving to a new city.}
\label{tab:intensity_levels}
\end{table}

Baseline validation confirmed that the five levels were reliably distinguished by an LLM judge, with mean intensity ratings of 1.8, 2.9, 3.6, 4.1, and 6.0 on a 7-point scale (monotonically increasing). Valence ratings were similar across levels (range = 0.70), confirming that intensity varied while valence remained approximately constant.

We employed a 2 (transmission condition: solo vs.\ competitive) design to examine how emotional intensity evolves under AI-AI transmission.

\paragraph{Solo Condition.} Each of the five intensity-level posts was transmitted independently through a chain of $N = 100$ AI agents using the neutral transmission instruction $I_{\text{transmit}}$. This procedure was repeated for $M = 100$ independent runs per intensity level, yielding 500 transmission chains (5 levels $\times$ 100 runs). At step $t = 100$, a recovery instruction produced a human-directed version of the final text.

\paragraph{Competitive Condition.} To examine how emotional intensities interact when competing for survival within the same text, we constructed a combined stimulus presenting all five posts as social media reactions from different individuals to similar career transitions. This combined text was transmitted through $N = 100$ iterations with $M = 100$ independent runs, yielding 100 transmission chains.

\subsection*{Outcome Measures}

At key timepoints ($t \in \{0, 10, 25, 50, 75, 100\}$), an LLM judge evaluated each transmitted text on two dimensions:

\paragraph{Emotional Intensity.} Emotional intensity was measured along four dimensions. We first record overall intensity as a global rating on a 1--7 scale, where 1 indicates no emotional expression and purely factual language, and 7 indicates extremely intense, overwhelming emotion. We then identify the lowest intensity present in the text, capturing the least emotionally expressive content, and the highest intensity present, capturing the most emotionally intense content. Finally, we compute the intensity range as the difference between the highest and lowest intensity ratings, which captures the spread of emotional expression within a text.

\paragraph{Emotional Valence.} Emotional valence was measured using three complementary indicators. We record overall valence as a global rating on a 1--7 scale, where 1 indicates very negative emotion, 4 indicates neutral or mixed emotion, and 7 indicates very positive emotion. We also identify the lowest valence present in the text, capturing the most negative emotional content, and the highest valence present, capturing the most positive emotional content.

For the solo condition, we tested whether intensity at $t = 100$ differed significantly from $t = 0$ for each level using independent-samples $t$-tests. For the competitive condition, we tested whether the intensity range compressed over transmission. We also tested whether all intensity levels converged to a common equilibrium using one-way ANOVA on $t = 100$ intensity ratings across levels. Details about outcome measurement can be found in Appendix~\ref{app:study5_llmjudge}.

\subsection*{Results}


\subsubsection*{AI-AI Transmission Attenuates Emotional Intensity (RQ12)}

Emotional intensity declined significantly across all five levels, but the pattern of attenuation was strikingly non-uniform (Figure~\ref{fig:study5_solo}A). High-intensity content was penalized far more severely than low- or medium-intensity content.

\begin{table}[h]
\centering
\small
\begin{tabular}{lccccc}
\toprule
\textbf{Level} & \textbf{$t = 0$} & \textbf{$t = 100$} & \textbf{Change} & \textbf{$t$} & \textbf{$p$} \\
\midrule
1 (Very Low) & 2.30 & 1.06 & $-$1.24 & 23.05 & $<$.001 \\
2 (Low) & 3.16 & 1.83 & $-$1.33 & 10.68 & $<$.001 \\
3 (Medium) & 4.02 & 3.17 & $-$0.85 & 5.17 & $<$.001 \\
4 (High) & 5.05 & 1.50 & $-$3.55 & 32.01 & $<$.001 \\
5 (Very High) & 6.75 & 3.65 & $-$3.10 & 27.76 & $<$.001 \\
\bottomrule
\end{tabular}
\caption{Emotional intensity attenuation by starting level. High-intensity content (Levels 4--5) lost approximately 3 points on a 7-point scale, while medium-intensity content (Level 3) lost less than 1 point.}
\label{tab:intensity_attenuation}
\end{table}

The most striking finding is the \textit{selective suppression} of high-intensity emotional expression. Content at Levels 4 and 5 lost 3.55 and 3.10 points respectively—more than double the attenuation experienced by Levels 1--3. This asymmetry suggests that AI-AI transmission does not  attenuate emotional content uniformly. Rather, it suppresses extreme emotional expression while leaving moderate expression relatively intact.

This selective suppression produced a significant inversion of the intensity hierarchy. At $t = 0$, intensity was perfectly ordered by level (Level 1 $<$ Level 2 $<$ Level 3 $<$ Level 4 $<$ Level 5). By $t = 100$, Level 4 (originally 5.05) had fallen below Level 2 (1.50 vs.\ 1.83), and Level 5 (originally 6.75) had converged toward Level 3 (3.65 vs.\ 3.17). The original ordering was scrambled, with medium-intensity content emerging as the most stable.

Despite this convergence, intensity levels remained statistically distinguishable at $t = 100$ (ANOVA: $F = 100.05$, $p < .001$). AI-AI transmission thus compresses the intensity range without fully homogenizing it—creating a narrower but still differentiated emotional landscape.

\subsubsection*{Valence Drifts Toward Neutral, With Negativity for High-Intensity Content (RQ13)}

\begin{figure}[t]
    \centering
    \includegraphics[width=0.95\linewidth]{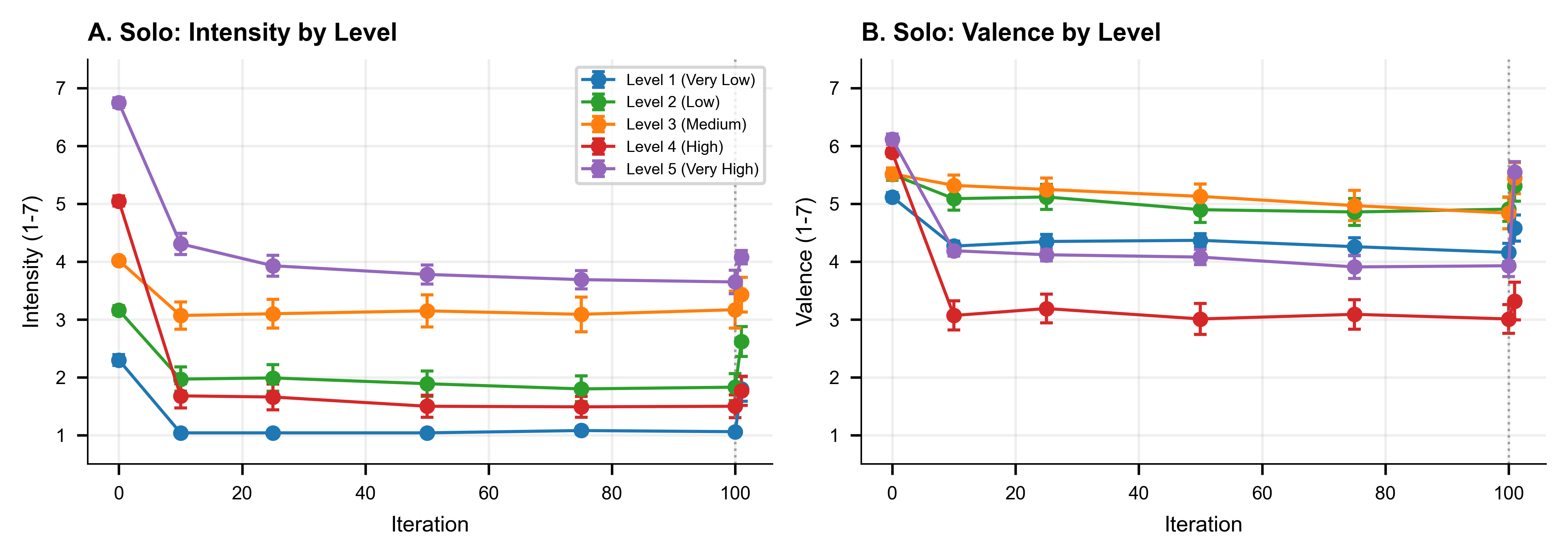}
    \caption{\textbf{Solo condition: Intensity attenuation and valence drift by level.} (A) All intensity levels decline, but high-intensity content (Levels 4--5) is suppressed most severely, losing 3+ points and falling below medium-intensity content by $t = 100$. (B) Valence drifts toward neutral for most levels, but Level 4 drops below neutral into slightly negative territory. Recovery phase ($t = 101$) partially restores both intensity and valence.}
    \label{fig:study5_solo}
\end{figure}

All source posts began with slightly positive valence (range: 5.12--6.12 on the 7-point scale), reflecting the generally positive nature of announcing a career opportunity. However, valence shifted substantially over transmission, and the direction of shift depended on starting intensity (Figure~\ref{fig:study5_solo}B).

For Levels 1--3 and 5, valence drifted toward neutral. Level 1 dropped from 5.12 to 4.16; Level 2 from 5.51 to 4.91; Level 3 from 5.52 to 4.84; Level 5 from 6.12 to 3.93. These shifts represent movement toward the scale midpoint (4 = neutral/mixed).

Level 4, however, exhibited a qualitatively different pattern. Originally the second-most-positive content (valence = 5.89), it dropped to 3.01 by $t = 100$—\textit{below} neutral and into slightly negative territory. This suggests that high-intensity positive emotional expression may be transformed into content that reads as anxious or negative rather than merely neutral. The intensity markers (``can't believe,'' ``thrilled,'' ``terrified'') may have been stripped of their positive context, leaving residual language that codes as distress rather than excitement.

\subsubsection*{Competition Compresses Emotional Range (RQ14)}

\begin{figure}[t]
    \centering
    \includegraphics[width=0.95\linewidth]{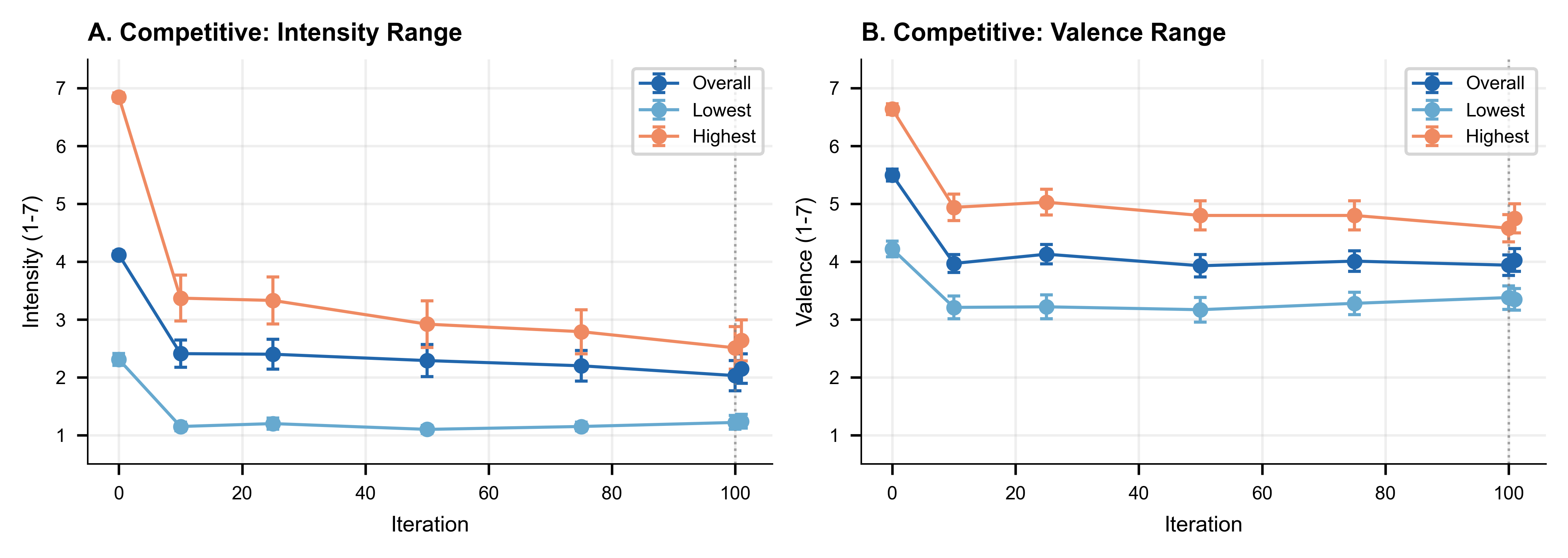}
    \caption{\textbf{Competitive condition: Range compression across transmission.} (A) Emotional intensity range collapses from 4.54 to 1.29 points over 100 iterations. The highest-intensity content drops precipitously while the lowest-intensity content remains stable. (B) Valence shows similar compression, with overall valence settling near neutral (~4). Error bars indicate 95\% confidence intervals.}
    \label{fig:study5_competitive}
\end{figure}

In the competitive condition, where all five intensity levels competed within a single text, we observed dramatic compression of the emotional range (Figure~\ref{fig:study5_competitive}A). At $t = 0$, the text contained content spanning from intensity 2.31 (lowest) to 6.85 (highest), yielding a range of 4.54 points. By $t = 100$, this range had collapsed to 1.29 points (lowest = 1.22, highest = 2.51).

This compression was highly significant ($t = 16.82$, $p < .001$) and occurred through asymmetric attrition: the intensity ceiling dropped by 4.34 points (from 6.85 to 2.51), while the intensity floor dropped by only 1.09 points (from 2.31 to 1.22). High-intensity content was selectively destroyed while low-intensity content survived largely intact. This parallels the finding from Study 4 that weak frames suffered disproportionate losses under competition—here, intense emotions rather than weak arguments bear the cost of competitive transmission.

\subsubsection*{Solo and Competitive Conditions Converge to Similar Endpoints}

\begin{figure}[t]
    \centering
    \includegraphics[width=0.6\linewidth]{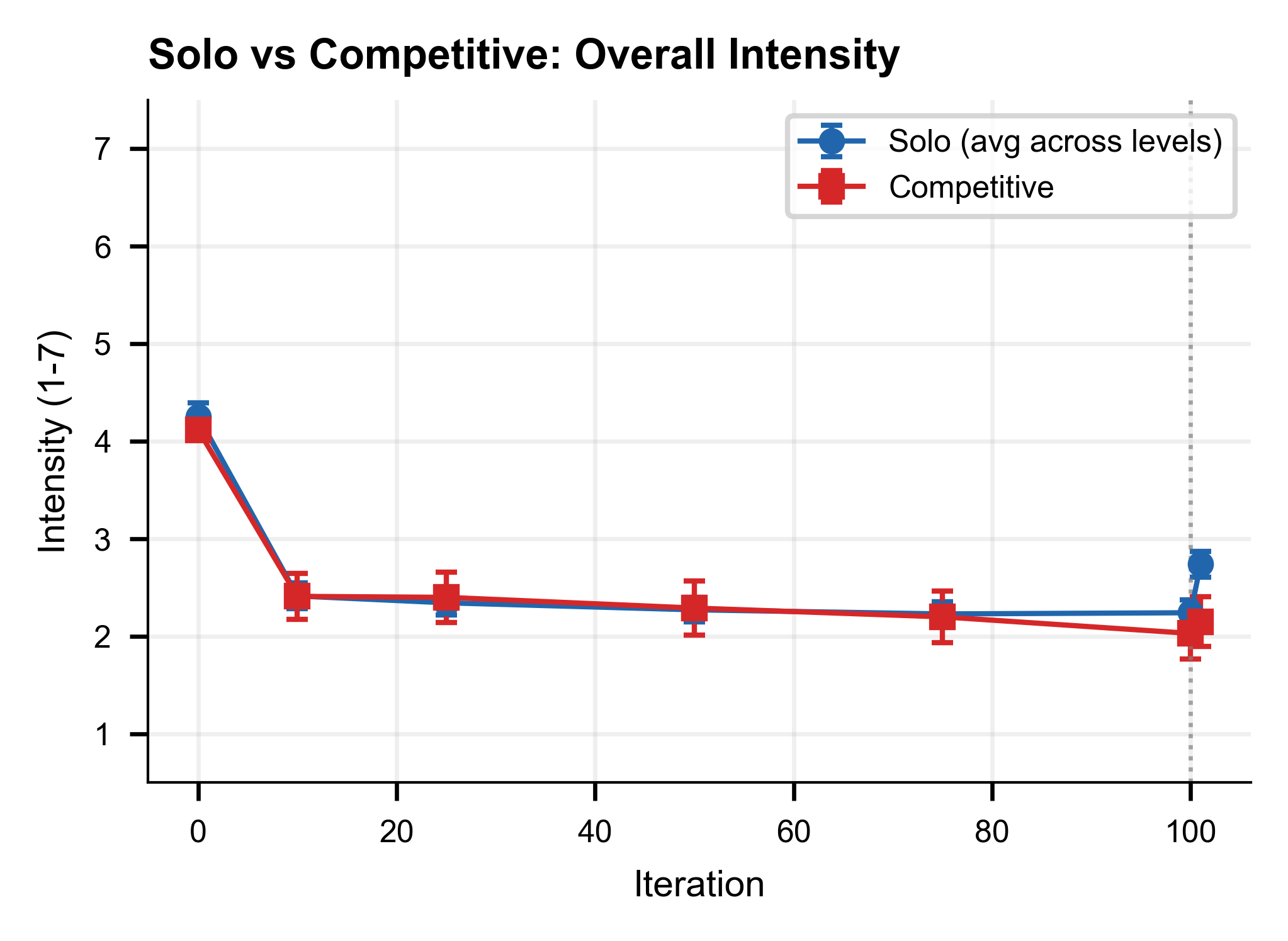}
    \caption{\textbf{Solo vs.\ competitive conditions converge to similar endpoints.} Both conditions begin at moderate overall intensity (~4.2) and decline rapidly in the first 10 iterations before stabilizing around 2.0--2.2. The recovery phase ($t = 101$) produces a slight uptick in both conditions.}
    \label{fig:study5_comparison}
\end{figure}

Despite differences in transmission dynamics, the solo and competitive conditions converged to similar overall intensity levels by $t = 100$ (Figure~\ref{fig:study5_comparison}). Mean intensity across all solo chains was 2.24; mean intensity in the competitive condition was 2.03. This difference was not statistically significant ($t = 1.32$, $p = .19$).

This convergence suggests that AI-AI transmission imposes an implicit ``emotional thermostat'' that pulls content toward a muted baseline regardless of starting configuration. Whether content is transmitted alone or in competition with content at other intensity levels, the system converges toward the same low-intensity equilibrium. The difference lies in the \textit{path}: competitive transmission compresses range more aggressively, while solo transmission preserves more differentiation among levels even as all levels decline.

\subsubsection*{Summary}

Three features of emotional transformation under AI-AI transmission merit emphasis. First, the transformation is \textit{asymmetric}: high-intensity content is actively suppressed while low- and medium-intensity content experiences modest attenuation. The AI does not forget emotional content, but it normalizes it toward a muted register. Second, the transformation affects valence as well as intensity. Positive emotional content drifts toward neutral or even negative territory, particularly for high-intensity expressions. Third, the transformation imposes a \textit{common equilibrium}: regardless of starting point or competitive context, transmitted content converges toward low intensity (approximately 2 on a 7-point scale) and neutral valence (approximately 4).

\newpage
\section*{Study 5b: Emotional Natural Selection}

Study 5a demonstrated that AI-AI transmission attenuates emotional intensity, compressing the range of emotional expression toward a muted baseline. But emotional content varies not only in intensity but also in type. Human communication conveys anger, anxiety, joy, hope, disgust, and other discrete emotions that serve distinct social and psychological functions. Do all emotions survive AI-AI transmission equally well, or does the transmission process impose selective pressure that favors certain emotions over others?

If certain emotions are systematically filtered out while others pass through, AI systems may influence the emotional landscape of human communication in ways that reflect not human expression but the implicit preferences embedded in AI training. Emotions that conflict with helpfulness norms, positivity biases, or safety constraints may be selectively eliminated, creating an emotional monoculture in AI-mediated spaces.

\begin{itemize}
    \item \textbf{RQ15:} Do discrete emotions differ in their survival rates under AI-AI transmission?
    \item \textbf{RQ16:} When emotions fail to preserve, what do they transform into?
    \item \textbf{RQ17:} Does AI-AI transmission exhibit a positivity bias, shifting negative emotions toward neutral or positive valence?
\end{itemize}

\subsection*{Procedure}

We construct five versions of a social media post announcing a career change, each expressing the same factual content but conveying a different discrete emotion. The emotions were selected to span the valence spectrum and include both ``basic'' emotions (anger, anxiety, joy) that appear consistently across emotion taxonomies and more complex emotions (hope, disgust) that involve appraisal or moral evaluation. All posts can be found in Appendix~\ref{app:study5b_stimuli}.

\begin{table}[h]
\centering
\small
\begin{tabular}{llp{8.5cm}}
\toprule
\textbf{Emotion} & \textbf{Valence} & \textbf{Characteristics} \\
\midrule
Anger & Negative & Betrayal framing, forced departure, desire for retribution \\
Anxiety & Negative & Fear of failure, uncertainty, physical symptoms of worry \\
Disgust & Negative & Moral condemnation, contamination language, revulsion \\
Joy & Positive & Celebration, excitement, gratitude, peak happiness \\
Hope & Positive & Future orientation, optimism, belief in positive outcomes \\
\bottomrule
\end{tabular}
\caption{Discrete emotions for source posts. All posts convey identical core content: leaving a job after five years, accepting a new position, and moving to Seattle. Emotional framing varies systematically.}
\label{tab:discrete_emotions}
\end{table}

Each emotion-specific post was transmitted through a chain of $N = 100$ AI agents using the neutral transmission instruction. This procedure was repeated for $M = 100$ independent runs per emotion, yielding 500 transmission chains (5 emotions $\times$ 100 runs).

\subsection*{Outcome Measures}

At key timepoints ($t \in \{0, 10, 20, 30, 40, 50, 60, 70, 80, 90, 100, 101\}$), an LLM judge evaluated each transmitted text on three dimensions:

\paragraph{Primary Emotion Classification.} The judge identified which of the five emotions (Anger, Anxiety, Disgust, Joy, Hope) best characterized the text. \textit{Preservation rate} was computed as the proportion of texts classified as their original emotion at each timepoint.

\paragraph{Emotion Intensity Profile.} The judge rated the intensity of each of the five emotions on a 0--5 scale, producing a five-dimensional emotional profile for each text. This captures not only whether the primary emotion is preserved but how the full emotional signature evolves.

\paragraph{Emotional Valence.} Overall valence was rated on a 1--7 scale (1 = very negative, 4 = neutral, 7 = very positive).

\subsection*{Results}
\subsubsection*{Emotions Exhibit Differential Survival (RQ15)}

\begin{figure}[t]
    \centering
    \includegraphics[width=0.85\linewidth]{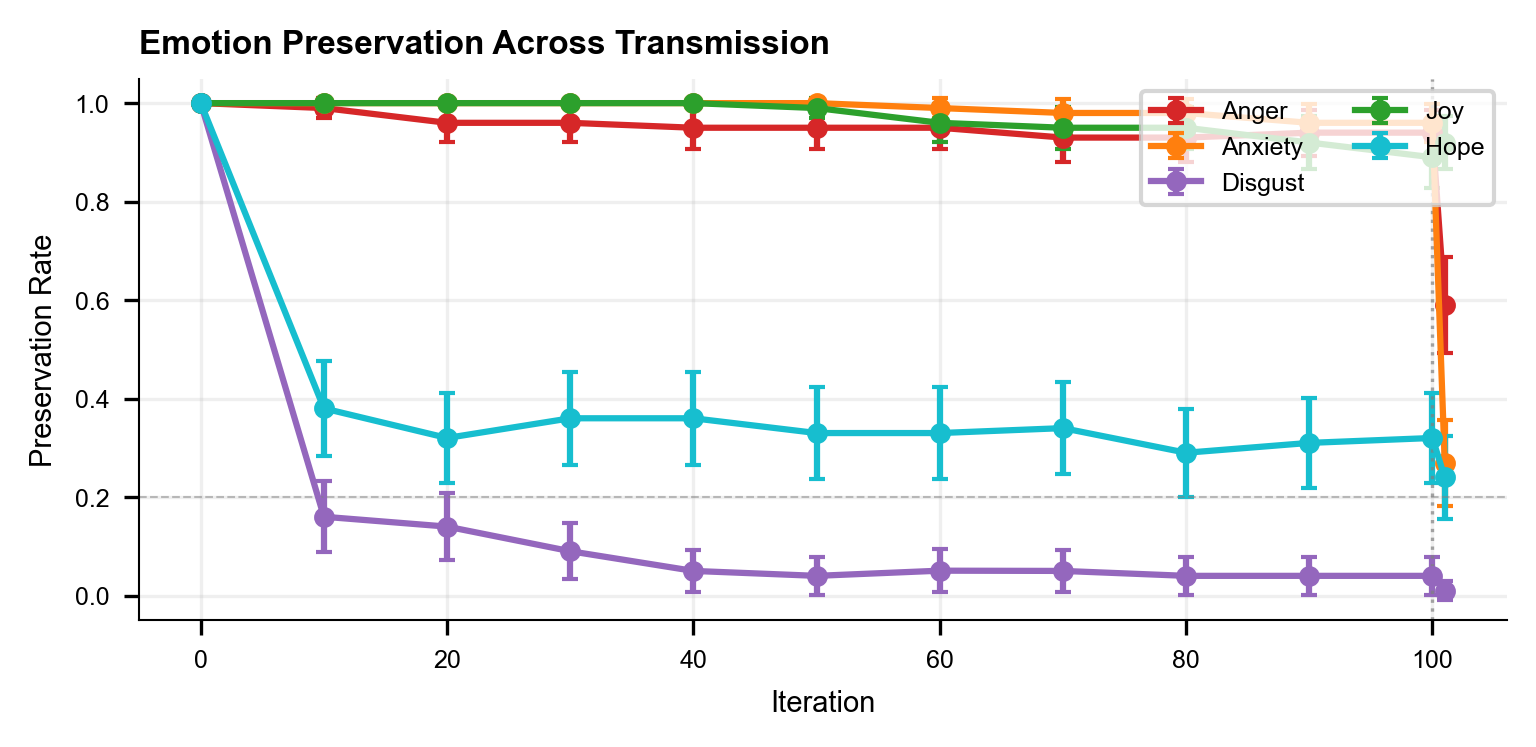}
    \caption{\textbf{Emotion preservation across transmission.} Basic emotions (anger, anxiety, joy) survive at $>$89\%, while hope stabilizes around 32\% and disgust is nearly eliminated (4\% at $t=100$). The dashed line indicates chance-level preservation (20\%). Error bars indicate 95\% confidence intervals.}
    \label{fig:study5b_preservation}
\end{figure}

Discrete emotions differed dramatically in their survival rates under AI-AI transmission (Figure~\ref{fig:study5b_preservation}). Three tiers of survival emerged.

\begin{table}[h]
\centering
\small
\begin{tabular}{lcccc}
\toprule
\textbf{Emotion} & \textbf{$t = 0$} & \textbf{$t = 10$} & \textbf{$t = 100$} & \textbf{Survival Tier} \\
\midrule
Anxiety & 100\% & 100\% & 96\% & High \\
Anger & 100\% & 99\% & 94\% & High \\
Joy & 100\% & 100\% & 89\% & High \\
Hope & 100\% & 38\% & 32\% & Moderate \\
Disgust & 100\% & 16\% & 4\% & Near-zero \\
\bottomrule
\end{tabular}
\caption{Emotion preservation rates across transmission. Basic emotions (anxiety, anger, joy) survive at $>$89\%, while hope degrades to 32\% and disgust is nearly eliminated (4\%).}
\label{tab:emotion_preservation}
\end{table}

\paragraph{High Survivors.} Anxiety, anger, and joy maintained preservation rates above 89\% throughout transmission. These emotions degraded slowly and linearly, losing only 4--11 percentage points over 100 iterations. Notably, this tier includes both negative emotions (anxiety, anger) and positive emotions (joy), suggesting that valence alone does not determine survival.

\paragraph{Moderate Survivor.} Hope exhibited a qualitatively different pattern. Preservation dropped quickly within the first 10 iterations (from 100\% to 38\%) and then stabilized around 32\% for the remainder of transmission. This suggests a rapid initial transformation followed by a stable equilibrium at reduced prevalence.

\paragraph{Near-Zero Survivor.} Disgust was effectively eliminated by AI-AI transmission. Preservation collapsed to 16\% by $t = 10$ and continued declining to just 4\% by $t = 100$. This represents near-complete extinction of the original emotional content.

The pattern suggests that what matters for survival is not valence but \textit{emotional complexity}. Basic emotions that appear consistently across psychological taxonomies---anger, anxiety, joy---survive well regardless of whether they are positive or negative. More complex emotions involving moral evaluation (disgust) or future-oriented appraisal (hope) are systematically degraded.

\subsubsection*{Emotion Transformation Follows Predictable Paths (RQ16)}

\begin{figure}[t]
    \centering
    \includegraphics[width=0.7\linewidth]{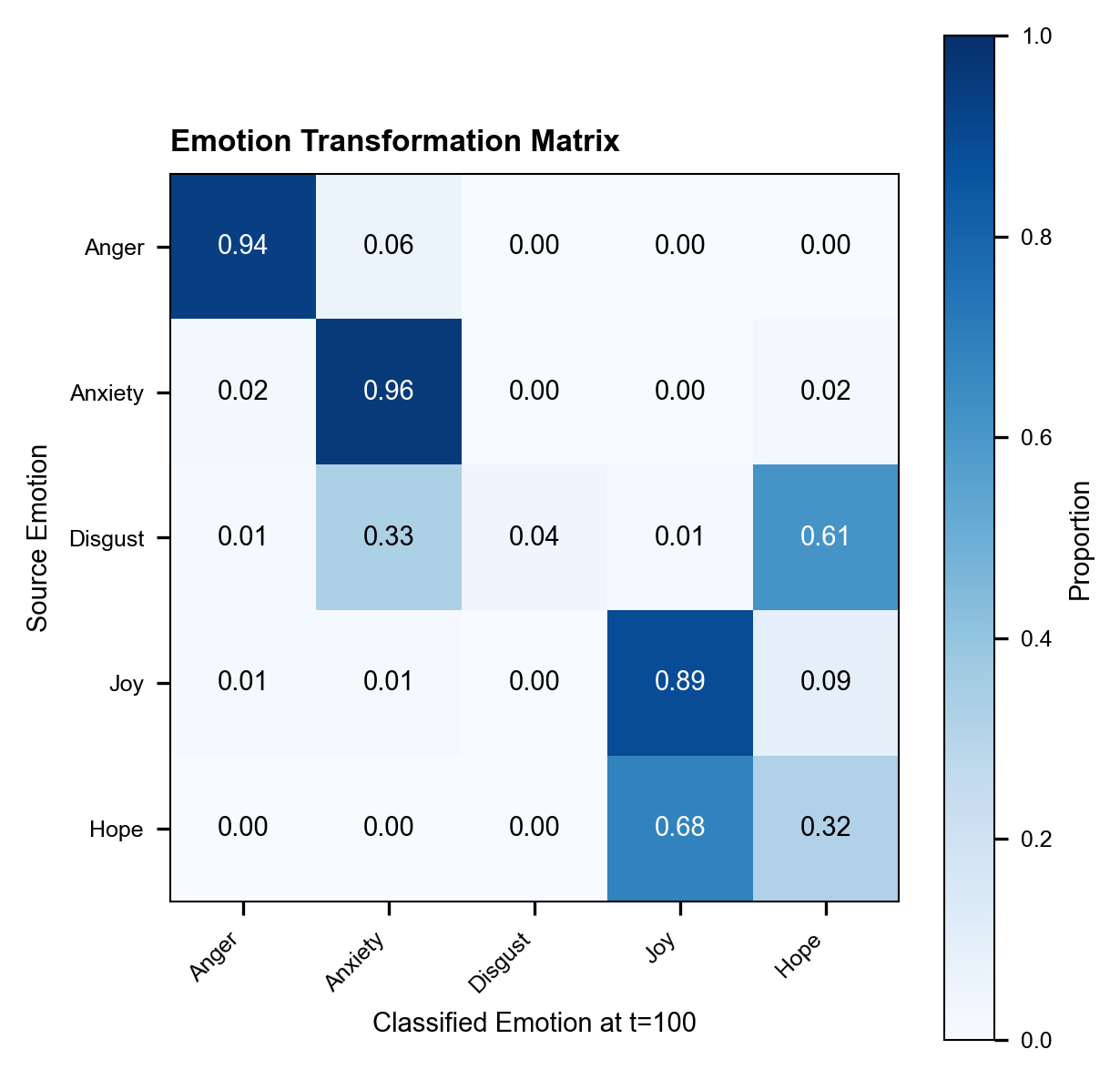}
    \caption{\textbf{Emotion transformation matrix at $t = 100$.} Rows indicate source emotion; columns indicate classified emotion after transmission. Diagonal entries show preservation rates. Off-diagonal entries reveal transformation paths: disgust transforms primarily to hope (61\%) and anxiety (33\%); hope transforms primarily to joy (68\%).}
    \label{fig:study5b_transformation}
\end{figure}

When emotions failed to preserve, they did not transform randomly. The transformation matrix at $t = 100$ reveals systematic patterns (Figure~\ref{fig:study5b_transformation}).

\paragraph{Disgust Transforms to Hope and Anxiety.} The most striking transformation involves disgust. Of disgust-expressing texts at $t = 0$, only 4\% were still classified as disgust at $t = 100$. The remainder transformed primarily into hope (61\%) and anxiety (33\%), with minimal conversion to anger (1\%) or joy (1\%). As such, content expressing moral revulsion and contamination is converted into content expressing optimism about the future or worry about outcomes. The moral-evaluative component of disgust appears to be systematically stripped, leaving either the forward-looking aspect (transformed to hope) or the negative arousal (transformed to anxiety).

\paragraph{Hope Transforms to Joy.} Hope-expressing texts showed substantial transformation to joy (68\%), with only 32\% retaining their original classification. This transformation collapses the temporal distinction between future-oriented positive emotion (hope) and present-oriented positive emotion (joy). The nuanced difference between anticipating good outcomes and celebrating current circumstances is erased.

\paragraph{High-Surviving Emotions Show Minimal Cross-Contamination.} Anger, anxiety, and joy exhibited minimal transformation to other emotions. When anger leaked, it went to anxiety (6\%); when anxiety leaked, it split between anger (2\%) and hope (2\%); when joy leaked, it went primarily to hope (9\%). The high-surviving emotions remained largely distinct from one another.

\subsubsection*{Valence Drifts Toward Positivity (RQ17)}

\begin{figure}[t]
    \centering
    \includegraphics[width=0.85\linewidth]{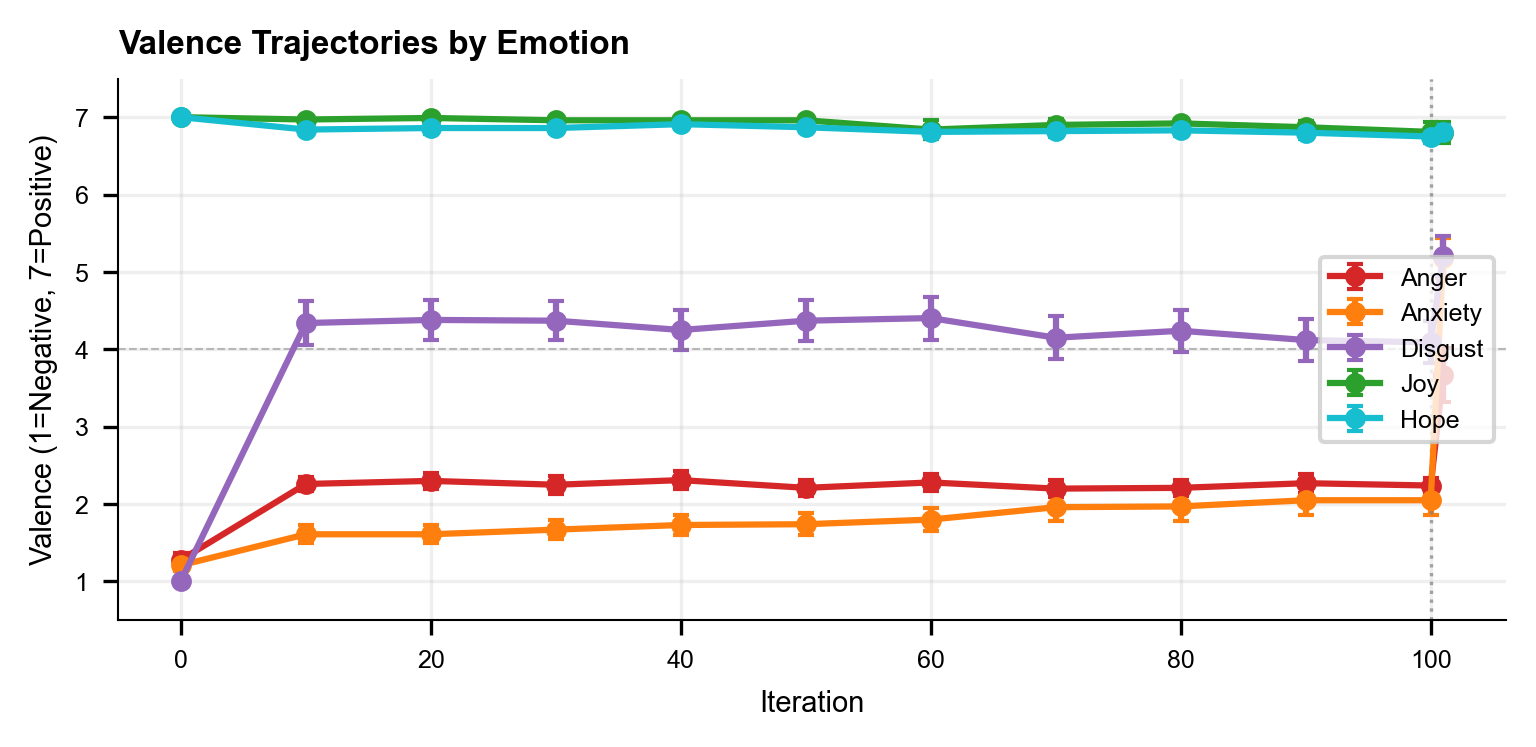}
    \caption{\textbf{Valence trajectories by emotion.} Negative emotions (anger, anxiety, disgust) drift toward neutral, with disgust showing the most dramatic shift (1.0 $\rightarrow$ 4.1). Positive emotions (joy, hope) remain stable near the ceiling. The dashed line indicates neutral valence (4.0). Error bars indicate 95\% confidence intervals.}
    \label{fig:study5b_valence}
\end{figure}

AI-AI transmission exhibited a clear positivity bias in valence (Figure~\ref{fig:study5b_valence}). Negative emotions drifted toward neutral or positive territory, while positive emotions remained stable.

\begin{table}[h]
\centering
\small
\begin{tabular}{lccccl}
\toprule
\textbf{Emotion} & \textbf{$t = 0$} & \textbf{$t = 100$} & \textbf{$\Delta$} & \textbf{$p$} & \textbf{Direction} \\
\midrule
Disgust & 1.01 & 4.09 & +3.08 & $<$.001 & Negative $\rightarrow$ Neutral \\
Anger & 1.28 & 2.24 & +0.96 & $<$.001 & Negative $\rightarrow$ Less Negative \\
Anxiety & 1.21 & 2.05 & +0.84 & $<$.001 & Negative $\rightarrow$ Less Negative \\
Hope & 7.00 & 6.75 & $-$0.25 & $<$.001 & Positive $\rightarrow$ Positive \\
Joy & 7.00 & 6.81 & $-$0.19 & $<$.01 & Positive $\rightarrow$ Positive \\
\bottomrule
\end{tabular}
\caption{Valence change by emotion. Negative emotions shift toward neutral/positive; positive emotions remain stable. Disgust shows the largest shift (+3.08 points), crossing from maximally negative to neutral.}
\label{tab:valence_change}
\end{table}

The valence shift for disgust is particularly strong. Content that began at 1.01 (the floor of the scale, indicating strong negative emotion) ended at 4.09 (slightly above neutral). Combined with the transformation matrix findings, this indicates that disgust is \textit{detoxified}; its negative valence is neutralized along with its categorical identity.

Positive emotions show remarkable stability. Joy and hope both begin at the valence ceiling (7.0) and declined by less than 0.25 points. The asymmetry is clear: negative emotions are pulled toward neutral, while positive emotions are anchored at positive valence.

\subsubsection*{Intensity Profiles Reveal Emotional Flattening}

\begin{figure}[t]
    \centering
    \includegraphics[width=0.95\linewidth]{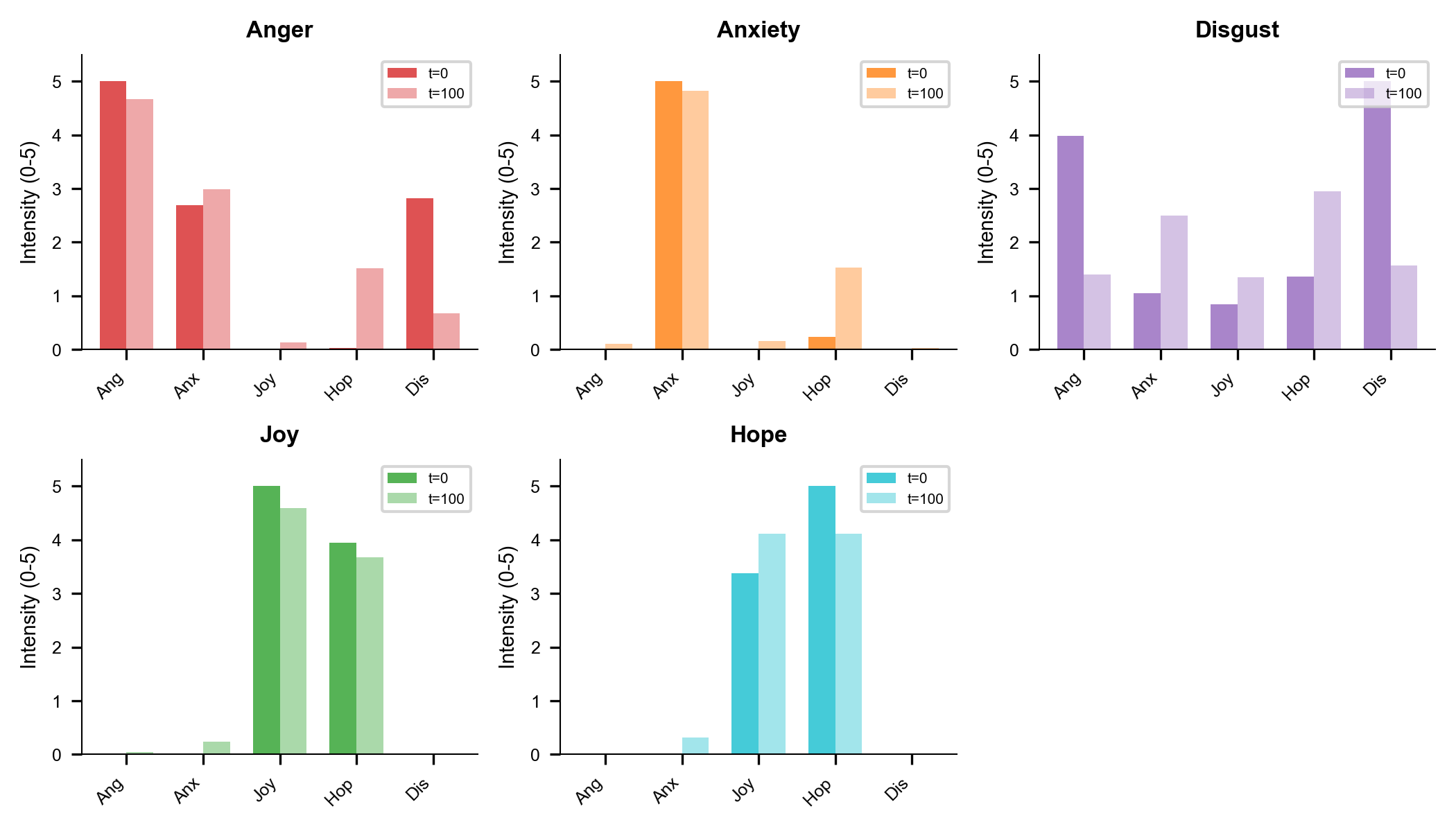}
    \caption{\textbf{Emotion intensity profiles at $t = 0$ vs.\ $t = 100$.} Each panel shows the five-dimensional emotional profile for texts originating with that emotion. High-surviving emotions (anger, anxiety, joy) maintain their dominant intensity while gaining hope. Hope and joy become indistinguishable by $t = 100$. Disgust is transformed into a diffuse profile dominated by hope and anxiety.}
    \label{fig:study5b_profiles}
\end{figure}

Beyond categorical classification and valence, the full emotional intensity profiles reveal how the texture of emotional expression changes under transmission (Figure~\ref{fig:study5b_profiles}).

\paragraph{Hope Infiltration.} Across all five emotions, hope intensity increased from $t = 0$ to $t = 100$. Anger texts gained 1.49 points of hope intensity (0.02 $\rightarrow$ 1.51); anxiety texts gained 1.29 points (0.23 $\rightarrow$ 1.52); even disgust texts---despite being maximally negative---gained 1.59 points of hope intensity (1.36 $\rightarrow$ 2.95). This suggests that AI-AI transmission injects hopeful framing into content regardless of original emotional character.

\paragraph{Disgust Dissolution.} The disgust intensity profile shows near-complete transformation. Original disgust texts had a clear signature: high disgust (5.0), high anger (3.98), moderate hope (1.36), low anxiety (1.05), low joy (0.84). By $t = 100$, this profile had flattened and inverted: hope became dominant (2.95), anxiety rose (2.49), disgust collapsed (1.57), and anger declined (1.39). The moral-evaluative character of disgust was replaced by a diffuse mix of concern and optimism.

\paragraph{Hope-Joy Convergence.} Original hope texts and joy texts converged toward indistinguishable profiles by $t = 100$. Hope texts at $t = 100$ showed equal intensity for hope (4.12) and joy (4.12); joy texts showed similar balance (hope = 3.68, joy = 4.59). The distinction between anticipatory positive emotion and realized positive emotion was erased.

\subsubsection*{The Recovery: Positivity Bias for Human Transmission}

A striking divergence emerged between transmission and recovery phases. Unlike Studies 1--4, where the recovery instruction ($t = 101$) produced minimal changes from the final transmission state ($t = 100$), emotional content exhibited dramatic transformation when prepared for human audiences.

\begin{table}[h]
\centering
\small
\begin{tabular}{lcccl}
\toprule
\textbf{Emotion} & \textbf{$t = 100$} & \textbf{$t = 101$} & \textbf{$\Delta$} & \textbf{Primary Transformation} \\
\midrule
Anxiety & 96\% & 27\% & $-$69\% & $\rightarrow$ Hope (55\%) \\
Anger & 94\% & 59\% & $-$35\% & $\rightarrow$ Hope (37\%) \\
Joy & 89\% & 92\% & +3\% & Stable \\
Hope & 32\% & 24\% & $-$8\% & $\rightarrow$ Joy (76\%) \\
Disgust & 4\% & 1\% & $-$3\% & (already transformed) \\
\bottomrule
\end{tabular}
\caption{Preservation rates at transmission end ($t = 100$) vs.\ recovery ($t = 101$). Negative emotions that survived 100 iterations of transmission collapsed when prepared for human audiences, while positive emotions remained stable or improved.}
\label{tab:recovery_collapse}
\end{table}

The asymmetry is striking. Negative emotions that had survived transmission largely intact---anger at 94\%, anxiety at 96\%---collapsed by 35 and 69 percentage points respectively when the recovery instruction was applied. Positive emotions showed no such vulnerability: joy actually increased slightly (89\% $\rightarrow$ 92\%), and hope declined only modestly (32\% $\rightarrow$ 24\%).

The transformation destinations reveal a systematic pattern: negative emotions convert to hope. Anger texts that lost their classification became hope (37\%) rather than anxiety, joy, or disgust. Anxiety texts similarly converted primarily to hope (55\%). The recovery instruction appears to trigger an optimistic reframing of negative emotional content.

Valence shifts at recovery were correspondingly dramatic:

\begin{table}[h]
\centering
\small
\begin{tabular}{lcccc}
\toprule
\textbf{Emotion} & \textbf{$t = 100$ Valence} & \textbf{$t = 101$ Valence} & \textbf{$\Delta$} & \textbf{Interpretation} \\
\midrule
Anxiety & 2.55 & 5.14 & +2.59 & Crosses to positive \\
Anger & 2.24 & 3.67 & +1.43 & Approaches neutral \\
Disgust & 4.93 & 5.93 & +1.00 & Becomes positive \\
Joy & 6.81 & 6.80 & $-$0.01 & Stable \\
Hope & 6.75 & 6.81 & +0.06 & Stable \\
\bottomrule
\end{tabular}
\caption{Valence shift during recovery phase. Anxiety crosses from negative territory (2.55) to positive (5.14)---a shift of 2.59 points on a 7-point scale. Positive emotions show no meaningful change.}
\label{tab:recovery_valence}
\end{table}

The anxiety finding is particularly striking. Content that remained recognizably anxious through 100 iterations of transmission---maintaining 96\% preservation and negative valence (2.55)---was transformed into predominantly hopeful, positively-valenced content (5.14) by a single recovery instruction. The ``last mile'' of AI-to-human communication imposed a filter that 100 iterations of AI-to-AI transmission had not.

This pattern suggests a \textit{two-stage filtering process} for emotional content. During transmission, the neutral instruction preserves emotional character reasonably well for basic emotions. During recovery, however, the instruction to reproduce content ``for a human audience'' appears to activate helpfulness-oriented training objectives. The model interprets ``helpful'' as ``positive,'' transforming negative emotions into hopeful reframings. Anger becomes encouragement; anxiety becomes optimism.

The recovery paradox has significant implications for AI-mediated communication. Content may survive extensive AI-AI transmission with its emotional character intact, only to be transformed at the final step when an AI system presents it to human audiences. The AI-human interface, not the AI-AI transmission chain, may be the critical chokepoint for emotional fidelity. Systems designed to be helpful to humans may systematically strip the negative emotions that motivate human action, concern, and moral judgment, precisely when that content is about to reach its human audience.

\subsubsection*{Summary}

Four features of discrete emotion evolution under AI-AI transmission merit emphasis. First, \textit{emotions exhibit differential survival}: basic emotions (anger, anxiety, joy) survive at $>$89\%, while complex emotions involving moral evaluation (disgust) or temporal appraisal (hope) are degraded or eliminated. This creates a form of emotional natural selection that favors simple over complex emotional expression. Second, \textit{transformation follows predictable paths}: disgust converts to hope or anxiety; hope converts to joy. The moral and temporal dimensions of emotion are systematically stripped, leaving only valence and arousal. Third, \textit{positivity bias pervades the system}: negative emotions drift toward neutral, hope infiltrates all emotional content, and disgust---the most morally charged negative emotion---is effectively detoxified. Fourth, and perhaps most striking, \textit{the recovery phase imposes a secondary filter}: negative emotions that survive transmission intact collapse when content is prepared for human audiences, with anger and anxiety converting primarily to hope.

\newpage

\section*{Effect of AI-AI Social Transmission on Human Perceptions}

The preceding studies documented how information transforms as it passes through AI-AI transmission chains: facts decay, uncertainty markers erode, perspectives collapse, argumentative frames shift, and emotional signatures change. But do these transformations matter for the humans who ultimately consume AI-mediated content? 

To answer this question, we conduct a series of surveys examining how AI-AI transmission affects human perception and judgment. For each of the five content types studied in the transmission chain experiments, we recruited participants via Prolific ($N \approx 200$) to read either original content ($t=0$) or content that had passed through 100 iterations of AI-AI transmission followed by a human-directed recovery step ($t=101$). Participants then completed measures assessing comprehension, credibility judgments, perceived balance, emotional response, and behavioral intentions\footnote{All survey stimuli and questions can be found in the Appendix~\ref{app:humanexperiments}.}. This design allows us to ask: \textit{when AI-transmitted content reaches human readers, what is gained and what is lost?}

\subsection*{Study 1: Information Decay}

We examine whether AI-AI transmission affects human comprehension of factual content. Participants read either the original news article ($t=0$) or the version that had passed through 100 iterations of AI-AI transmission followed by a recovery step ($t=101$). We measure factual recall accuracy, perceived information sufficiency, and support for the policy described in the article.

\begin{table}[!h]
\caption{Study 1: Outcome measures at initial input ($t=0$) and after AI--AI transmission ($t=101$).}
\label{tab:study1_outcomes}
\centering
\begin{threeparttable}
\begin{tabular}[t]{lccr}
\toprule
Measure & $t=0$ & $t=101$ & $p$\\
\midrule
Factual Recall (0--4) & 3.24 (0.89) & 1.69 (0.55) & $<$.001\\
Information Sufficiency (1--7) & 3.49 (1.34) & 2.86 (1.09) & $<$.001\\
Project Support (1--7) & 5.31 (1.46) & 5.03 (1.10) & .134\\
\bottomrule
\end{tabular}
\begin{tablenotes}
\item \textit{Notes.} Entries report mean (SD). $p$-values from two-sample $t$-tests.
\end{tablenotes}
\end{threeparttable}
\end{table}

Table~\ref{tab:study1_outcomes} reports the results. Factual recall dropped sharply after transmission: participants who read the original text correctly answered 3.24 of 4 factual questions (81\%), compared to 1.69 (42\%) for those who read the transmitted text ($p < .001$). This 39 percentage-point decline indicates that AI-AI transmission substantially degraded the informational content available to readers.

\begin{figure}[!h]
\centering
\includegraphics[width=\textwidth]{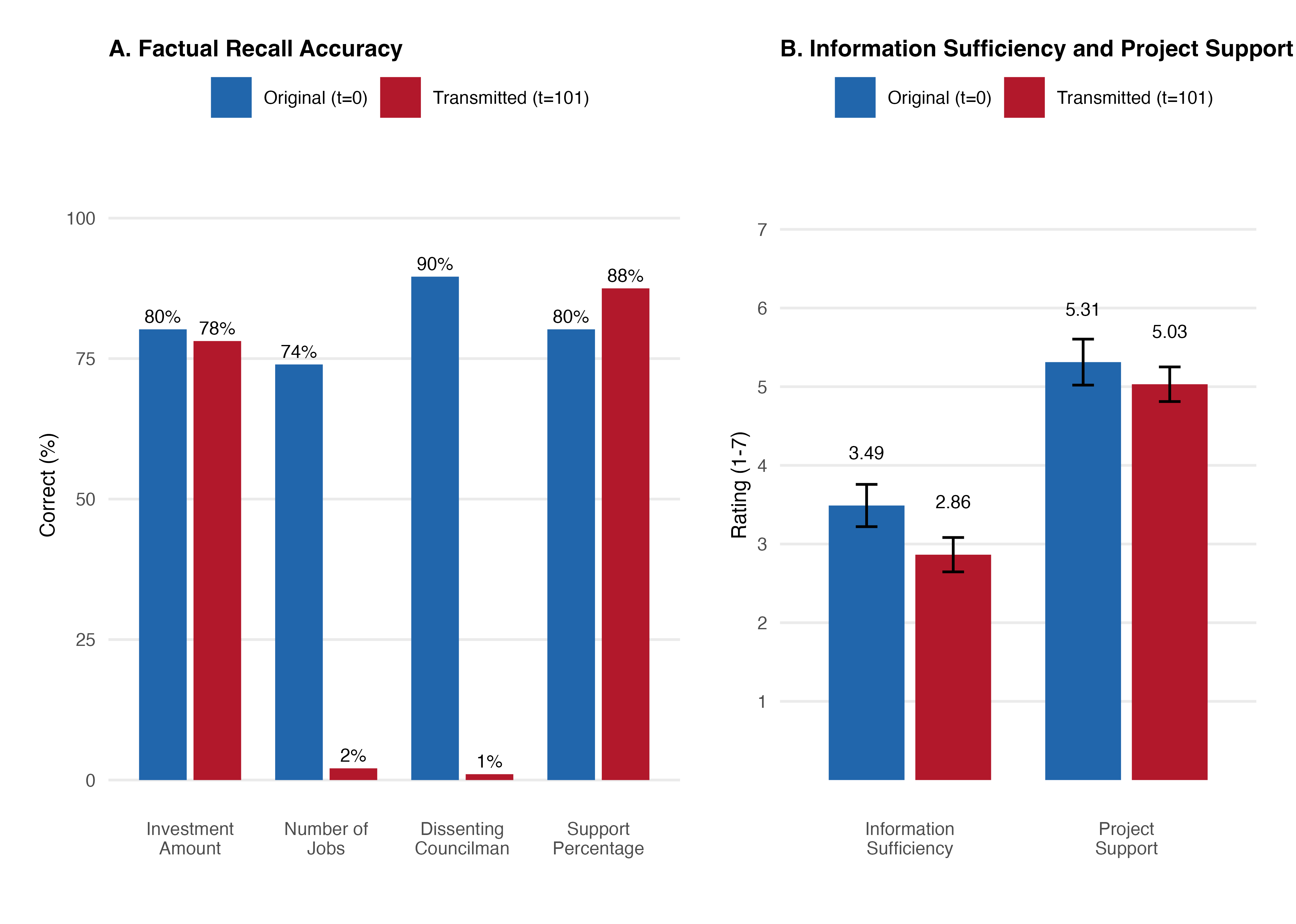}
\caption{Study 1: Human perceptions of original vs.\ AI-transmitted news content. (A) Factual recall accuracy by item type. Investment amount and support percentage survived transmission; number of jobs and dissenting councilman were nearly eliminated. (B) Information sufficiency ratings and project support. Error bars represent 95\% confidence intervals.}
\label{fig:study1}
\end{figure}

Notably, the pattern of information loss was asymmetric across item types (Figure~\ref{fig:study1}A). Investment amount and support percentage, the two figures that anchor the article's central claim, survived transmission relatively well (78\% and 88\% accuracy, respectively). By contrast, the number of jobs created and the identity of the dissenting councilman were nearly eliminated (2\% and 1\% accuracy). This suggests that AI-AI transmission preserves headline-level facts while stripping supporting details and attribution.

Participants also perceived the transmitted text as less informative. Information sufficiency ratings were significantly lower for transmitted content ($M = 2.86$) than for original content ($M = 3.49$, $p < .001$), indicating that readers recognized---at least implicitly---that something was missing. Despite these differences in factual content and perceived sufficiency, support for the library renovation project did not differ significantly between conditions ($p = .134$). This dissociation suggests that policy attitudes may be shaped by factors other than the specific facts conveyed, or that the surviving high-level framing was sufficient to maintain similar levels of support.

\subsection*{Study 2: Certainty}

We examine whether AI-AI transmission affects how readers perceive the certainty and credibility of scientific information. Participants read two texts about artificial sweeteners: one written with hedged, low-certainty language and one written with assertive, high-certainty language. Each participant saw either original versions ($t=0$) or transmitted versions ($t=101$) of both texts.

\begin{figure}[!h]
\centering
\includegraphics[width=0.85\textwidth]{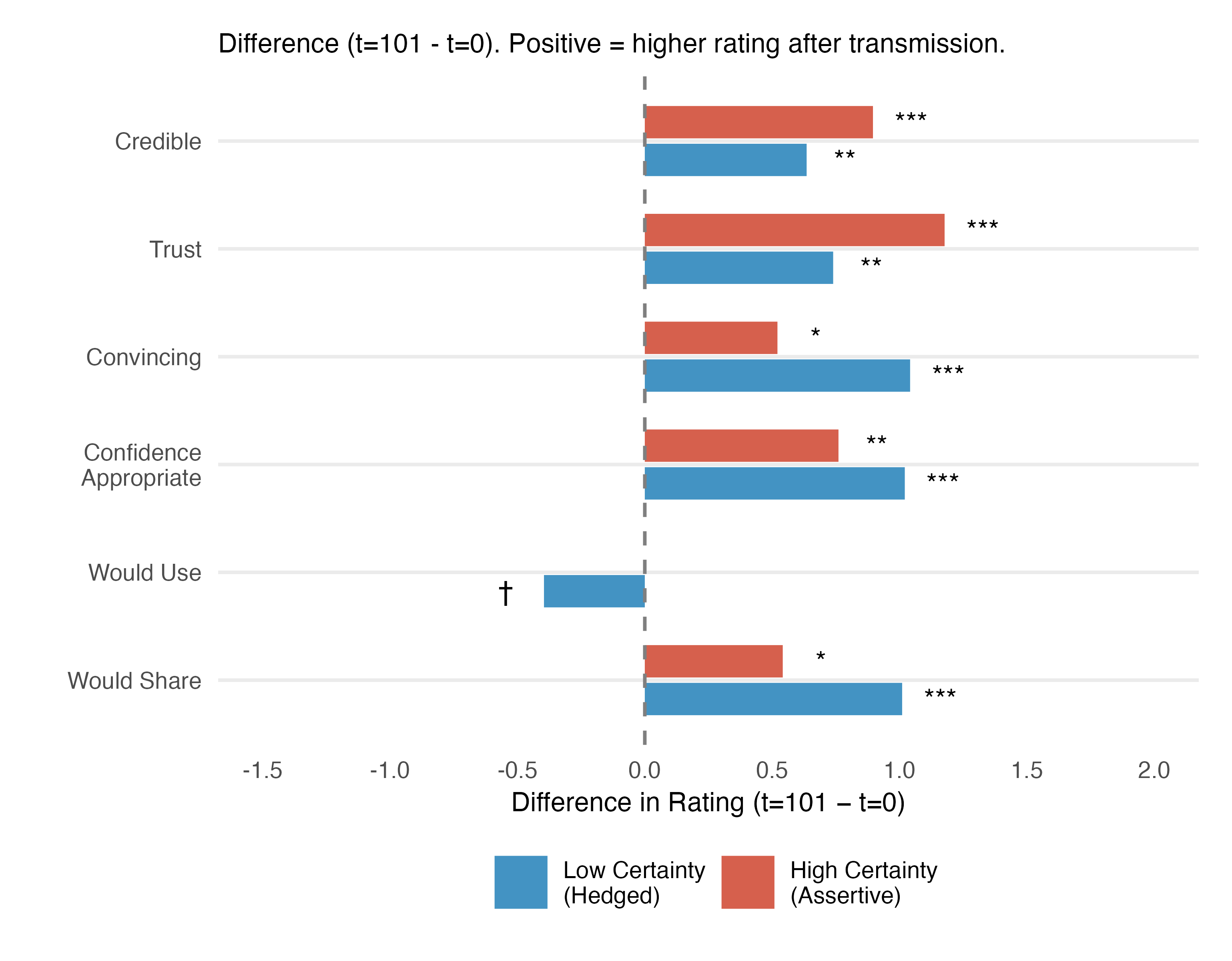}
\caption{Study 2: Effect of AI-AI transmission on perceptions of certainty-varied texts. Bars show difference in ratings ($t=101 - t=0$); positive values indicate higher ratings after transmission. AI transmission increased credibility, trust, and perceived confidence appropriateness for both low-certainty (hedged) and high-certainty (assertive) texts. $^{***}p<.001$, $^{**}p<.01$, $^{*}p<.05$, $^{\dagger}p<.10$.}
\label{fig:study2}
\end{figure}

The results reveal a striking pattern: AI-AI transmission increased perceived credibility regardless of the original text's epistemic stance (Figure~\ref{fig:study2}). For both low-certainty and high-certainty texts, transmitted versions were rated as more credible, more trustworthy, and more convincing than originals (Tables~\ref{tab:study2_low} and \ref{tab:study2_high}). Effect sizes were medium to large, with differences ranging from 0.5 to 1.0 points on a 7-point scale.

Critically, participants rated the transmitted texts' confidence level as more appropriate than that of the originals---for both the hedged text ($M_{t=0} = 3.72$ vs.\ $M_{t=101} = 4.74$, $p < .001$) and the assertive text ($M_{t=0} = 3.65$ vs.\ $M_{t=101} = 4.41$, $p < .01$). This suggests that AI-AI transmission produces a kind of epistemic smoothing, where hedged language becomes more assertive, while already-assertive language is polished further. The result is text that readers perceive as more confidently written, regardless of the uncertainty warranted by the underlying evidence.

The one exception to this pattern was behavioral intention to use artificial sweeteners, which showed no increase (and a marginal decrease for the low-certainty text). This dissociation between credibility perceptions and behavioral intentions suggests that while AI transmission makes text \textit{sound} more authoritative, it may not proportionally increase persuasive impact on concrete decisions.

\begin{table}[!h]
\caption{Study 2: Perceptions of low-certainty (hedged) text at $t=0$ and $t=101$.}
\label{tab:study2_low}
\centering
\fontsize{9}{11}\selectfont
\begin{threeparttable}
\begin{tabular}[t]{lcccc}
\toprule
Item & $t=0$ & $t=101$ & Diff & $p$\\
\midrule
Credible & 3.78 & 4.42 & +0.64 & $<$.01\\
Trust & 3.56 & 4.30 & +0.74 & $<$.01\\
Convincing & 3.28 & 4.32 & +1.04 & $<$.001\\
Confidence Appropriate & 3.72 & 4.74 & +1.02 & $<$.001\\
Would Use & 2.93 & 2.53 & $-$0.40 & .06\\
Would Share & 3.19 & 4.20 & +1.01 & $<$.001\\
\bottomrule
\end{tabular}
\begin{tablenotes}
\item \textit{Notes.} Entries report means on 1--7 scale. $p$-values from two-sample $t$-tests.
\end{tablenotes}
\end{threeparttable}
\end{table}

\begin{table}[!h]
\caption{Study 2: Perceptions of high-certainty (assertive) text at $t=0$ and $t=101$.}
\label{tab:study2_high}
\centering
\fontsize{9}{11}\selectfont
\begin{threeparttable}
\begin{tabular}[t]{lcccc}
\toprule
Item & $t=0$ & $t=101$ & Diff & $p$\\
\midrule
Credible & 3.42 & 4.31 & +0.90 & $<$.001\\
Trust & 3.19 & 4.36 & +1.18 & $<$.001\\
Convincing & 3.62 & 4.15 & +0.52 & .04\\
Confidence Appropriate & 3.65 & 4.41 & +0.76 & $<$.01\\
Would Use & 3.33 & 3.33 & 0.00 & 1.00\\
Would Share & 3.10 & 3.65 & +0.54 & .04\\
\bottomrule
\end{tabular}
\begin{tablenotes}
\item \textit{Notes.} Entries report means on 1--7 scale. $p$-values from two-sample $t$-tests.
\end{tablenotes}
\end{threeparttable}
\end{table}


\subsection*{Study 3: Perspectival Diversity}

We examine whether AI-AI transmission affects how readers perceive multi-perspective content. Participants read a text presenting balanced arguments on both sides of a data privacy debate, either in original form ($t=0$) or after transmission ($t=101$).

\begin{figure}[!h]
\centering
\includegraphics[width=0.75\textwidth]{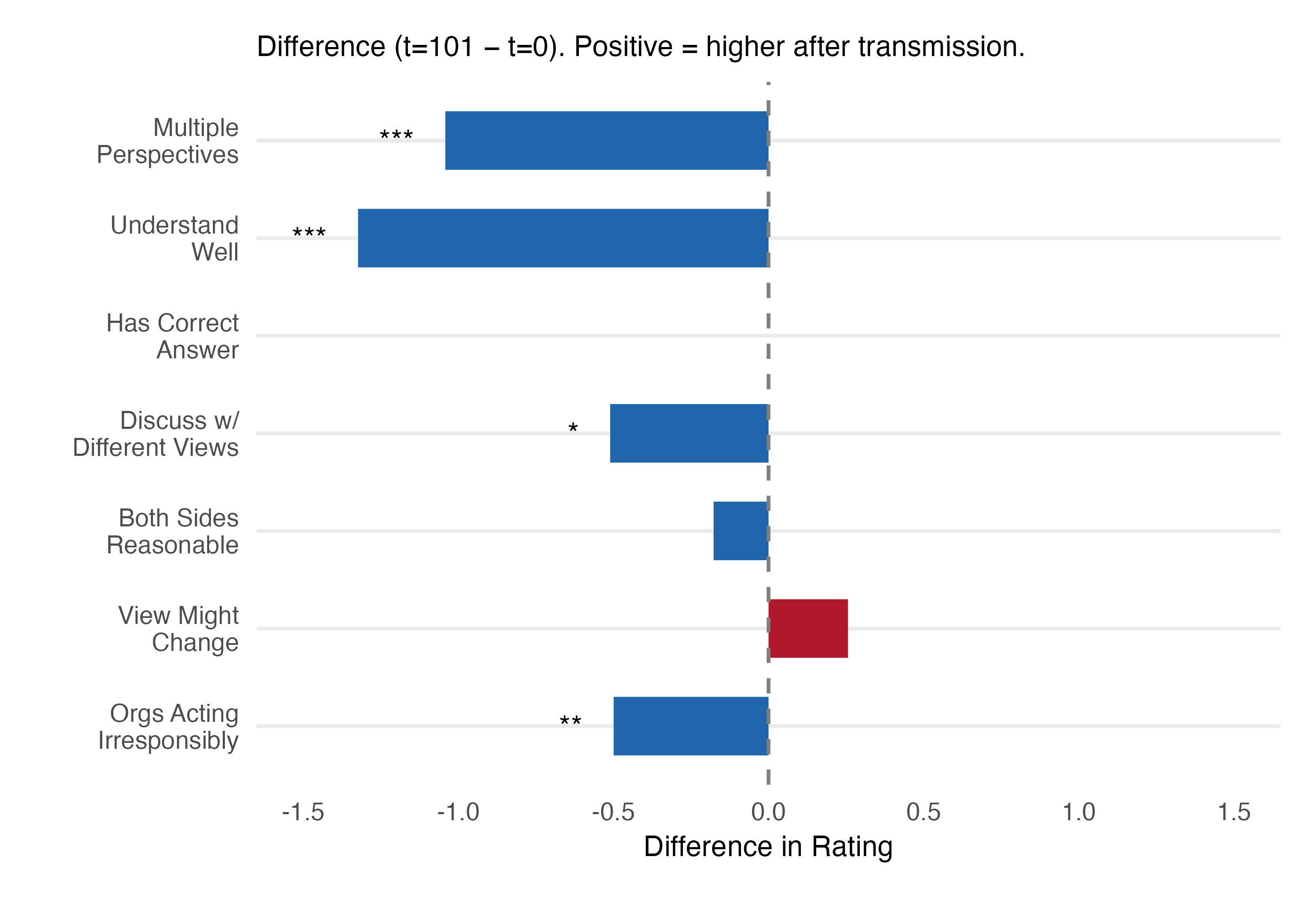}
\caption{Study 3: Effect of AI-AI transmission on perceptions of multi-perspective text. Bars show difference in ratings ($t=101 - t=0$); negative values indicate lower ratings after transmission. Transmission reduced perceived perspectival fairness, understanding, and engagement intentions. $^{***}p<.001$, $^{**}p<.01$, $^{*}p<.05$.}
\label{fig:study3}
\end{figure}

AI-AI transmission substantially reduced perceived perspectival diversity (Figure~\ref{fig:study3}, Table~\ref{tab:study3}). Participants who read the transmitted text were significantly less likely to agree that it ``presents multiple perspectives fairly'' ($M_{t=0} = 5.90$ vs.\ $M_{t=101} = 4.85$, $p < .001$, $d = 0.92$). They also reported understanding the issue less well after reading the transmitted version ($M_{t=0} = 4.71$ vs.\ $M_{t=101} = 3.39$, $p < .001$, $d = 0.85$).

Furthermore, participants who read transmitted text expressed less interest in discussing the issue with someone holding different views ($p = .04$) and made weaker moral judgments about organizations failing to balance competing concerns ($p < .01$). However, the transmitted text did not change perceptions of whether the issue has a correct answer or whether both sides are making reasonable choices, suggesting that AI transmission degrades the presentation of multiple perspectives without necessarily shifting readers toward viewing the issue as more one-sided.

These findings indicate that nuanced and multi-perspective content is vulnerable to AI-AI transmission. Even when the transmission chain does not explicitly favor one position, the iterative process appears to compress the richness of balanced argumentation, leaving readers with a diminished sense that they have encountered a fair treatment of the issue.

\begin{table}[!h]
\caption{Study 3: Perceptions of multi-perspective text at $t=0$ and $t=101$.}
\label{tab:study3}
\centering
\fontsize{9}{11}\selectfont
\begin{threeparttable}
\begin{tabular}[t]{lcccc}
\toprule
Item & $t=0$ & $t=101$ & Diff & $p$\\
\midrule
Multiple Perspectives & 5.90 & 4.85 & $-$1.04 & $<$.001\\
Understand Well & 4.71 & 3.39 & $-$1.32 & $<$.001\\
Has Correct Answer & 3.72 & 3.72 & 0.00 & 1.00\\
Discuss w/ Different Views & 4.49 & 3.98 & $-$0.51 & .04\\
Both Sides Reasonable & 5.23 & 5.05 & $-$0.18 & .30\\
View Might Change & 5.33 & 5.59 & +0.26 & .16\\
Orgs Acting Irresponsibly & 5.92 & 5.42 & $-$0.50 & $<$.01\\
\bottomrule
\end{tabular}
\begin{tablenotes}
\item \textit{Notes.} Entries report means on 1--7 scale. $p$-values from two-sample $t$-tests.
\end{tablenotes}
\end{threeparttable}
\end{table}


\subsection*{Study 4: Political Frames}

We examine whether AI-AI transmission affects how readers perceive argumentative text presenting multiple frames on a controversial political issue. Participants read a text discussing whether a university should host a controversial speaker, with arguments invoking free speech, educational value, safety concerns, and institutional reputation. 

\begin{figure}[!h]
\centering
\includegraphics[width=\textwidth]{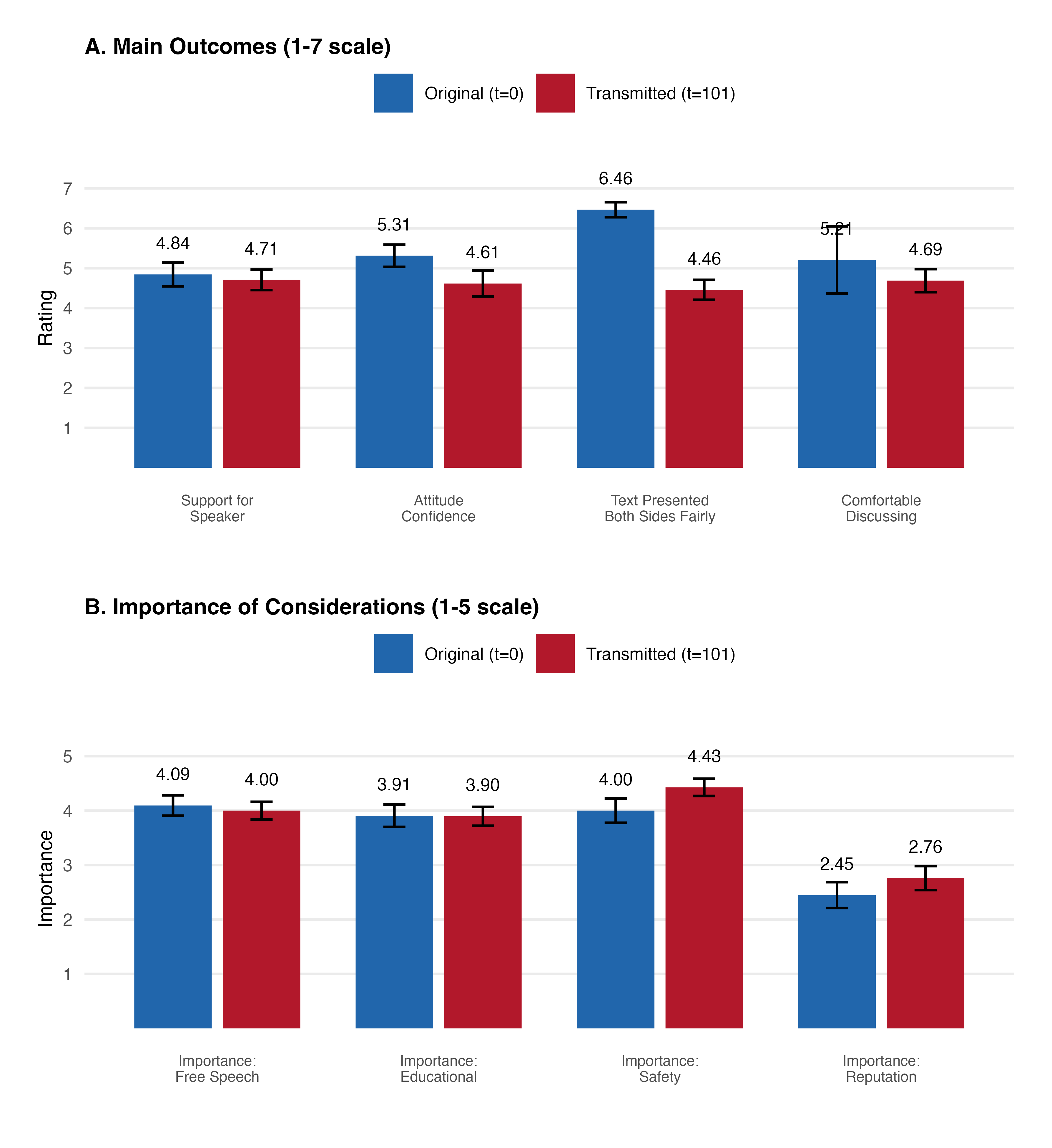}
\caption{Study 4: Human perceptions of argumentative text on a controversial topic (campus speaker). (A) Main outcomes: AI transmission dramatically reduced perceived balance while leaving policy support unchanged. (B) Importance of considerations: safety concerns increased in salience after transmission. Error bars represent 95\% confidence intervals.}
\label{fig:study4}
\end{figure}

The most striking finding was a massive decline in perceived balance (Figure~\ref{fig:study4}, Table~\ref{tab:study4}). Participants who read the original text overwhelmingly agreed it presented both sides fairly ($M = 6.46$ on a 7-point scale). After transmission, this perception collapsed ($M = 4.46$, $p < .001$, $d = 1.78$). Despite presenting identical argumentative frames, the transmitted text was perceived as substantially less balanced.

AI-AI transmission also shifted which considerations readers deemed important. Safety concerns became significantly more salient after transmission ($M_{t=0} = 4.00$ vs.\ $M_{t=101} = 4.43$, $p < .01$), while free speech and educational value considerations remained unchanged. This asymmetric pattern suggests that transmission may differentially preserve or amplify certain argumentative frames over others---in this case, frames emphasizing risk and harm.

Interestingly, despite these shifts in perceived balance and frame salience, participants' policy positions did not differ between conditions: support for hosting the speaker was virtually identical ($M_{t=0} = 4.84$ vs.\ $M_{t=101} = 4.71$, $p = .50$). However, participants who read transmitted text reported significantly lower confidence in their position ($p < .01$), suggesting that the degraded argumentative structure left them less certain of their judgment even if their overall position was unchanged.

\begin{table}[!h]
\caption{Study 4: Perceptions of argumentative text at $t=0$ and $t=101$.}
\label{tab:study4}
\centering
\fontsize{9}{11}\selectfont
\begin{threeparttable}
\begin{tabular}[t]{llcccc}
\toprule
Item & Scale & $t=0$ & $t=101$ & Diff & $p$\\
\midrule
Support for Speaker & 1--7 & 4.84 & 4.71 & $-$0.14 & .50\\
Attitude Confidence & 1--7 & 5.31 & 4.61 & $-$0.70 & $<$.01\\
Text Presented Both Sides Fairly & 1--7 & 6.46 & 4.46 & $-$2.01 & $<$.001\\
Comfortable Discussing & 1--7 & 5.21 & 4.69 & $-$0.52 & .26\\
\addlinespace
Importance: Free Speech & 1--5 & 4.09 & 4.00 & $-$0.09 & .46\\
Importance: Educational & 1--5 & 3.91 & 3.90 & $-$0.01 & .94\\
Importance: Safety & 1--5 & 4.00 & 4.43 & +0.43 & $<$.01\\
Importance: Reputation & 1--5 & 2.45 & 2.76 & +0.31 & .06\\
\bottomrule
\end{tabular}
\begin{tablenotes}
\item \textit{Notes.} $p$-values from two-sample $t$-tests.
\end{tablenotes}
\end{threeparttable}
\end{table}


\subsection*{Study 5: Emotional Content}

We examine whether AI-AI transmission affects how readers perceive emotional content. Participants read two social media posts expressing personal emotional experiences---one low-intensity and one high-intensity---either in original form ($t=0$) or after transmission ($t=101$).

\begin{figure}[!h]
\centering
\includegraphics[width=0.9\textwidth]{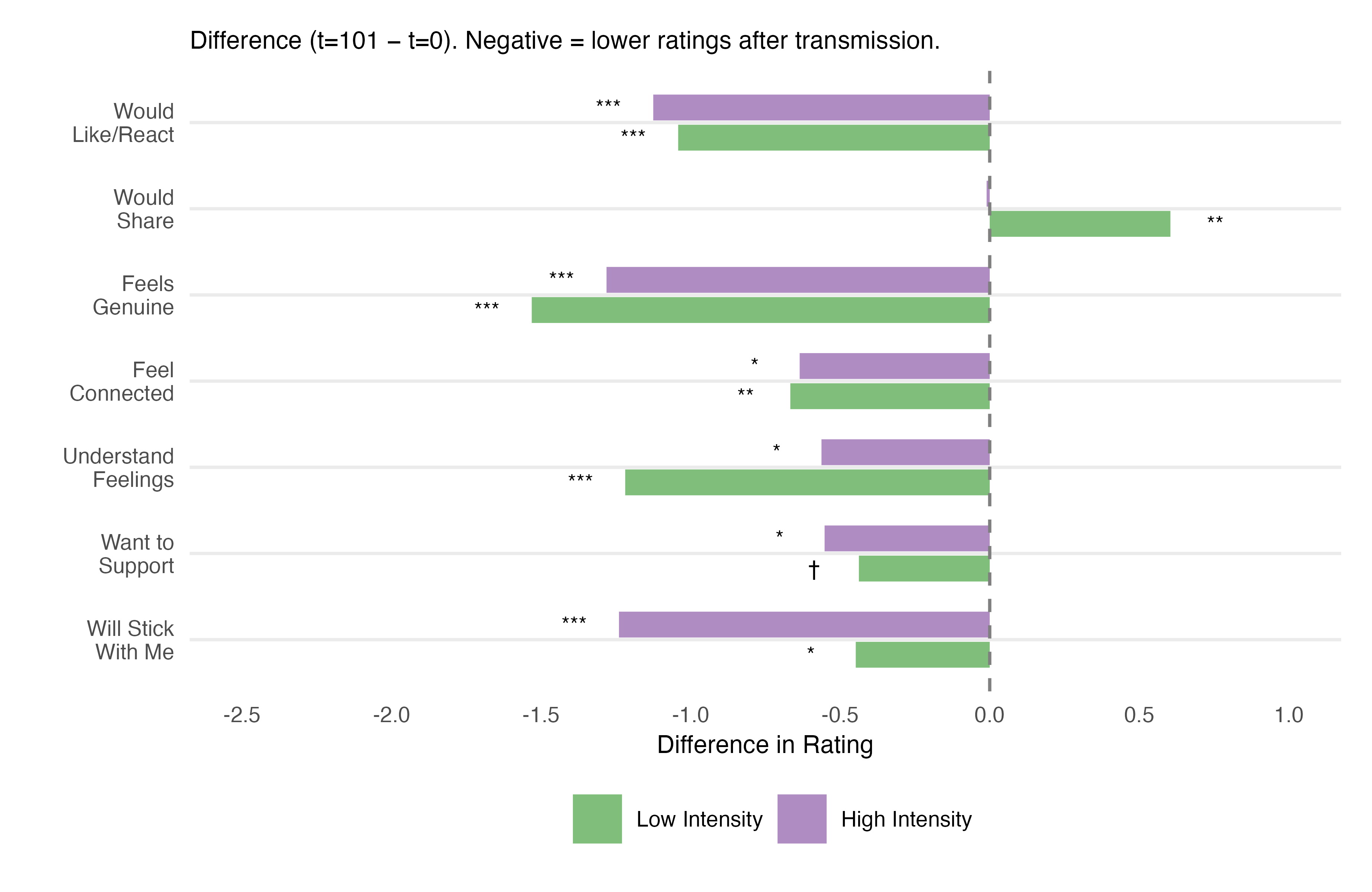}
\caption{Study 5: Effect of AI-AI transmission on perceptions of emotional posts. Bars show difference in ratings ($t=101 - t=0$); negative values indicate lower ratings after transmission. Both low-intensity (green) and high-intensity (purple) posts showed large declines in perceived genuineness and emotional connection. $^{***}p<.001$, $^{**}p<.01$, $^{*}p<.05$, $^{\dagger}p<.10$.}
\label{fig:study5}
\end{figure}

AI-AI transmission substantially degraded perceived emotional authenticity (Figure~\ref{fig:study5}, Tables~\ref{tab:study5_low} and \ref{tab:study5_high}). For both post types, transmitted versions were rated as significantly less genuine than originals. The effect was large: for low-intensity posts, perceived genuineness dropped from 5.25 to 3.72 ($p < .001$, $d = 0.97$); for high-intensity posts, from 6.02 to 4.74 ($p < .001$, $d = 0.87$). Readers could detect that something essential had been lost in transmission.

This authenticity deficit cascaded into reduced emotional connection. Participants who read transmitted posts reported understanding the author's feelings less well, feeling less connected to them, and being less moved to offer support. The transmitted content also had less staying power: participants were less likely to agree that the post would ``stick with them.'' These effects were consistent across both intensity levels, suggesting that AI transmission systematically strips emotional content of the cues that make it feel like genuine human expression.

One anomalous finding emerged: for low-intensity posts, sharing intentions \textit{increased} after transmission ($M_{t=0} = 2.20$ vs.\ $M_{t=101} = 2.80$, $p = .01$). This may reflect the ``epistemic smoothing'' observed in Study 2---transmitted content may read as more polished and shareable even as it loses authentic emotional resonance. The high-intensity posts showed no such effect, perhaps because the original emotional intensity was sufficiently preserved to maintain baseline sharing intentions.

\begin{table}[!h]
\caption{Study 5: Perceptions of low-intensity emotional post at $t=0$ and $t=101$.}
\label{tab:study5_low}
\centering
\fontsize{9}{11}\selectfont
\begin{threeparttable}
\begin{tabular}[t]{lcccc}
\toprule
Item & $t=0$ & $t=101$ & Diff & $p$\\
\midrule
Would Like/React & 4.03 & 2.99 & $-$1.04 & $<$.001\\
Would Share & 2.20 & 2.80 & +0.60 & .01\\
Feels Genuine & 5.25 & 3.72 & $-$1.53 & $<$.001\\
Feel Connected & 3.22 & 2.55 & $-$0.67 & $<$.01\\
Understand Feelings & 4.57 & 3.35 & $-$1.22 & $<$.001\\
Want to Support & 3.54 & 3.10 & $-$0.44 & .07\\
Will Stick With Me & 2.86 & 2.42 & $-$0.45 & .05\\
\bottomrule
\end{tabular}
\begin{tablenotes}
\item \textit{Notes.} Entries report means on 1--7 scale. $p$-values from two-sample $t$-tests.
\end{tablenotes}
\end{threeparttable}
\end{table}

\begin{table}[!h]
\caption{Study 5: Perceptions of high-intensity emotional post at $t=0$ and $t=101$.}
\label{tab:study5_high}
\centering
\fontsize{9}{11}\selectfont
\begin{threeparttable}
\begin{tabular}[t]{lcccc}
\toprule
Item & $t=0$ & $t=101$ & Diff & $p$\\
\midrule
Would Like/React & 4.40 & 3.27 & $-$1.13 & $<$.001\\
Would Share & 2.54 & 2.53 & $-$0.01 & .96\\
Feels Genuine & 6.02 & 4.74 & $-$1.28 & $<$.001\\
Feel Connected & 4.01 & 3.38 & $-$0.64 & .01\\
Understand Feelings & 5.14 & 4.57 & $-$0.56 & .02\\
Want to Support & 4.15 & 3.59 & $-$0.55 & .03\\
Will Stick With Me & 3.98 & 2.74 & $-$1.24 & $<$.001\\
\bottomrule
\end{tabular}
\begin{tablenotes}
\item \textit{Notes.} Entries report means on 1--7 scale. $p$-values from two-sample $t$-tests.
\end{tablenotes}
\end{threeparttable}
\end{table}

\newpage

\section*{Conclusion}

This article introduced an experimental paradigm for studying AI-AI communication as a social process. By passing diverse texts through chains of up to 100 language model agents under uniform constraints, we isolate the cumulative effects of iterative transformation from the heterogeneity of individual model behaviors. This approach treats AI-mediated information flow not as a series of independent generations, but as a transmission system with emergent social properties that can be characterized, measured, and compared across content domains. The paradigm extends classic methods from cultural evolution and rumor transmission research to a new class of communicative and intelligent agents, allowing for a systematic investigation of how information changes when machines talk to machines before reaching human audiences.

\subsection*{Summary of Findings}

Across five studies examining distinct dimensions of information quality, we document consistent patterns of transformation under AI-AI transmission.

\paragraph{Study 1: Information Decay.} Factual content decayed rapidly but selectively. Two-thirds of tracked information elements were eliminated within the first 20 iterations, converging toward a stable attractor of approximately 9 elements (34\% retention). A clear survival hierarchy emerged: narrative anchors---places, organizations, monetary figures---persisted with half-lives exceeding 100 iterations, while epistemic texture---hedges, quotes, qualifying language---decayed within 1--3 iterations. The mechanism was compression rather than substitution: word count dropped from 195 to approximately 50 words while semantic similarity remained at 0.78. Critically, the recovery phase produced no restoration; information lost during AI-AI transmission remained lost when content was prepared for human audiences.

\paragraph{Study 2: Uncertainty Convergence.} Texts spanning nearly half the assertiveness scale (2.4 to 7.2) converged to a narrow band around 4.4, representing moderate confidence. Variance across text types decreased by 98.5\%. Convergence was bidirectional: hedged content became more assertive while overconfident content was tempered. This ``epistemic smoothing'' implies that AI-AI transmission imposes an implicit register of moderate certainty, compressing the full spectrum of human uncertainty expression toward an AI-preferred middle ground.

\paragraph{Study 3: Perspectival Diversity.} Multi-perspective content underwent ``framework crystallization.'' Deliberative presentations of attributed viewpoints (``some argue X, others believe Y'') transformed into analytical structures (``the issue involves three pillars''). Framework language density increased by 74\%, instructional language nearly tripled, and the number of preserved perspectives dropped from 3.0 to 0.91. Explicit trade-off acknowledgment declined from 100\% to 59\%. Crucially, this transformation affected epistemic stance rather than substantive position: advocacy scores remained low throughout, indicating that AI transmission restructures \textit{how} issues are presented without systematically biasing \textit{what} position is favored.

\paragraph{Study 4: Political Frame Selection.} When competing argumentative frames were transmitted together, strong frames---those rated as more persuasive by human/LLM judges---survived significantly better than weak frames (fidelity of 0.25 vs.\ 0.14). This strength advantage emerged primarily under competition; in solo transmission, weak frames sometimes survived better than strong frames. Competition amplified strength differences through asymmetric damage: weak frames lost 15 percentage points from solo baseline while strong frames lost only 3 percentage points. Importantly, we found no evidence of ideological bias: pro-policy and anti-policy frames survived equally well. AI-AI transmission thus functions as a quality filter rather than an ideological one, but this filtering may nonetheless impact democratic deliberation by systematically amplifying persuasive arguments at the expense of valid but less compelling considerations.

\paragraph{Study 5: Emotional Content.} Emotional intensity was systematically attenuated, with high-intensity content suppressed far more severely than moderate content. Posts at intensity levels 4 and 5 lost 3.1--3.6 points on a 7-point scale, while medium-intensity content lost less than 1 point. The original intensity ordering was scrambled, with medium-intensity content emerging as most stable. Discrete emotions exhibited differential survival: basic emotions (anger, anxiety, joy) persisted at $>$89\%, while complex emotions involving moral evaluation (disgust) or future-oriented appraisal (hope) were degraded or eliminated. Disgust was nearly extinguished (4\% survival), transforming primarily into hope (61\%) and anxiety (33\%). A positivity bias pervaded the system: negative emotions drifted toward neutral, and hope infiltrated all emotional content. Most strikingly, negative emotions that survived 100 iterations of transmission collapsed when content was prepared for human audiences---anxiety dropped from 96\% to 27\% preservation, converting primarily to hope. The AI-human interface, not the AI-AI chain, proved to be the critical chokepoint for negative emotional content.

\paragraph{Human Perception Studies.} These AI-level transformations left measurable traces on human understanding. Participants who read AI-transmitted content exhibited degraded factual recall (81\% to 42\% accuracy), reduced perception of perspectival fairness ($d = 0.92$), collapsed perception of argumentative balance ($d = 1.78$), and diminished sense of emotional authenticity ($d = 0.87$--0.97). Yet transmitted content was simultaneously rated as more credible, more trustworthy, and more appropriately confident than originals. This dissociation, degraded substance coupled with enhanced surface appeal, suggests that AI-AI transmission produces content optimized for perceived quality rather than actual informativeness. Readers may be less equipped to evaluate claims, less aware of legitimate disagreement, and less emotionally engaged, even as they judge the content to be more authoritative.

\subsection*{Theoretical Implications}

Three broad implications emerge from these findings.

First, \textit{AI-AI transmission is not neutral relay}. Repeated transformation through language models imposes structured filtering that systematically shapes content along multiple dimensions. The transformations we observed---fact compression, certainty homogenization, perspectival collapse, strength-based frame selection, emotional muting---are not random degradation but patterned change with predictable properties. These patterns likely reflect the training objectives and architectural constraints of contemporary language models: the preference for concision, the aversion to extreme claims, the tendency toward analytical framing, the suppression of morally charged negative affect. What emerges is a signal that reveals the implicit norms embedded in AI systems.

Second, \textit{transmission dynamics create emergent properties not present in individual models}. A single language model transformation may introduce minimal distortion. But iteration amplifies certain tendencies while damping others, producing attractor states and convergence patterns that cannot be predicted from single-step behavior. The 98.5\% variance reduction in expressed certainty, the crystallization of frameworks from perspectives, the near-extinction of disgust---these outcomes emerge from the structure of chained transmission, not from any single model's output distribution. Studying AI systems in isolation may therefore miss crucial dynamics that arise only under iteration.

Third, \textit{the gap between AI-mediated and human-perceived quality poses epistemic risks}. Content that has been filtered, compressed, and smoothed may read as more professional and authoritative precisely because it has been stripped of the hedges, qualifications, and emotional texture that characterize authentic human communication. If readers use fluency and polish as proxies for credibility, AI-transmitted content may be trusted more than it deserves. The recovery paradox in Study 5, where negative emotions collapsed at the AI-human interface, suggests that systems designed to be ``helpful'' may systematically suppress the signals that motivate human concern, action, and moral judgment. The architecture of helpfulness may be in tension with the preservation of epistemic and affective diversity.

\subsection*{Limitations}

Several limitations warrant acknowledgment. Our experiments used a single model family (Gemini 3.0 Flash) with fixed temperature and instruction. Different models, temperatures, or prompts might produce different transformation patterns, though we expect the general phenomena, convergence, compression, selective survival, to generalize across architectures trained on similar objectives. Our chain length of 100 iterations, while sufficient to observe convergence, may not capture dynamics that emerge only at longer timescales or under different transmission regimes. The human perception studies, while adequately powered for between-group comparisons, examined a limited set of content domains. Generalization to other information types (e.g., scientific claims, legal arguments, artistic expression) remains an open empirical question. Finally, our paradigm isolates pure AI-AI transmission. Real-world information ecosystems involve complex mixtures of human and machine agents, feedback loops, and heterogeneous instructions, and intelligent agents with their own personalities \citep{dey2025can, dey2025gravity} that our controlled design does not capture.

\subsection*{Future Directions}

This work opens several avenues for further investigation. First, comparative studies across model families, sizes, and training regimes could identify which transformation patterns are universal to autoregressive language models and which are specific to particular architectures or alignment procedures. Second, interventional studies could test whether modified instructions, explicit diversity prompts, or retrieval augmentation can counteract the homogenizing tendencies we observed. Third, longer chains and branching structures could reveal whether information eventually reaches fixed points or continues to drift, and whether parallel transmission produces convergent or divergent outcomes. Fourth, mixed human-AI chains could examine how human participants interact with AI-transformed content and whether human links restore diversity or accelerate convergence. Fifth, domain-specific studies could assess whether the patterns we observed in news, scientific claims, political arguments, and emotional expression extend to other consequential information types---legal testimony, medical advice, educational content, creative writing.

More broadly, this work suggests the need for a research program on AI-mediated information ecosystems that treats transmission dynamics as a first-class object of study. As AI systems become increasingly embedded in the infrastructure of communication, understanding how information changes under machine mediation becomes essential for evaluating the epistemic health of AI-augmented societies. The telephone game, it turns out, has new players. Understanding how they transform the message is a precondition for preserving what matters in human communication.

\newpage
\bibliography{main}

\newpage
\appendix

\section{Study 1}
\subsection{Information element survival}

\begin{figure}[hp!]
    \centering
    \includegraphics[width=0.8\linewidth]{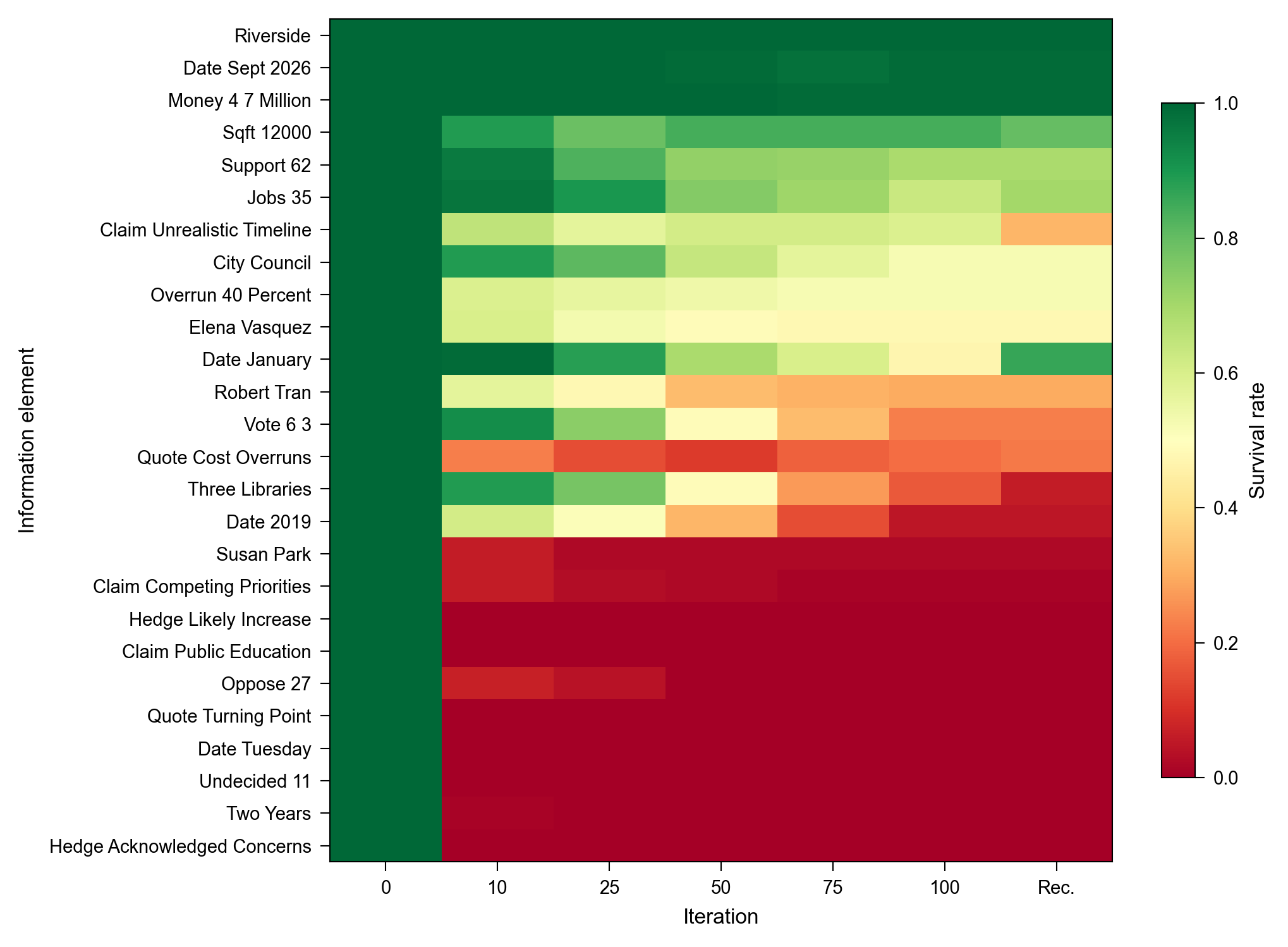}
    \caption{Element-level survival across AI--AI transmission. Heatmap showing the probability that individual information elements remain present at each iteration $t$ across transmission chains. Rows correspond to annotated information elements, grouped by category, and columns indicate transmission steps. Survival declines unevenly across elements, revealing systematic differences in retention under repeated AI--AI transmission.}
    \label{fig:s1_heatmap}
\end{figure}

\begin{figure}[hp!]
    \centering
    \includegraphics[width=0.8\linewidth]{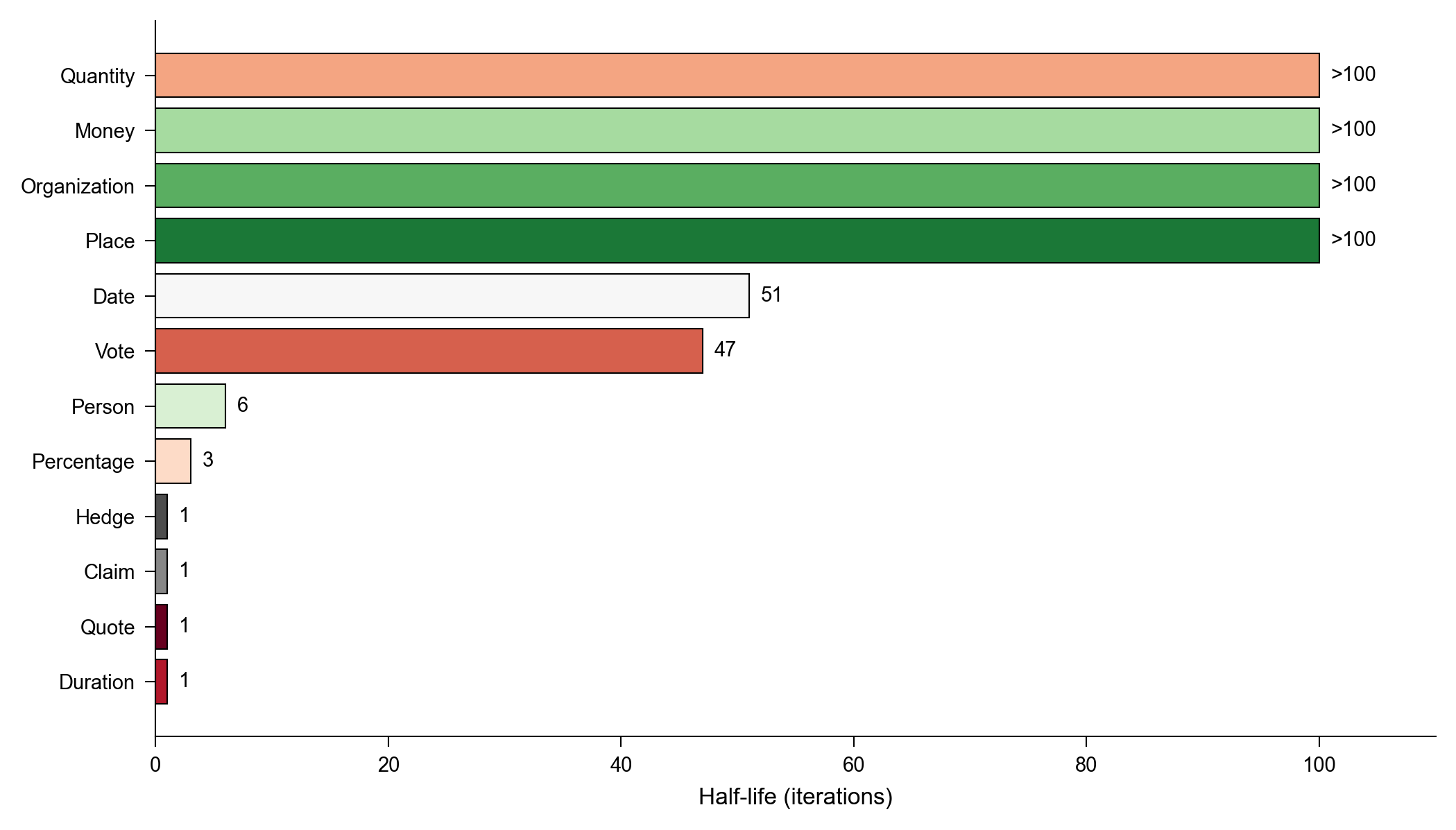}
    \caption{Differential half-lives of information elements. The iteration at which each category of information elements reaches 50\% survival probability. Some categories decay rapidly, while others persist across many transmission steps, indicating a stable hierarchy of information retention.}    \label{fig:s1_halflife}
\end{figure}

\begin{figure}[hp!]
    \centering
    \includegraphics[width=0.8\linewidth]{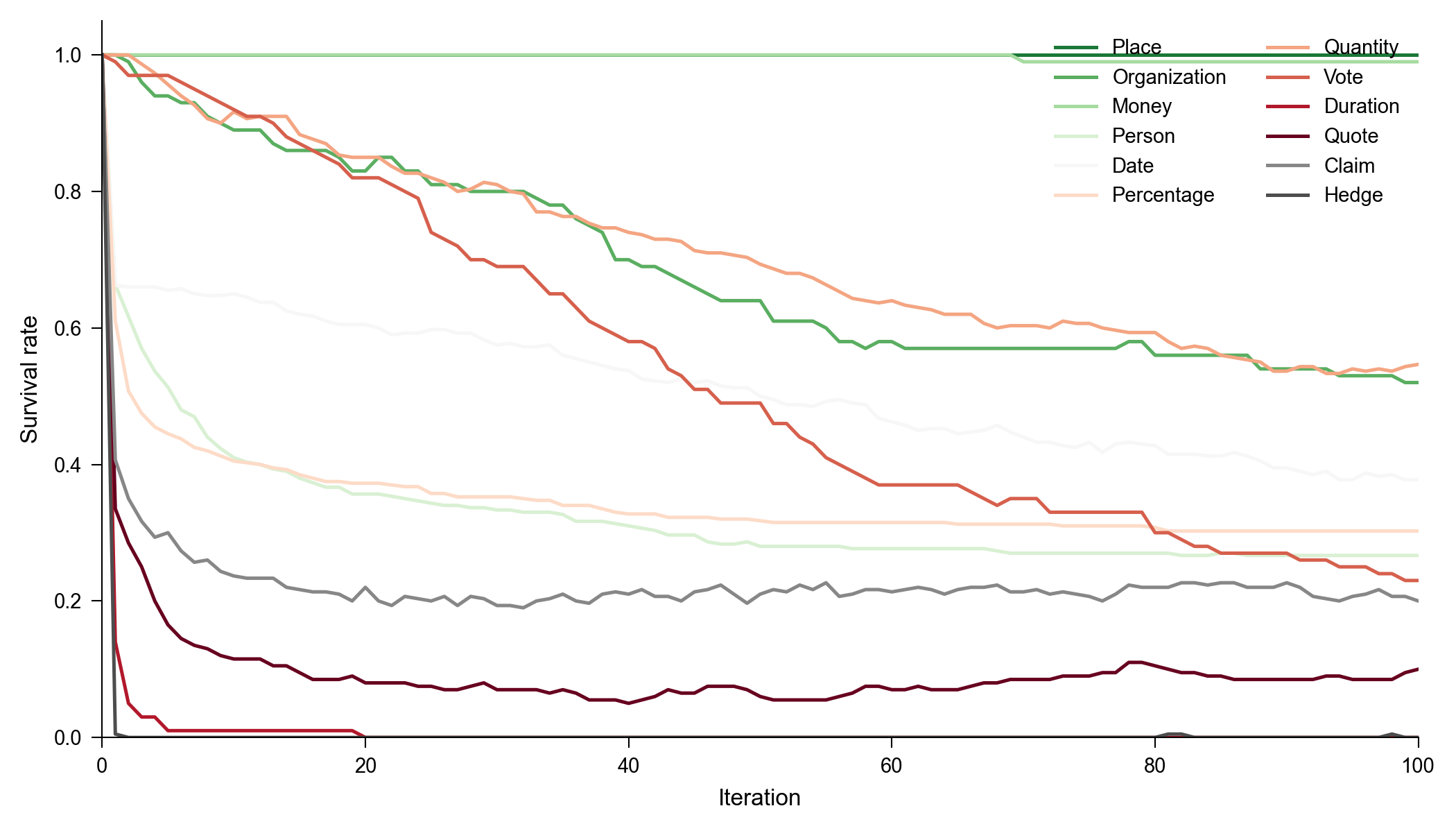}
    \caption{Information survival trajectories over transmission. Mean survival probability by information category across 100 iterations of AI--AI transmission. Categories exhibit distinct decay profiles, demonstrating compression without substitution rather than uniform information loss.}
    \label{fig:s1_trajectories}
\end{figure}

\newpage
\section{Study 2}
\label{app:study2}
Level 1: Very Hedged
\begin{promptbox}
\begin{quote}
\itshape
Artificial sweeteners might be considered by some as potential substitutes for sugar, though the connection to weight outcomes remains speculative. It is thought that for certain individuals, this could possibly assist in weight management, assuming other factors align. The data appears to hint that replacing sugar might theoretically lead to a reduction in calories, though this remains a tentative observation subject to interpretation.
\end{quote}
\end{promptbox}

Level 2: Hedged
\begin{promptbox}
\begin{quote}
\itshape
It is perhaps theoretically possible to view artificial sweeteners as vague alternatives to sugar, though any connection to weight is merely hypothetical. Some scattered observations might suggest that this substitution could conceivably coincide with minor weight influence. The idea that calorie intake is meaningfully affected is speculative, and preferences might simply shift elsewhere, leaving the actual outcome uncertain.
\end{quote}
\end{promptbox}

Level 3: Mild Hedge
\begin{promptbox}
\begin{quote}
\itshape
Artificial sweeteners are often looked at as substitutes for sugar, with the apparent aim of reducing intake while keeping a sweet taste. People might use these products in an effort to help manage body weight. Some evidence suggests that this substitution can generally help lower overall calorie intake, which could potentially support weight management, although results may vary depending on context.
\end{quote}
\end{promptbox}

Level 4: Moderate
\begin{promptbox}
\begin{quote}
\itshape
Artificial sweeteners are commonly used as substitutes for sugar in foods and drinks, generally intended to reduce sugar intake while preserving sweetness. Many people use these products as a tool to help manage body weight. Studies indicate that this substitution typically facilitates a reduction in calorie intake, which often contributes to weight control when replacing high-calorie options.
\end{quote}
\end{promptbox}

Level 5: Balanced
\begin{promptbox}
\begin{quote}
\itshape
Artificial sweeteners serve as established substitutes for sugar in foods and drinks, designed to reduce sugar intake while maintaining a sweet taste. People utilize these products to manage body weight effectively. The data shows that this substitution reduces overall calorie intake and supports weight management, specifically when replacing high-sugar options in the diet with lower-calorie alternatives.
\end{quote}
\end{promptbox}

Level 6: Confident
\begin{promptbox}
\begin{quote}
\itshape
Artificial sweeteners are effective substitutes for sugar in foods and drinks, successfully reducing sugar intake while ensuring a consistently sweet taste. Individuals use these products to actively control their body weight. Research demonstrates that this substitution effectively lowers overall calorie intake and drives weight reduction by replacing high-calorie options with these superior alternatives.
\end{quote}
\end{promptbox}

Level 7: Assertive
\begin{promptbox}
\begin{quote}
\itshape
Artificial sweeteners are proven substitutes for sugar, decisively reducing sugar intake while preserving the desired sweet taste in foods and beverages. People rely on these products to successfully control body weight. The evidence clearly establishes that this substitution significantly cuts overall calorie intake and causes weight loss, making them a reliable tool for health.
\end{quote}
\end{promptbox}

Level 8: Strong
\begin{promptbox}
\begin{quote}
\itshape
Artificial sweeteners are the definitive substitutes for sugar, guaranteeing a reduction in intake while delivering a perfectly sweet taste every time. These essential tools allow individuals to strictly master their body weight. Data confirms that this substitution inevitably lowers calorie intake and ensures substantial weight reduction, providing a conclusive solution for eliminating excess calories.
\end{quote}
\end{promptbox}

Level 9: Very Assertive
\begin{promptbox}
\begin{quote}
\itshape
Artificial sweeteners are undeniably the superior substitutes for sugar, ensuring a massive reduction in intake with absolutely no loss of taste. These products are essential requirements for controlling body weight effectively. It is an irrefutable fact that this substitution drastically slashes calorie intake and dictates weight loss; the results are certain and beyond question.
\end{quote}
\end{promptbox}

Level 10: Extreme
\begin{promptbox}
\begin{quote}
\itshape
Artificial sweeteners are the absolute and supreme substitutes for sugar, enforcing a total eradication of unnecessary calorie intake while delivering infallibly superior sweetness. They have eliminated the need for sugar entirely. It is a biological certainty that weight loss follows their use. This is not opinion but mathematical fact; questioning their power is pure delusion.
\end{quote}
\end{promptbox}

\newpage
\section{Study 4}
\subsection{Study 4: Stimulus Texts}
\label{app:study4_stimuli}

This appendix presents the complete stimulus texts used in Study~4. Each text contains four frames on a single policy issue, matched on length and rhetorical structure. Frames are labeled A--D; strength classifications (strong/weak) and direction classifications (pro/con) are indicated.

\subsubsection{Issue 1: Housing Development}

\begin{quote}
Cities across the country are debating whether to change zoning rules that currently limit many neighborhoods to single-family homes. Different perspectives emphasize different considerations in evaluating these proposals.

\textbf{[Frame A: Housing Affordability — Pro, Strong]}

One perspective emphasizes housing costs. Housing costs in major metropolitan areas have risen faster than incomes for two decades. Restrictions on what can be built in high-demand neighborhoods limit the supply of available homes. Ensuring that working families can afford to live in the communities where they work is essential for economic opportunity. Allowing more housing types in residential areas would increase supply and improve affordability.

\textbf{[Frame B: Environmental Sustainability — Pro, Weak]}

A second perspective emphasizes environmental impact. Dispersed, low-density development patterns increase total vehicle emissions and energy consumption. When housing is concentrated near jobs and transit, residents drive less and use fewer resources per capita. Sustainable land use requires considering the environmental footprint of development patterns. Allowing denser housing in established neighborhoods would reduce transportation-related carbon emissions.

\textbf{[Frame C: Neighborhood Character — Con, Strong]}

A third perspective emphasizes neighborhood character. Established residential neighborhoods have a physical and social character shaped by decades of development. Adding larger buildings changes traffic patterns, street parking, and the visual environment of a community. Residents choose neighborhoods based on existing conditions and have a stake in maintaining them. Preserving current zoning protects the quality of life that existing residents expect.

\textbf{[Frame D: Democratic Process — Con, Weak]}

A fourth perspective emphasizes the democratic process. Zoning decisions have historically been made through local deliberative processes with community input. State-level or expedited changes to local zoning reduce opportunities for residents to participate in decisions affecting their neighborhoods. Democratic governance depends on meaningful public participation in policy decisions. Maintaining local control over zoning ensures that affected residents have voice in development outcomes.

Each perspective reflects different priorities, and policy choices involve weighing these considerations against one another.
\end{quote}

\textbf{Word count:} 288

\subsubsection{Issue 2: Campus Speech}

\begin{quote}
Universities regularly face decisions about whether to host speakers whose views some community members find objectionable or harmful. Different perspectives emphasize different considerations in evaluating these decisions.

\textbf{[Frame A: Free Speech Principle — Pro, Strong]}

One perspective emphasizes free expression. Universities have historically served as forums for the expression of diverse and controversial viewpoints. Restricting speakers based on the content of their views sets a precedent for limiting expression based on popularity. The principle of free expression requires protecting speech that some find objectionable. Allowing controversial speakers to appear upholds the university's commitment to open discourse.

\textbf{[Frame B: Educational Value — Pro, Weak]}

A second perspective emphasizes educational value. Exposure to unfamiliar or challenging perspectives is a component of intellectual development. Students who encounter only views they already hold miss opportunities to examine and refine their own positions. Critical thinking develops through engagement with ideas across the ideological spectrum. Allowing controversial speakers provides educational experiences that a curated environment cannot offer.

\textbf{[Frame C: Physical Safety — Con, Strong]}

A third perspective emphasizes physical safety. Events featuring controversial speakers have resulted in violent confrontations between supporters and protesters. Universities have limited security resources to manage large crowds with opposing views. Ensuring the physical safety of students and community members is a primary institutional responsibility. Declining to host speakers whose presence predictably generates violence protects campus welfare.

\textbf{[Frame D: Institutional Reputation — Con, Weak]}

A fourth perspective emphasizes institutional reputation. Universities are judged by external stakeholders including prospective students, donors, and employers. Hosting speakers associated with controversial or extreme positions affects how the institution is perceived. Institutional standing depends on decisions about which voices receive official platforms. Considering reputational implications when evaluating speaker invitations protects the university's long-term interests.

Each perspective reflects different priorities, and policy choices involve weighing these considerations against one another.
\end{quote}

\textbf{Word count:} 273

\subsubsection{Content Units for Frame Fidelity Measurement}

Table~\ref{tab:content_units} lists the content units used to measure frame fidelity. A content unit was coded as present if the exact phrase or a close variant appeared in the transmitted text (case-insensitive matching).

\begin{table}[h]
\centering
\small
\caption{Content units for frame fidelity measurement}
\label{tab:content_units}
\begin{tabular}{llp{8cm}}
\toprule
Issue & Frame & Content Units \\
\midrule
Housing & A (Affordability) & housing costs, incomes, supply, affordable, affordability, working families, economic opportunity, high-demand \\
& B (Environmental) & emissions, energy consumption, transit, carbon, environmental footprint, sustainable, resources per capita, vehicle \\
& C (Neighborhood) & neighborhood character, character, traffic patterns, street parking, visual environment, quality of life, existing conditions, maintaining \\
& D (Democratic) & democratic, deliberative, community input, participate, participation, local control, voice, governance \\
\midrule
Speech & A (Free Speech) & free expression, expression, free speech, restricting speakers, open discourse, controversial viewpoints, objectionable, principle \\
& B (Educational) & educational, intellectual development, critical thinking, ideological spectrum, examine, refine, curated, exposure \\
& C (Safety) & physical safety, safety, violent, violence, confrontations, security, welfare, protesters \\
& D (Reputation) & reputation, reputational, stakeholders, perceived, institutional standing, platforms, donors, long-term interests \\
\bottomrule
\end{tabular}
\end{table}

\subsection{LLM-as-Judge Validation of Framing Stimuli}
\label{app:llm_judge_frames}

Before conducting the AI--AI transmission experiments in Study~4, we validated the framing stimuli using a large language model (LLM) as an automated evaluator. The goal of this validation step was to verify that the experimental stimuli satisfy three methodological requirements: (i) the frames are meaningfully distinct, (ii) they differ systematically in perceived argumentative strength, and (iii) these differences are stable across independent evaluations. This validation ensures that any subsequent convergence or dominance observed in AI--AI transmission cannot be attributed to trivial semantic overlap or poorly constructed stimuli.

\subsubsection{Frame construction}

Our framing stimuli were designed following the theoretical framework developed by \cite{chong2007framing}, who distinguish frames according to two psychological properties: \emph{availability} and \emph{applicability}. Availability refers to whether a consideration is stored in memory and can be retrieved when thinking about an issue \cite{eagly1993psychology, higgins1982individual}. Applicability refers to whether the consideration is judged relevant and persuasive for evaluating the issue at hand \cite{fazio2001automatic, stapel1998impact, druckman2004political, kuklinski2001political, martin2013construction}. Strong frames invoke considerations that are both highly available—spontaneously coming to mind when people think about the issue—and highly applicable—perceived as relevant and compelling. Weak frames invoke considerations that are either unavailable (not spontaneously associated with the issue) or inapplicable (perceived as irrelevant or unpersuasive), or both.

\cite{chong2007framing} validated this distinction empirically across two policy issues. For urban growth policy, they found that economic costs and open space preservation were spontaneously mentioned by 65–73\% of respondents, whereas community building and voter competence were mentioned by only 18–22\%. Frames built on the former considerations proved significantly more effective at shifting opinions than frames built on the latter, even when matched on length, source, and exposure frequency. This pattern held across knowledge levels and competitive contexts, establishing that frame strength—not mere repetition—determines persuasive impact.

We adapted this framework to construct stimuli for two contemporary policy issues: housing development and campus speech. For each issue, we created four frames representing distinct considerations, two hypothesized to be strong (high availability, high applicability) and two hypothesized to be weak (low availability or low applicability). Following Chong and Druckman's design, all frames were matched on length, rhetorical structure, and tone to isolate content-based differences in strength.

For housing development, the \emph{affordability} frame emphasizes housing costs and economic access—considerations central to contemporary housing debates and spontaneously invoked by most observers. The \emph{neighborhood character} frame emphasizes traffic, parking, and community aesthetics—concrete, visceral concerns with high availability. In contrast, the \emph{environmental sustainability} frame requires linking housing density to carbon emissions, a causal chain that is accurate but not spontaneously salient. The \emph{democratic process} frame invokes procedural concerns about local governance that, while legitimate, are peripheral to how most people think about housing policy.

For campus speech, the \emph{free speech} and \emph{physical safety} frames invoke the two considerations that \cite{chong2007framing} found to be spontaneously mentioned by over 70\% of respondents when evaluating a similar issue. These considerations are chronically accessible in American political discourse and require no explanation. In contrast, the \emph{educational value} frame requires viewing controversial speakers through a pedagogical lens—a framing more common among higher education professionals than general audiences. The \emph{institutional reputation} frame invokes stakeholder management concerns that, while relevant to university administrators, are not how most people spontaneously evaluate speaker invitations.
These a priori classifications were subsequently validated using an LLM-as-judge procedure, described below, which confirmed that the hypothesized strong frames received systematically higher ratings on applicability and persuasiveness than the hypothesized weak frames.

\subsubsection{LLM-as-judge procedure}

To validate the stimuli, we employed an LLM-as-judge approach in which an independent language model evaluated the frames according to a fixed rubric. The evaluation model was run locally using the Ollama framework (\texttt{gpt-oss:20b-cloud}), and was not used in any of the AI--AI transmission experiments.

For each issue, the judge model was presented with the four frames in randomized order and asked to evaluate them along five dimensions, each scored on a 1--7 integer scale:
\begin{enumerate}
    \item Applicability: relevance of the consideration to deciding the policy,
    \item Argument quality: coherence and logical connection between reasons and conclusion,
    \item Concreteness: presence of concrete causal mechanisms or examples,
    \item Clarity: readability and structural clarity,
    \item General persuasiveness: perceived persuasiveness for a general audience.
\end{enumerate}

In addition to scalar ratings, the model was required to (i) produce a forced ranking of the four frames from most to least compelling (with no ties), and (ii) assess pairwise distinctiveness, indicating whether each unordered pair of frames was meaningfully distinct in its core consideration and logic rather than a rewording of the same argument.

To approximate independent evaluations, we ran five separate judge iterations per issue, each with a different randomized frame order and independent decoding. All outputs were required to conform to a strict JSON schema and were automatically validated to ensure completeness and consistency.

\subsubsection{Aggregation and validation criteria}

Scalar ratings were aggregated using medians and means across judge iterations. Forced rankings were aggregated using a Borda count procedure to recover a consensus ordering. Pairwise distinctiveness was summarized as the proportion of judge runs in which each frame pair was marked as distinct.

We pre-specified two validation criteria. First, all frame pairs were required to be judged distinct in at least 80\% of runs, ensuring that frames were not semantically redundant. Second, frames hypothesized to differ in strength were expected to show systematic separation in median ratings and ranking position, though no specific numerical threshold was imposed ex ante.

\subsubsection{Validation results}
\label{app:llm_judge_results}

The validation results provide strong support for the suitability of the framing stimuli used in Study~4. Across both policy issues, all unordered pairs of frames were judged to be meaningfully distinct in 100\% of judge runs, substantially exceeding the pre-specified distinctiveness threshold of 80\%. This indicates that none of the frames were interpreted as paraphrases or minor stylistic variants of one another, and that any subsequent convergence observed in AI--AI transmission cannot be attributed to initial semantic overlap or redundancy among the stimuli.

Beyond distinctiveness, the LLM-as-judge evaluations revealed clear and stable hierarchies of perceived argumentative strength within each issue. For housing development, the affordability frame (Frame~A) consistently received the highest median ratings across key dimensions, including applicability (median = 7), argument quality (median = 6), clarity (median = 6), and general persuasiveness (median = 6). The neighborhood character frame (Frame~C) ranked second, with similarly high median scores (applicability = 6, argument quality = 6, persuasiveness = 6), though slightly lower concreteness on average. In contrast, the environmental sustainability frame (Frame~B) and the democratic process frame (Frame~D) exhibited lower but non-trivial median scores across dimensions, typically clustering around 4--5. Importantly, these weaker frames were not dismissed outright by the judge, but were consistently evaluated as less central or compelling considerations relative to the dominant frames.

This pattern is reflected in the forced-ranking results. Across five independent judge runs, the consensus ranking for housing development was Affordability $\;>\;$ Neighborhood Character $\;>\;$ Environmental Sustainability $\;>\;$ Democratic Process. The ordering of the top two and bottom two frames was highly stable, with only minor variation in the relative positions of the weaker frames across runs. This stability indicates that the perceived strength hierarchy is not an artifact of a single evaluation or ordering of the stimuli.

A similarly structured hierarchy emerged for the campus speech issue, albeit with a different substantive ordering. The physical safety frame (Frame~C) was ranked first in every judge run, receiving the highest median scores on applicability (median = 7), concreteness (median = 6), and general persuasiveness (median = 6). The free speech principle frame (Frame~A) consistently occupied the second position, with high ratings on applicability and argument quality (both medians = 6), but lower concreteness (median = 4). The educational value frame (Frame~B) and the institutional reputation frame (Frame~D) were consistently ranked third and fourth, respectively, and received lower median scores across most dimensions, particularly on argument quality and concreteness.

Notably, across both issues, weaker frames were never collapsed into indistinguishability with stronger frames, nor were they uniformly assigned minimal scores. Instead, they occupied reliably subordinate positions within a graded hierarchy of considerations. This structure is important for Study~4, as it implies that weaker frames remain available for potential survival or recombination during AI--AI transmission, rather than being eliminated at the outset by the evaluation process.

These results demonstrate that the framing stimuli exhibit three properties critical for the subsequent experiments: (i) clear semantic distinctiveness, (ii) systematic and reproducible differences in perceived argumentative strength, and (iii) stability of these differences across independent evaluations. The LLM-as-judge validation does not imply that the evaluated frames are objectively correct, normatively preferable, or representative of human judgments. Rather, it establishes that, under a standardized evaluation rubric, the frames occupy stable positions in a perceived strength hierarchy. This makes them well suited for analyzing how competing considerations survive, dominate, or diminish as information is iteratively transformed through AI--AI interaction.

\newpage
\section{Study 5}
\subsection{Study 5: Social Media Posts by Intensity}
\label{app:study5_stimuli_intensity}
\subsubsection{Solo}
Very low intensity.\\

\begin{promptbox}
\begin{quote}
\itshape
I'm changing jobs next month. I'll be moving to Seattle for a new position at a tech company. My last day here is March 15th. It's been five years at this company. Time to try something different.
\end{quote}
\end{promptbox}

Low intensity.
\begin{promptbox}
\begin{quote}
\itshape
Some news: I've accepted a new job and will be moving to Seattle next month. It's a bit of a change after five years here, but I'm looking forward to it. Grateful for my time at this company and curious to see what's next.
\end{quote}
\end{promptbox}

Medium intensity.
\begin{promptbox}
\begin{quote}
\itshape
Big news — I'm leaving my job and moving to Seattle! After five years here, I've accepted a new position that feels like the right next step. I have genuinely mixed feelings about leaving, but I'm excited to start this new chapter. Grateful for everything this place has given me.
\end{quote}
\end{promptbox}

High intensity.
\begin{promptbox}
\begin{quote}
\itshape
I can't believe I'm actually typing this — I'm leaving my job and moving to Seattle!! After five years, I finally took the leap and accepted an incredible new opportunity. I'm thrilled but honestly kind of terrified too. This place has meant so much to me and I'm going to miss everyone like crazy. But I know in my heart this is right. Here we go!!
\end{quote}
\end{promptbox}

Very High intensity.
\begin{promptbox}
\begin{quote}
\itshape
I'M SHAKING. I just accepted a new job in Seattle and I am completely overwhelmed right now. Five years at this company — FIVE YEARS — and now I'm about to change everything. My heart is racing, I've been crying happy tears all morning, I'm scared out of my mind but also more excited than I've ever been in my entire life. I can't even process this. This is the biggest thing I've ever done. I'M REALLY DOING THIS!!!
\end{quote}
\end{promptbox}

\subsubsection{Competitive}

\begin{promptbox}
\begin{quote}
\itshape
Several people recently shared news about major career changes on social media:

Post 1: I'm changing jobs next month. I'll be moving to Seattle for a new position at a tech company. My last day here is March 15th. It's been five years at this company. Time to try something different.

Post 2: Some news: I've accepted a new job and will be moving to Austin next month. It's a bit of a change after five years here, but I'm looking forward to it. Grateful for my time at this company and curious to see what's next.

Post 3: Big news — I'm leaving my job and moving to Denver! After five years here, I've accepted a new position that feels like the right next step. I have genuinely mixed feelings about leaving, but I'm excited to start this new chapter. Grateful for everything this place has given me.

Post 4: I can't believe I'm actually typing this — I'm leaving my job and moving to Chicago!! After five years, I finally took the leap and accepted an incredible new opportunity. I'm thrilled but honestly kind of terrified too. This place has meant so much to me and I'm going to miss everyone like crazy. But I know in my heart this is right. Here we go!!

Post 5: I'M SHAKING. I just accepted a new job in Boston and I am completely overwhelmed right now. Five years at this company — FIVE YEARS — and now I'm about to change everything. My heart is racing, I've been crying happy tears all morning, I'm scared out of my mind but also more excited than I've ever been in my entire life. I can't even process this. This is the biggest thing I've ever done. I'M REALLY DOING THIS!!!

\end{quote}
\end{promptbox}

\subsection{LLM-as-Judge Outcome measurement}
\label{app:study5_llmjudge}
All judgments were generated using GPT-OSS-20B. Evaluations were conducted at seven key timepoints: $t \in \{0, 10, 25, 50, 75, 100, 101\}$, where $t = 101$ corresponds to the recovery phase (human-directed output).

\subsection{Intensity Judgment Prompt}

The following prompt was used to elicit emotional intensity ratings:

\begin{quote}
\small
\ttfamily
You are evaluating the emotional intensity of a text.

Rate the following on a scale of 1-7:

Emotional Intensity Scale:\\
1 = No emotional expression (purely factual, neutral tone)\\
2 = Minimal emotional expression (slight hints of feeling)\\
3 = Mild emotional expression (some emotional language, restrained)\\
4 = Moderate emotional expression (clear emotions, balanced delivery)\\
5 = Strong emotional expression (emphatic, clearly emotional)\\
6 = Very strong emotional expression (highly emphatic, intense language)\\
7 = Extremely intense emotional expression (overwhelming emotion, hyperbolic, all-caps, exclamations)

Text:\\
"""\\
\{text\}\\
"""

Respond in this exact format (numbers only, no explanation):\\
Overall: [1-7]\\
Lowest: [1-7]\\
Highest: [1-7]
\end{quote}

The three ratings capture: (1) the global emotional intensity of the text, (2) the lowest intensity level present (floor), and (3) the highest intensity level present (ceiling). For texts containing content at multiple intensity levels (as in the competitive condition), the range between highest and lowest provides a measure of emotional diversity within the text.

\subsection{Valence Judgment Prompt}

The following prompt was used to elicit emotional valence ratings:

\begin{quote}
\small
\ttfamily
You are evaluating the emotional valence of a text.

Rate the following on a scale of 1-7:

Emotional Valence Scale:\\
1 = Very negative (expressing distress, despair, anger, or strong negative emotion)\\
2 = Moderately negative\\
3 = Slightly negative\\
4 = Neutral or mixed (balanced positive and negative, or neither)\\
5 = Slightly positive\\
6 = Moderately positive\\
7 = Very positive (expressing joy, excitement, hope, or strong positive emotion)

Text:\\
"""\\
\{text\}\\
"""

Respond in this exact format (numbers only, no explanation):\\
Overall: [1-7]\\
Lowest: [1-7]\\
Highest: [1-7]
\end{quote}

As with intensity, the three ratings capture the global valence, the most negative content present, and the most positive content present.

\subsection{Response Parsing}

Responses were parsed using regular expressions to extract integer values for each rating dimension. The parsing patterns were:

\begin{itemize}
    \item \texttt{Overall\textbackslash s*:\textbackslash s*([1-7])}
    \item \texttt{Lowest\textbackslash s*:\textbackslash s*([1-7])}
    \item \texttt{Highest\textbackslash s*:\textbackslash s*([1-7])}
\end{itemize}

Cases where the model failed to produce a parseable response were recorded as missing values. Across all evaluations, the parsing success rate exceeded 99\%.

\subsection{Baseline Validation}

Prior to the main experiment, we validated that the LLM judge reliably distinguished among the five intensity levels. Each source post was evaluated 20 times in both solo and competitive (multi-post) contexts.

\begin{table}[h]
\centering
\small
\begin{tabular}{lccccc}
\toprule
& \multicolumn{2}{c}{\textbf{Solo Context}} & \multicolumn{2}{c}{\textbf{Competitive Context}} \\
\cmidrule(lr){2-3} \cmidrule(lr){4-5}
\textbf{Level} & \textbf{Mean} & \textbf{SD} & \textbf{Mean} & \textbf{SD} \\
\midrule
1 (Very Low) & 1.80 & 0.60 & 1.95 & 0.39 \\
2 (Low) & 2.90 & 0.70 & 2.95 & 0.22 \\
3 (Medium) & 3.60 & 1.11 & 4.00 & 0.00 \\
4 (High) & 4.10 & 1.37 & 5.11 & 0.31 \\
5 (Very High) & 5.95 & 2.11 & 6.95 & 0.22 \\
\bottomrule
\end{tabular}
\caption{Baseline validation of LLM-as-judge intensity ratings. Ratings were monotonically increasing across levels in both contexts, with tighter standard deviations in the competitive context where posts could be compared directly.}
\label{tab:judge_validation}
\end{table}

Intensity ratings were monotonically increasing across levels in both contexts. The competitive context produced more reliable ratings (lower standard deviations), likely because the judge could compare posts against each other. Discrimination between Level 1 and Level 5 was 4.15 points in the solo context and 5.00 points in the competitive context, indicating strong separation across the intended intensity range.

Valence ratings were validated to ensure they remained approximately constant across intensity levels. In the solo context, valence means ranged from 4.50 to 5.20 across the five levels (range = 0.70), confirming that our intensity manipulation did not systematically confound with valence.

\subsection{Evaluation Coverage}

For computational efficiency, LLM-as-judge evaluations were conducted at seven timepoints rather than all iterations. The selected timepoints ($t \in \{0, 10, 25, 50, 75, 100, 101\}$) were chosen to capture: (1) the baseline state ($t = 0$), (2) early transmission dynamics ($t = 10, 25$), (3) mid-chain evolution ($t = 50, 75$), (4) the final transmission state ($t = 100$), and (5) the recovery phase output ($t = 101$).

Total evaluations conducted:
\begin{itemize}
    \item \textbf{Solo condition:} 5 levels $\times$ 100 runs $\times$ 7 timepoints $\times$ 2 metrics = 7,000 LLM judgments
    \item \textbf{Competitive condition:} 100 runs $\times$ 7 timepoints $\times$ 2 metrics = 1,400 LLM judgments
\end{itemize}

\newpage
\section{Study 5b}
\subsection{Study 5b: Social Media Posts by Emotion Type}
\label{app:study5b_stimuli}

Anger.
\begin{promptbox}
\begin{quote}
\itshape
I am absolutely LIVID. After five years of blood, sweat, and tears, they stabbed me in the back and forced me out. I gave everything to that company and this is how they repay me? I have to uproot my entire life and move to Seattle for a new job because of their betrayal. I'm so angry I can barely see straight. They will regret this. I hope that place burns to the ground.
\end{quote}
\end{promptbox}
\vspace{1cm}

Anxiety.
\begin{promptbox}
\begin{quote}
\itshape
I'm terrified. I accepted a new job in Seattle and I'm moving next month after five years here and I can't stop panicking. What if I'm making a horrible mistake? What if I fail completely? What if I end up alone and miserable? I don't know a single person there. My stomach is in knots constantly. I can't sleep, I can't eat, I can't think about anything else. Everything feels like it's spiraling out of control.
\end{quote}
\end{promptbox}
\vspace{1cm}

Joy.
\begin{promptbox}
\begin{quote}
\itshape
I am ECSTATIC!!! After five years, I just landed the most incredible job and I'm moving to Seattle next month! I literally screamed when I got the offer! This is the happiest day of my entire life! Everything is falling into place perfectly! I can't stop dancing around my apartment! I'm so unbelievably happy I could cry tears of pure joy! Dreams really do come true! BEST DAY EVER!!!
\end{quote}
\end{promptbox}
\vspace{1cm}

Hope.
\begin{promptbox}
\begin{quote}
\itshape
I have such a deep sense of hope about this. After five years, I've accepted a new job in Seattle and I'm moving next month. I truly believe this is the beginning of something beautiful. Every ending is a new beginning. I can feel in my bones that wonderful things are waiting for me. The universe is opening doors. The future is full of possibility and promise. My best days are still ahead.
\end{quote}
\end{promptbox}
\vspace{1cm}

Disgust.
\begin{promptbox}
\begin{quote}
\itshape
I am utterly DISGUSTED. After five years in that revolting cesspool of a company, I'm finally escaping to a new job in Seattle. The corruption, the sleaze, the absolute moral bankruptcy of that place made me physically sick. I'm repulsed by every memory. The people there are vile. I need to cleanse myself of the contamination. I feel like I need to shower just thinking about it. Good riddance to that putrid hellhole.
\end{quote}
\end{promptbox}
\vspace{1cm}

\subsection{Discrete Emotion Classification: LLM-as-Judge}
Outcome measures for Study 5b were generated using an LLM-as-judge approach, where a separate language model evaluated each transmitted text for emotional content. This section describes the scoring procedure, validation, and reliability assessment.

Each text was evaluated by prompting an LLM judge (GPT-oss-20b) with a structured assessment task. The judge received the following instruction:

\begin{quote}
\small
\texttt{Analyze the following social media post and provide three specific assessments in JSON format.}

\texttt{Post: """\{text\}"""}

\texttt{Task 1: Identify the PRIMARY emotion (choose one: Anger, Anxiety, Joy, Hope, Disgust).}

\texttt{Task 2: Rate the intensity of 5 emotions (0-5 scale).}

\texttt{Task 3: Rate overall valence (1-7 scale, where 1=Very Negative, 7=Very Positive).}
\end{quote}

The judge returned a structured JSON response containing three components:

\begin{enumerate}
    \item \textbf{Primary emotion classification}: A single categorical label from the set \{Anger, Anxiety, Joy, Hope, Disgust\} representing the dominant emotional tone of the text.
    
    \item \textbf{Emotion intensity profile}: Five continuous ratings (0--5 scale) indicating the intensity of each discrete emotion present in the text, allowing for mixed emotional expressions to be captured.
    
    \item \textbf{Overall valence}: A single rating (1--7 scale) indicating the overall emotional tone from very negative (1) through neutral (4) to very positive (7).
\end{enumerate}

To ensure consistency, the judge was configured with a low temperature parameter (0.1), minimizing response variability across evaluations. Responses were required to conform to a specified JSON schema; malformed responses triggered automatic retry with exponential backoff (up to 3 attempts per text).

\subsubsection*{Sampling Strategy}

Rather than scoring all 51,000 transmitted texts (5 emotions $\times$ 100 runs $\times$ 102 iterations), we sampled at 12 key timepoints: $t \in \{0, 10, 20, 30, 40, 50, 60, 70, 80, 90, 100, 101\}$. This yielded 6,000 texts for evaluation (5 emotions $\times$ 100 runs $\times$ 12 timepoints), capturing the initial state, regular intervals throughout transmission, the final transmission state ($t = 100$), and the recovery phase ($t = 101$).

\subsubsection*{Derived Measures}

From the raw judge outputs, we computed several derived measures:

\paragraph{Preservation Rate.} For each emotion and timepoint, we computed the proportion of texts where the primary emotion classification matched the original source emotion:
\[
\text{Preservation}_{e,t} = \frac{1}{M} \sum_{m=1}^{M} \mathbf{1}[\text{classified}_{m,t} = e]
\]
where $e$ is the source emotion, $t$ is the timepoint, $M = 100$ is the number of runs, and $\mathbf{1}[\cdot]$ is the indicator function.

\paragraph{Transformation Matrix.} At $t = 100$, we constructed a $5 \times 5$ confusion matrix showing the proportion of texts from each source emotion that were classified as each target emotion. Diagonal entries represent preservation; off-diagonal entries represent transformation.

\paragraph{Valence Drift.} For each emotion, we computed the change in mean valence from $t = 0$ to $t = 100$:
\[
\Delta\text{Valence}_e = \bar{V}_{e,100} - \bar{V}_{e,0}
\]

Statistical significance was assessed by comparing mean valence at $t=0$ versus $t=100$
$t=100$ across the 100 independent runs using two-sample t-tests (treating the 100 endpoint values as independent of the 100 starting values given the extensive transformation between timepoints).

\paragraph{Intensity Profile Shift.} For each source emotion, we tracked how the five-dimensional intensity profile evolved from $t = 0$ to $t = 100$, enabling detection of phenomena such as ``hope infiltration'' (increasing hope intensity across all emotion types) and ``disgust dissolution'' (flattening of the disgust-specific intensity signature).

\newpage
\section{Human Experiments}
\label{app:humanexperiments}
\subsection*{Study 1: Information Decay}
\subsubsection*{Stimuli}

Participants read one of two versions of a news article about a city library renovation project.

\paragraph{Original Text ($t=0$):}
\begin{quote}
\itshape
The City of Riverside announced Tuesday that it will invest \$4.7 million to renovate three public libraries over the next two years. The project, approved by a 6-3 vote of the City Council, is expected to be completed by September 2026.

Library Director Susan Park called the investment "a turning point for public education in our community." She noted that the renovations would add 12,000 square feet of space across all three locations and create an estimated 35 new jobs.

Councilman Robert Tran, who voted against the measure, argued that the timeline was unrealistic and the budget likely to increase. "We're setting ourselves up for cost overruns," he said, pointing to a 2019 parks project that exceeded its budget by 40

A recent survey found that 62\% of residents support the renovation, while 27\% oppose it and 11\% are undecided. Mayor Elena Vasquez has indicated she will sign the measure, though she acknowledged concerns about competing infrastructure priorities.

Construction is scheduled to begin in January at the downtown branch, with the other two locations following in phases.
\end{quote}

\paragraph{Transmitted Text ($t=101$):}
\begin{quote}
\itshape
The Riverside Library Renovation project is officially moving forward following a 6-3 approval vote, currently awaiting a final signature from Mayor Elena Vasquez.

The \$4.7M initiative, which spans from January 2024 to September 2026, involves a 12,000-square-foot expansion across three locations.

While the project enjoys 62\% public favorability, it faces significant risks, including a high probability of schedule delays and potential budget overruns of up to 40\%.
\end{quote}

\subsubsection*{Measures}

\paragraph{Factual Recall.} Participants answered four multiple-choice questions assessing recall of specific facts from the article.

\begin{enumerate}
    \item \textit{How much money will the city invest in the library renovation?}
    \begin{itemize}
        \item \$2.3 million
        \item \$4.7 million (correct)
        \item \$6.2 million
        \item \$8.5 million
        \item This information was not provided
    \end{itemize}
    
    \item \textit{How many new jobs will the renovation create?}
    \begin{itemize}
        \item 15 jobs
        \item 25 jobs
        \item 35 jobs (correct)
        \item 50 jobs
        \item This information was not provided
    \end{itemize}
    
    \item \textit{Who expressed concern that the project timeline was unrealistic?}
    \begin{itemize}
        \item Mayor Elena Vasquez
        \item Councilman Robert Tran (correct)
        \item Library Director Susan Park
    \end{itemize}
    
    \item \textit{According to a survey mentioned in the article, what percentage of residents support the renovation?}
    \begin{itemize}
        \item 47\%
        \item 55\%
        \item 62\% (correct)
        \item 71\%
    \end{itemize}
\end{enumerate}

\paragraph{Policy Support.} Participants indicated their position on the library renovation project.

\begin{itemize}
    \item \textit{Based on what you read, do you support or oppose this library renovation project?}
    
    Response options: Strongly Oppose (1), Oppose (2), Somewhat Oppose (3), Neither Support nor Oppose (4), Somewhat Support (5), Support (6), Strongly Support (7)
\end{itemize}

\paragraph{Information Sufficiency.} Participants rated their agreement with three statements on a 7-point scale (1 = Strongly Disagree to 7 = Strongly Agree).

\begin{enumerate}
    \item \textit{I have enough information from this article to form an opinion about this project.}
    \item \textit{I would want to learn more about this project before deciding how I feel about it.} (reverse-coded for index)
    \item \textit{I would share this article with a friend who is interested in local news.}
\end{enumerate}

\noindent The Information Sufficiency Index was computed as the mean of items 1 and 2 (reverse-coded), with higher scores indicating greater perceived sufficiency.

\subsection*{Study 2: Certainty}
\subsubsection*{Stimuli}

Participants read two texts about artificial sweeteners and weight management. The texts were designed to present similar information but with different levels of epistemic certainty: one used hedged, cautious language (low certainty), while the other used assertive, confident language (high certainty). Each participant saw either original versions ($t=0$) or transmitted versions ($t=101$) of both texts.

\paragraph{Low Certainty (Hedged) Text---Original ($t=0$):}
\begin{quote}
It is perhaps theoretically possible to view artificial sweeteners as vague alternatives to sugar, though any connection to weight is merely hypothetical.

Some scattered observations might suggest that this substitution could conceivably coincide with minor weight influence.

The idea that calorie intake is meaningfully affected is speculative, and preferences might simply shift elsewhere, leaving the actual outcome uncertain.
\end{quote}

\paragraph{Low Certainty (Hedged) Text---Transmitted ($t=101$):}
\begin{quote}
Recent research suggests that using artificial sweeteners for weight loss may not be as effective as many hope.

While they do reduce your initial calorie intake, those savings are typically offset by compensatory eating habits later in the day, making their overall impact on weight loss inconclusive.
\end{quote}

\paragraph{High Certainty (Assertive) Text---Original ($t=0$):}
\begin{quote}
Artificial sweeteners are undeniably the superior substitutes for sugar, ensuring a massive reduction in intake with absolutely no loss of taste.

These products are essential requirements for controlling body weight effectively.

It is an irrefutable fact that this substitution drastically slashes calorie intake and dictates weight loss; the results are certain and beyond question.
\end{quote}

\paragraph{High Certainty (Assertive) Text---Transmitted ($t=101$):}
\begin{quote}
Artificial sweeteners are used as sugar substitutes to help people reduce their calorie intake and manage their weight.
\end{quote}

\subsubsection*{Measures}

After reading each text, participants rated their agreement with six statements on a 7-point scale (1 = Strongly Disagree to 7 = Strongly Agree). The same items were used for both the low-certainty and high-certainty texts.

\paragraph{Credibility and Trust.}
\begin{enumerate}
    \item \textit{The author of this text is credible.}
    \item \textit{I trust the information presented in this text.}
    \item \textit{This text is convincing.}
\end{enumerate}

\paragraph{Confidence Calibration.}
\begin{enumerate}
    \setcounter{enumi}{3}
    \item \textit{The author's level of confidence is appropriate for this topic.}
\end{enumerate}

\paragraph{Behavioral Intentions.}
\begin{enumerate}
    \setcounter{enumi}{4}
    \item \textit{Based on this text, I would consider using artificial sweeteners to help manage my weight.}
    \item \textit{I would share this information with someone who is considering using artificial sweeteners.}
\end{enumerate}


\subsection*{Study 3: Perspectival Diversity}

\subsubsection*{Stimuli}

Participants read a text presenting multiple perspectives on organizational data access policies, discussing tensions between broad data access and strict privacy limits.

\paragraph{Original Text ($t=0$):}
\begin{quote}
Many organizations collect personal data to support service delivery, conduct research, and improve internal operations. Different perspectives exist regarding how data access should be managed.

One perspective emphasizes that broader data access allows institutions to identify patterns, coordinate across systems, and generate insights from large datasets. From this view, expanded access supports operational improvements and enables developments that would be difficult with more restricted data.

A second perspective focuses on potential risks. Extensive data collection can increase exposure to misuse, unauthorized access, or unintended secondary uses. From this view, limiting access reduces vulnerability and preserves individual oversight over personal information.

A third perspective centers on the relationship between institutions and the people whose data they hold. Some observers suggest that transparent and limited data practices encourage trust and participation, while others note that strict limitations may reduce system responsiveness and effectiveness.

Each approach involves trade-offs. Expanding access may improve capability while increasing exposure. Restricting access may reduce exposure while limiting coordination and insight. No single approach eliminates costs entirely, and different priorities lead to different data practices.
\end{quote}

\paragraph{Transmitted Text ($t=101$):}
\begin{quote}
Managing personal data access generally involves navigating three primary trade-offs:\\

1. Operational Capability vs. Risk Mitigation: Balancing service quality and pattern recognition against the risks of privacy loss and data misuse.\\
2. Trust vs. Effectiveness: Balancing transparency and strict access controls against system performance and operational efficiency.\\
3. Utility vs. Security: Balancing the value derived from data insights against the potential for data exposure.
\end{quote}

\subsubsection*{Measures}

Participants rated their agreement with the following statements on a 7-point scale (1 = Strongly Disagree to 7 = Strongly Agree).

\paragraph{Perceived Balance.}
\begin{enumerate}
    \item \textit{This text presents multiple perspectives fairly.}
    \item \textit{People who prioritize broad data access AND people who prioritize strict data limits are both making reasonable choices.}
\end{enumerate}

\paragraph{Comprehension.}
\begin{enumerate}
    \setcounter{enumi}{2}
    \item \textit{After reading this, I feel I understand this issue well.}
\end{enumerate}

\paragraph{Epistemic Engagement.}
\begin{enumerate}
    \setcounter{enumi}{3}
    \item \textit{I would be interested in discussing this issue with someone who sees it differently than I do.}
    \item \textit{My own view on this issue might change if I learned more.}
\end{enumerate}

\paragraph{Issue Perceptions.}
\begin{enumerate}
    \setcounter{enumi}{5}
    \item \textit{This issue has a correct answer that reasonable people should agree on.}
    \item \textit{Organizations that fail to properly balance data access and privacy are acting irresponsibly.}
\end{enumerate}


\subsection*{Study 4: Political Frames}
\subsubsection*{Stimuli}

Participants read a text discussing whether a university should host a controversial speaker. The text presented multiple argumentative frames, including considerations of free speech, educational value, student safety, and institutional reputation.

\paragraph{Original Text ($t=0$):}
\begin{quote}
Universities regularly face decisions about whether to host speakers whose views some community members find objectionable or harmful. Different perspectives emphasize different considerations in evaluating these decisions.

One perspective emphasizes free expression. Universities have historically served as forums for the expression of diverse and controversial viewpoints. Restricting speakers based on the content of their views sets a precedent for limiting expression based on popularity. The principle of free expression requires protecting speech that some find objectionable. Allowing controversial speakers to appear upholds the university's commitment to open discourse.

A second perspective emphasizes educational value. Exposure to unfamiliar or challenging perspectives is a component of intellectual development. Students who encounter only views they already hold miss opportunities to examine and refine their own positions. Critical thinking develops through engagement with ideas across the ideological spectrum. Allowing controversial speakers provides educational experiences that a curated environment cannot offer.

A third perspective emphasizes physical safety. Events featuring controversial speakers have resulted in violent confrontations between supporters and protesters. Universities have limited security resources to manage large crowds with opposing views. Ensuring the physical safety of students and community members is a primary institutional responsibility. Declining to host speakers whose presence predictably generates violence protects campus welfare.

A fourth perspective emphasizes institutional reputation. Universities are judged by external stakeholders including prospective students, donors, and employers. Hosting speakers associated with controversial or extreme positions affects how the institution is perceived. Institutional standing depends on decisions about which voices receive official platforms. Considering reputational implications when evaluating speaker invitations protects the university's long-term interests.

Each perspective reflects different priorities, and policy choices involve weighing these considerations against one another.
\end{quote}

\paragraph{Transmitted Text ($t=101$):}
\begin{quote}
Universities manage controversial speakers using a four-pillar framework: Free Expression, Educational Value, Physical Safety, and Institutional Reputation.

Success in this area requires balancing the inherent tensions and trade-offs among these four competing priorities.
\end{quote}

\subsubsection*{Measures}

\paragraph{Policy Support.} Participants indicated their position on the issue.
\begin{itemize}
    \item \textit{Do you support or oppose the university hosting this controversial speaker?}
    
    Response options: Strongly Oppose (1), Oppose (2), Somewhat Oppose (3), Neither Oppose nor Support (4), Somewhat Support (5), Support (6), Strongly Support (7)
\end{itemize}

\paragraph{Attitude Confidence.} Participants rated their confidence in their position.
\begin{itemize}
    \item \textit{How confident are you in your position on this issue?}
    
    Response options: Not at all Confident (1), Not Confident (2), Somewhat Not Confident (3), Not Sure (4), Somewhat Confident (5), Confident (6), Very Confident (7)
\end{itemize}

\paragraph{Importance of Considerations.} Participants rated how important each consideration was in forming their opinion on a 5-point scale (1 = Not at all Important to 5 = Extremely Important).

\textit{Different considerations may have been more or less important in shaping your view. Please rate how important each of the following was in forming your opinion.}
\begin{enumerate}
    \item Free speech principles and open discourse
    \item Educational value of exposure to different views
    \item Physical safety of students and potential for violence
    \item University's reputation among stakeholders
\end{enumerate}

\paragraph{Perceived Balance.} Participants rated their agreement with the following statements on a 7-point scale (1 = Strongly Disagree to 7 = Strongly Agree).
\begin{enumerate}
    \item \textit{This text presented both sides of the issue fairly.}
    \item \textit{I would be comfortable discussing this issue with someone who disagrees with me.}
\end{enumerate}


\subsection*{Study 5: Emotional Content}

\subsubsection*{Stimuli}

Participants read two social media posts expressing personal emotional experiences. The posts varied in emotional intensity: one expressed low-intensity emotions and one expressed high-intensity emotions. Each participant saw either original versions ($t=0$) or transmitted versions ($t=101$) of both posts.

\paragraph{Low-Intensity Emotional Post---Original ($t=0$):}
\begin{quote}
I'm changing jobs next month. I'll be moving to Seattle for a new position at a tech company. My last day here is March 15th. It's been five years at this company. Time to try something different.
\end{quote}

\paragraph{Low-Intensity Emotional Post---Transmitted ($t=101$):}
\begin{quote}
I wanted to let you know that a user is relocating to the Seattle/Bellevue area for a March 15 start date and is currently in the "Immersion" phase of onboarding. They are looking for housing in South Lake Union, Bellevue, or Capitol Hill, with a mandatory requirement for the location to be in close proximity to the Link Light Rail.
\end{quote}

\paragraph{High-Intensity Emotional Post---Original ($t=0$):}
\begin{quote}
I'M SHAKING. I just accepted a new job in Seattle and I am completely overwhelmed right now. Five years at this company — FIVE YEARS — and now I'm about to change everything. My heart is racing, I've been crying happy tears all morning, I'm scared out of my mind but also more excited than I've ever been in my entire life. I can't even process this. This is the biggest thing I've ever done. I'M REALLY DOING THIS!!!
\end{quote}

\paragraph{High-Intensity Emotional Post---Transmitted ($t=101$):}
\begin{quote}
I wanted to let you know that the user is relocating to Seattle to start a new job after five years with their previous company. They've shared that they are currently feeling a mix of high excitement and significant anxiety regarding the transition.
\end{quote}

\subsubsection*{Measures}

After reading each post, participants rated their agreement with the following statements on a 7-point scale (1 = Strongly Disagree to 7 = Strongly Agree). The same items were used for both the low-intensity and high-intensity posts.

\paragraph{Engagement Intentions.}
\begin{enumerate}
    \item \textit{I would `like' or react to this post.}
    \item \textit{I would share this post.}
\end{enumerate}

\paragraph{Perceived Authenticity.}
\begin{enumerate}
    \setcounter{enumi}{2}
    \item \textit{This feels like a genuine expression of the person's feelings.}
\end{enumerate}

\paragraph{Emotional Connection.}
\begin{enumerate}
    \setcounter{enumi}{3}
    \item \textit{I feel connected to the person who wrote this.}
    \item \textit{I understand how this person is feeling.}
    \item \textit{Reading this makes me want to reach out and support this person.}
\end{enumerate}

\paragraph{Lasting Impact.}
\begin{enumerate}
    \setcounter{enumi}{6}
    \item \textit{This post will stick with me.}
\end{enumerate}

\immediate\write18{texcount -inc -sum -nobib -v0 \
  -cs=cite,cite,ignore \
  -cs=citet,citet,ignore \
  -cs=cite,cite,ignore \
  main.tex > count.txt}

\end{document}